\documentclass[12pt]{article}
\usepackage{amsmath,amssymb,amsthm,amscd,dsfont}
\usepackage{epsfig}
\usepackage[usenames]{color}
\newcommand{\Real}{\mathds{R}\mathrm{e}}

\newcommand{\e}{{\rm e}}

\newcommand{\bea}{\begin{eqnarray}}
\newcommand{\eea}{\end{eqnarray}}
\newcommand{\beq}{\begin{equation}}
\newcommand{\eeq}{\end{equation}}

\newcommand{\be}{\begin{equation}}
\newcommand{\ee}{\end{equation}}
\newcommand{\ba}{\begin{aligned}}
\newcommand{\ea}{\end{aligned}}
\newcommand{\ben}{\begin{eqnarray}\displaystyle}
\newcommand{\een}{\end{eqnarray}}

\newcommand{\sectiono}[1]{\section{#1}\setcounter{equation}{0}}

\newdimen\tableauside\tableauside=1.0ex
\newdimen\tableaurule\tableaurule=0.4pt
\newdimen\tableaustep
\def\phantomhrule#1{\hbox{\vbox to0pt{\hrule height\tableaurule width#1\vss}}}
\def\phantomvrule#1{\vbox{\hbox to0pt{\vrule width\tableaurule height#1\hss}}}
\def\sqr{\vbox{%
  \phantomhrule\tableaustep
  \hbox{\phantomvrule\tableaustep\kern\tableaustep\phantomvrule\tableaustep}%
  \hbox{\vbox{\phantomhrule\tableauside}\kern-\tableaurule}}}
\def\squares#1{\hbox{\count0=#1\noindent\loop\sqr
  \advance\count0 by-1 \ifnum\count0>0\repeat}}
\def\tableau#1{\vcenter{\offinterlineskip
  \tableaustep=\tableauside\advance\tableaustep by-\tableaurule
  \kern\normallineskip\hbox
    {\kern\normallineskip\vbox
      {\gettableau#1 0 }%
     \kern\normallineskip\kern\tableaurule}%
  \kern\normallineskip\kern\tableaurule}}
\def\gettableau#1{\ifnum#1=0\let\next=\null\else
\squares{#1}\let\next=\gettableau\fi\next}

\begin{document}
\begin{titlepage}


\vskip .6cm

\centerline{\Large \bf Thermodynamics of theories with sixteen
supercharges}

\vspace*{1.0ex}

\centerline{\Large \bf in non-trivial vacua}



\medskip

\vspace*{4.0ex}

\centerline{\rm Gianluca Grignani$^{a}$, Luca Griguolo$^{b}$,
Nicola Mori$^{c}$ and Domenico Seminara$^{c}$ }

\vspace*{4.0ex}

\centerline{ \rm $^a$ \it Dipartimento di Fisica and Sezione
I.N.F.N., Universit\`a di Perugia,}\centerline{\it Via A. Pascoli
I-06123, Perugia, Italia} \centerline{\tt grignani@pg.infn.it}

\vspace*{1.4ex}

\centerline{ \rm $^b$ \it Dipartimento di  Fisica, Universit\`a  di
Parma, INFN Gruppo Collegato di Parma,} \centerline{\rm\it  Parco
Area delle Scienze 7/A, 43100 Parma, Italy} \centerline{\tt
griguolo@fis.unipr.it}

\vspace*{1.4ex}

\centerline{ \rm $^c$ \it Dipartimento di Fisica, Polo Scientifico
Universit\`a di Firenze,} \centerline{\rm\it  INFN Sezione di
Firenze Via  G. Sansone 1, 50019 Sesto Fiorentino, Italy}
\centerline{\tt mori@fi.infn.it, seminara@fi.infn.it}

\vspace*{3ex} \centerline{\bf Abstract} \vspace*{3ex} We study the
thermodynamics of maximally supersymmetric $U(N)$ Yang-Mills theory
on $\mathds{R}\times S^2$ at large $N$. The model arises as a
consistent truncation of ${\cal N}=4$ super Yang-Mills on
$\mathds{R}\times S^3$ and as the continuum limit of the plane-wave
matrix model expanded around the $N$ spherical membrane vacuum. The
theory has an infinite number of classical BPS vacua, labeled by a
set of monopole numbers, described by dual supergravity solutions.
We first derive the Lagrangian and its supersymmetry transformations
as a deformation of the usual dimensional reduction of ${\cal N}=1$
gauge theory in ten dimensions. Then we compute the partition
function in the zero 't Hooft coupling limit in different monopole
backgrounds and with chemical potentials for the $R$-charges. In the
trivial vacuum we observe a first-order Hagedorn transition
separating a phase in which the Polyakov loop has vanishing
expectation value from a regime in which this order parameter is
non-zero, in analogy with the four-dimensional case. The picture
changes in the monopole vacua due to the structure of the fermionic
effective action. Depending on the regularization procedure used in
the path integral, we obtain two completely different behaviors,
triggered by the absence or the appearance of a Chern-Simons term.
In the first case we still observe a first-order phase transition,
with Hagedorn temperature depending on the monopole charges. In the
latter the large $N$ behavior is obtained by solving a unitary
multi-matrix model with a peculiar logarithmic potential, the system
does not present a phase transition and it always appears in a
``deconfined'' phase.

\vfill

\end{titlepage}

\newpage

\tableofcontents

\sectiono{Introduction}
In the context of the AdS/CFT correspondence
\cite{Maldacena:1997re,Gubser:1998bc,Witten:1998qj,Itzhaki:1998dd}
an interconnected family of theories with sixteen supercharges has
been recently studied \cite{Lin:2005nh}. They all have a mass gap
and a discrete spectrum of excitations. These theories can be
obtained from consistent truncations of ${\cal N}=4$ super
Yang-Mills on $\mathds{R}\times S^3$ and have many BPS vacua.
Remarkably, smooth gravity solutions corresponding to all these
vacua can be described rather explicitly. At large 't Hooft
coupling some properties of the dual string theory have also been
examined according to the pioneering proposal
of~\cite{Lin:2004nb}.

From the gauge theoretical point of view it seems particulary appealing to investigate the properties
of one specific theory belonging to this class, namely the maximally supersymmetric $U(N)$ Yang-Mills
theory on $\mathds{R}\times S^2$. This theory already appeared in~\cite{Maldacena:2002rb} where it
arises from the fuzzy sphere vacuum (membrane vacuum) of the plane-wave matrix model by taking a
large $N$ limit that removes the fuzzyness. The model can also be constructed from the familiar
${\cal N}=4$ SYM theory by truncating the free-field spectrum on $\mathds{R}\times S^3$ to states
that are invariant under $U(1)_L\subset SU(2)_L$, where $SU(2)_L$ is one of the $SU(2)$ factors in
the $SO(4)$ rotation group of the three-sphere. Geometrically this corresponds to a dimensional
reduction of the four-dimensional supersymmetric theory along the $U(1)$ fiber of $S^3$ seen as an
Hopf fibration over $S^2$. The resulting model lives in one dimension less and maintains
supersymmetry through a rather interesting mechanism. The particular dimensional reduction breaks the
natural $SO(7)$ $R$-charge symmetry to $SO(6)$, singling out one of the seven scalars of the
maximally supersymmetric Yang-Mills theory, which then behaves differently from the others. It
combines with the gauge fields to form a peculiar Chern-Simons-like term that is crucial to preserve
the sixteen supercharges, balancing the appearance of mass terms for fermions and scalars. The BPS
vacua are generated by the same term that allows to combine the field strength and the scalar into a
perfect square whose zero-energy configurations are determined by $N$ integers $n_1,...,n_N$
associated to monopole numbers on the sphere.

The model represents an interesting example of a supersymmetric non-conformal gauge theory, with
smooth gravitational dual and non-trivial vacuum structure, defined on a compact space. The last
feature is particulary appealing in the study of the thermal properties of the theory. Recently the
thermodynamics of large $N$ theories on compact spaces has attracted much attention. On compact
spaces the Gauss's law restricts physical states to gauge singlets. Consequently, even at weak 't
Hooft coupling the theories are in a confining phase at low temperature and undergo a deconfinement
transition at a critical temperature. For example, the partition function of ${\mathcal N}=4$ super
Yang-Mills theory on $\mathds{R}\times S^3$ was computed at large $N$ and small coupling
in~\cite{Sundborg:1999ue,Aharony:2003sx,Spradlin:2004pp}. It was shown that the free energy is of
order ${\cal O}(1)$ at low temperature and of order ${\cal O}(N^2)$ above a critical temperature. At
strictly zero 't~Hooft coupling the transition is a first-order Hagedorn-like transition. At small
coupling a first or a second order transition is expected, depending on the particular matter content
of the theory. The computation in the $\mathcal{N}=4$ maximally supersymmetric case has never been
performed but in~\cite{Kim:2003rz}
 it was argued that the maximally supersymmetric plane-wave deformation of Matrix theory and
$\mathcal{N}=4$ SYM should show similar behavior, including thermodynamics. The plane wave matrix
model is a theory with sixteen supercharges and it was argued in~\cite{Maldacena:2002rb} to be dual
to a little string theory compactified on $S^5$. For a small sphere, this theory is weakly coupled
and one may study the little string theory thermodynamics rather explicitly~\cite{Furuuchi:2003sy}.
The phase transition for this model was shown to remain first order in~\cite{Hadizadeh:2004bf}
indicating that this might also be the case for $\mathcal{N}=4$ SYM. This was shown by computing the
relevant parts of the effective potential for the Polyakov loop operator to three loop
order~\cite{Hadizadeh:2004bf}. With the same procedure it was shown in~\cite{Aharony:2005bq} that
also for pure Yang-Mills the phase transition remains first-order up to three loops. The phase
transition at weak coupling is basically driven by a Hagedorn-like behavior of the spectrum in the
confining phase, suggesting a possible relationship with the dual description of large $N$ gauge
theories in terms of strings. For ${\cal N}=4$ the relevant string theory lives on an asymptotic AdS
space and, at large 't~Hooft coupling, the deconfinement phase transition corresponds to a
Hawking-Page transition \cite{Hawking:1982dh,Witten:1998zw}. The thermal AdS space dominates at low
temperature and the AdS-Schwarzschild black hole is the relevant saddle-point in the high-temperature
regime. The original proposal presented in~\cite{Sundborg:1999ue,Aharony:2003sx} to connect the phase
transitions at small coupling on compact spaces with the gravitational/stringy physics stimulated a
large number of investigations. Lower-dimensional theories on tori were examined in
\cite{Aharony:2004ig,Aharony:2005ew}, while the inclusion of chemical potentials for the $R$-charges
was discussed in \cite{Yamada:2006rx,Harmark:2006di} and, more recently, pure Yang-Mills theory on
$S^2$ \cite{Papadodimas:2006jd} was found to have a second order phase transition at small 't Hooft
coupling.

In this paper we study the thermodynamics of ${\cal N}=8$ super Yang-Mills theory on
$\mathds{R}\times S^2$. We first derive the Lagrangian and its supersymmetry transformations as a
deformation of the usual dimensional reduction of ${\cal N}=1$ gauge theory in ten dimensions.
Actually our procedure will generate a larger class of three-dimensional theory: according to the
particular choice of the generalized Killing spinor equation we obtain also theories on AdS$_3$ with
peculiar Chern-Simons couplings. Then we compute the ${\cal N}=8$ partition function in the zero 't
Hooft coupling limit, for different monopole vacua. In the trivial vacuum we observe a first-order
Hagedorn transition separating a phase in which the Polyakov loop has vanishing expectation value
from a regime in which this order parameter is non-zero, in complete analogy with the
four-dimensional case. The Hagedorn temperature is also obtained in the presence of chemical
potentials for the $R$-charges. Discussions on the dual gravitational picture~\cite{Lin:2005nh} and
the possibility of matching the gauge theory Hagedorn transition with a stringy Hagedorn transition,
by exploiting for example a decoupling limit as
in~\cite{Harmark:2006di,Harmark:2006ta,Harmark:2006ie,Harmark:2007px} postponed to a forthcoming
investigation.

The situation is very different in the non-trivial monopole vacua.
The original $U(N)$ gauge group is broken to a direct product
$U(N_1)\times U(N_2)\times..U(N_k)$ and the constituent fields
transform, in general, under bifundamental representations of
$U(N_I)\times U(N_J)$. Because of the Gauss's law on a compact
manifold, however, the only allowed excitations are $SU(N_I)\times
SU(N_J)$ singlets. Different selection rules are instead possible
for the $U(1)$ charges in three dimensions, depending on the
definition of the fermionic Fock vacuum in the presence of
background monopoles \cite{Borokhov:2002ib}. The appearance of
fermionic zero-modes makes possible, in general, to assign a
non-trivial charge to the Fock vacuum, as clearly explained in
\cite{Jackiw:1975fn}. In the path-integral formalism this
corresponds to precise choices in regularizing fermionic functional
determinants which might produce Chern-Simons terms in the effective
action. In our case the different possibilities are clearly
manifested in the matrix model describing the partition function. We
recall that, in the trivial vacuum, the thermal partition function
is reduced to an integral over a single $U(N)$ matrix
\cite{Sundborg:1999ue,Aharony:2003sx}
\begin{equation}
\mathcal{Z}(\beta)=\int[dU]\exp\Bigl[-S_{eff}(U)\Bigr]
\end{equation}
where $U=e^{i\beta\alpha}$ ($\alpha$ is the zero mode of the gauge
field $A_0$ on $S^2\times S^1$ and $\beta=1/T$ the inverse of the
temperature). In the non-trivial monopole vacuum
$\mathcal{Z}(\beta)$ is given instead by a multi-matrix model over a
set of unitary matrices $U_I(N_I)$, $i=1,2,..k$, reflecting the
breaking of the $U(N)$ gauge group. More importantly the effective
action $S_{eff}(U_I)$, at zero 't Hooft coupling, can be modified by
the presence of logarithmic terms $NQ_I \mathrm{Tr}\log(U_I)$ that
implement selection rules on the $U(1)$ charges. The large $N$
analysis is highly affected by these new interactions: they
contribute at order $N^2$ and can drive the relevant saddle-point
always at a non-zero value of the Polyakov loop. Unitary matrix
model of the kind we encountered in our analysis have been
previously considered in the eighties
\cite{Green:1981mx,Rossi:1982vw}, but with an important difference:
in those studies the coefficient weighting the logarithmic term Tr
$\log(U)$ in the action was taken independent on $N$. Conversely the
large $N$ saddle-points were not modified by its presence, being
determined by the rest of the action. In our case, instead, we have
to cope with a linear dependence on $N$ and we cannot simply borrow
those results. We have therefore performed an entirely new large $N$
analysis of these kind of models, starting from an exact
differential equation of the Painlev\'e type that describes the
finite $N$ partition function \cite{FW2002}.

The paper is organized as follows. In section~2 we construct the
supersymmetric Yang-Mills theory on $\mathds{R}\times S^2$ using a
different strategy with respect to~\cite{Lin:2005nh}
and~\cite{Maldacena:2002rb} (see also \cite{Ishiki:2006rt} for a
careful derivation of the Hopf reduction and \cite{Ishii:2007ex} for
an extension to more general fiber bundles). We start from ${\cal
N}=1$ super Yang-Mills theory in ten dimensions and consider its
dimensional reduction on $\mathds{R}\times S^2$. We find the
relevant Killing spinors that generate the rigid supersymmetry,
generalizing to our case the approach developed
in~\cite{Blau:2000xg}. We further determine the deformations of the
original ten dimensional Lagrangian and of the supersymmetry
transformations ensuring the global invariance of the action.
Interestingly, using the same strategy it is possible to construct
two other maximally supersymmetric gauge theories on
three-dimensional curved spacetimes, living both on AdS$_3$ and
differing from the theory introduced in~\cite{Maldacena:2002rb} in
the structure of the Chern-Simons terms. In section~3 we briefly
examine the BPS vacua of the model, we comment on their
gravitational description and the related instanton solutions.

We then turn to study the thermodynamics at zero 't Hooft coupling. Following the analysis
in~\cite{Sundborg:1999ue,Aharony:2003sx}, we obtain the partition function of the theory in a generic
vacuum, in terms of matrix integrals. In section~4 we present the results of the relevant functional
determinants in the background of a gauge flat-connection and of a monopole potential, recovering the
appropriate single-particle partition functions for scalars, spinors and vectors. Careful
$\zeta$-function evaluations are deferred to the appendices. We discuss the emerging, on the monopole
background, of new logarithmic terms in the effective action, directly related, in this formalism, to
the appearance of fermionic zero-modes. We explain their dependence on the regularization procedure
and remark their interplay with a typical three-dimensional phenomenon, the induction of Chern-Simons
terms. We interpret their effect as a part of the projection into singlets of the gauge group, as
required by the Gauss's~law. Section~5 is devoted to discuss the large $N$ thermodynamics in the
trivial vacuum. We determine the critical temperature at which the first-order phase transition takes
place and we generalize the result to the case of non vanishing chemical potentials for the
$R$-charges. Finally, in sections~\ref{Thermobps1} and \ref{Thermobps2}, we study the large $N$
theory on the non-trivial monopole backgrounds: we consider a large class of vacua, characterized by
the set of integers $n_1,..,n_k$ and large $N$ degeneracies $N_1,..,N_k$. According to the discussion
of section 4, we study two different choices for the logarithmic terms, within our regularization
procedure. First, in section \ref{Thermobps1}, we discuss the ``uncharged" case, that amounts to make
a particular choice of branch cuts, in the $\zeta$-function regularization procedure
\cite{Deser:1997nv,Deser:1997gp}, that cancels the Chern-Simons like contributions. In turn we get a
non-vanishing Casimir energy, depending explicitly on the monopole background. The resulting unitary
multi-matrix model is an obvious generalization of the trivial case. We find again a first-order
phase transition, with an Hagedorn temperature explicitly depending on the monopole numbers. We
discuss also some particular class of vacua, characterized by large monopole charges, whose Hagedorn
temperature approaches the one of the theory on $S^3/\mathbb{Z}_k$ in trivial vacuum. In
section~\ref{Thermobps2} we discuss the opposite situation of a ``maximally" charged fermionic
vacuum: we have a non-trivial modification of the unitary multi-matrix model due to appearance of the
new logarithmic terms and vanishing Casimir energy. For the sake of clarity we will restrict our
discussion to a particular simple background $(n,n,..,n,-n,-n..,-n)$. We show the existence of a
non-trivial saddle-point for the effective action for a wide range of temperatures starting from
zero, within the assumption that we can disregard higher windings contributions in this regime. This
implies that the theory is always in a ``deconfined" phase. We have to face the problem of computing
the free energy and the phase structure of the matrix model \beq \mathcal{Z}(\beta,p)=\int DU
\exp\left(\beta N( \mathrm{Tr}(U)+\mathrm{Tr}(U^{\dagger }))\right)\det(U)^{N p}, \eeq that is a
non-trivial deformation of the familiar Gross-Witten model~\cite{Gross:1980he}. Its large $N$
behavior is carefully studied in section~7.1 , obtaining the exact free energy in terms of the
solution of a fourth-order algebraic equation: we prove that there is no phase transition as long as
$p\neq 0$, in contrast with the usual $p=0$ case, that appears as a singular point in the parameter
space. In section~7.2 we use the results of our analysis to derive a set of saddle-point equations
for the partition function which describes the ``deconfined" phase. The disappearance of the
confining regime is consistent with the known results on finite temperature 2+1 dimensional gauge
theories where, once a topological mass (a Chern-Simons term) is turned on, there cannot be a phase
transition~\cite{Grignani:1995iv,Grignani:1995hx,Grignani:1996ft}.
 In section~8 we briefly
draw our conclusions and discuss future directions. Several
appendices are devoted to technical aspects and to an alternative
derivation of the partition functions. In appendix A we report some
details on supersymmetry transformations. In appendix B we give the
details of the computation of functional determinants. In appendix C
we recover the results for the single-particle partition functions
from those of the parent ${\mathcal N}=4$ theory by explicitly
constructing the projector into the $U(1)$ invariant modes. We also
check the consistency of our results with those
of~\cite{Hikida:2006qb}, where the theory on $\mathds{R}\times
S^3/\mathds{Z}_k$ has been studied. Appendix D is instead focused on
some technical aspects, related to the solution of the large $N$
matrix integrals.

\sectiono{ Lagrangian and supersymmetry  on $\mathds{R}\times S^2$ from $D=10$}
\label{sectD10D3}

There are many ways to construct the Lagrangian of the gauge
theory with sixteen supercharges on $\mathds{R}\times S^2$ and its
supersymmetry transformations. For instance,
in~\cite{Maldacena:2002rb} this theory was obtained from the
plane-wave matrix model action expanded around the $k$-membrane
vacuum in the large $N$ limit. Subsequently, in~\cite{Lin:2005nh}
it was derived as a $U(1)$ truncation of the spectrum of the
${\mathcal N}=4$ gauge theory on $\mathds{R}\times S^3$. Since
here we  shall be mainly concerned with the field theoretical
features of this ${\mathcal N}=8$ model, we shall follow a more
conventional (and maybe pedagogical) approach: the Lagrangian and
its supersymmetry transformations  will be derived as a
\textit{deformation} of the standard toroidal compactification of
${\cal N}=1$ gauge theory in ten dimensions.

We first consider the theory on the flat Minkowski space in three
dimensions, $\mathds{M}_{(1,2)}$. The ${\cal N}=8$ theory in this
case is  the  straightforward dimensional  reduction of the ${\cal
N}=1$ theory in $D=10$. The most convenient and compact way to
present its Lagrangian is to maintain the ten-dimensional notation
and to write (see appendix \ref{ConvSusy} for a summary of our
conventions\footnote{In general we shall omit the trace over the
gauge generator in our equations, unless it is source of
confusion.}) \beq \label{susy1} \mathcal{L}^{^{(0)}}  =
-\frac{1}{2}F_{MN}F^{MN} + i\overline{\psi}\Gamma^M D_M \psi. \eeq
All the fields in (\ref{susy1}) only depend on the space-time
coordinates $(x^0,x^1,x^2)$. In particular, from the
three-dimensional point of view, the gauge field $A_M$ contains
the reduced gauge field $A_\mu$ and seven scalars
$(\phi_m)=(\phi_3,\phi_4,\cdots,\phi_9)\equiv(\phi_3,\phi_{\overline{m}})$.
The flat ten dimensional space-time metric is diagonal and it has
the factorized structure $\mathds{T}^7\times\mathds{M}_{(1,2)}$.

Our goal is now to promote the supersymmetric theory in the flat
2+1-dimensional space-time to a supersymmetric theory on the
curved space $\mathds{R}\times S^2$. It is useful to keep a
ten-dimensional notation where the above space-time is viewed as a
submanifold embedded in $\mathds{T}^7\times \mathds{R}\times S^2$
with the metric \beq \label{susy2}
ds^2=-dt^2+R^2(d\theta^2+\sin^2\theta
d\varphi^2)+\sum_{i=1}^7d\eta_i^2\ . \eeq Here the coordinates
$\theta$ and $\varphi$ span the sphere $S^2$ of radius $R$, while
the internal angular coordinates $\eta_i$ parameterize the torus
$\mathds{T}^7$. The action (\ref{susy1}) in the background
(\ref{susy2}) is still meaningful once we introduce the
appropriate dependence on the vielbein and the spin-connections in
the covariant derivatives. The real issue is whether this theory
will have any supersymmetry. The action (\ref{susy1}) on flat
space is invariant under the usual supersymmetry transformations
written in terms of a constant arbitrary spinor $\epsilon$
\begin{equation} \label{susy6}
\begin{split}
\delta^{^{(0)}} A_M &= -2i\overline{\psi}\Gamma_M\epsilon,\\
\delta^{^{(0)}} \psi &= F_{MN} \Gamma^{MN} \epsilon\ .
\end{split}
\end{equation} Constant spinors  however do not exist, in general,
on a curved space. For a space-time of the type (\ref{susy2}), the
notion of a constant spinor should be replaced with that of a
\textsl{Killing spinor}~\cite{Blau:2000xg}. Its specific definition
may depend on the detail of the geometry, but, for us, it will be a
spinor satisfying an equation of the type \beq \label{susy3}
\nabla_\mu \epsilon =K_\mu^{\ \nu}\Gamma_\nu \Gamma^{123} \epsilon\
, \eeq where the Greek indices run only over the three-dimensional
space-time since the transverse coordinates $\eta_i$ are flat and we
can always choose $\epsilon$ to be a constant along these
directions. In (\ref{susy3}) we have also inserted an additional
dependence on the $\Gamma$ matrices through a monomial factor
$\Gamma^{123}$ \footnote{The direction $(1,2)$ span the tangent
space to the sphere $S^2$, while the index $3$ is along the first of
the compactified dimensions.}. This has double role: (a) it makes
(\ref{susy3}) compatible with the ten-dimensional chirality
conditions; (b) it generates, as we shall see, the relevant massive
deformations for our fields. Finally the tensor $K_\mu^{\ ~\nu}$
expresses an additional freedom in constructing the Killing spinors.
In a curved space, there is in fact no \textsl{a priori} reason to
treat  all the coordinates symmetrically. In the $\mathds{R}\times
S^2$ curved space-time geometry there is a natural splitting between
space and time and thus it is quite natural to weight them
differently by choosing \beq \label{susy4} K_\mu^{\ ~\nu}=\alpha
\left[\left(\delta_\mu^\nu+ k_\mu k^\nu\right)- \mathcal{B}k_\mu
k^\nu\right], \eeq where $k_\mu$ is the time-like Killing vector of
(\ref{susy2}) and $\alpha, \mathcal{B}$ are two arbitrary
parameters. The parameter $\alpha$ is fixed by imposing the
necessary integrability condition (the
first)~\cite{vanNieuwenhuizen:1983wu}, which arises from the
commutator $[\nabla_\mu,\nabla_\nu] \epsilon$. This can be either
expressed in terms of the space-time curvature scalar
$\mathcal{R}=2/R^2$ or, through (\ref{susy3}), in terms of $K_\mu^{\
~\nu}$ and consequently of $\alpha$. We thus get for $\alpha$ \beq
\label{susy5} \alpha=\frac{1}{2 R}\ . \eeq The parameter
$\mathcal{B}$, instead, remains free and it will be determined in
the following.

The variation of the action (\ref{susy1}) with respect to the
supersymmetry  transformations (\ref{susy6}) written in terms of a
non-constant supersymmetry parameter $\epsilon$ does not vanish.
Terms depending on the covariant derivatives of $\epsilon$
(\ref{susy3}) are in fact generated (see appendix A for
conventions and more details) \beq \label{susy7}
\begin{split}
\delta^{^{(0)}}\mathcal{L}^{^{(0)}} =& 2\Real\{ i\overline{\psi} F_{MN}
\Gamma^\mu \Gamma^{MN} \nabla_\mu \epsilon\}\\
=&2\Real\{ i\mathcal{B}\alpha\overline{\psi}[\Gamma^{ij} F_{ij} -2\Gamma^{0i}F_{0i} +2\Gamma^{jm} D_j
\phi_m -2 \Gamma^{0m} D_0
\phi_m -ig \Gamma^{mn}[\phi_m, \phi_n]]\Gamma^{123}\epsilon\\
& +i\alpha\overline{\psi}[-2\Gamma^{ij} F_{ij}+4 \Gamma^0 D_0 \phi_m
-2ig \Gamma^{mn}[\phi_m, \phi_n] ]\Gamma^{123}\epsilon\}.
\end{split}
\eeq where in the second equality we have used (\ref{susy3}) and
(\ref{susy4}). This undesired variation can be compensated by
adding the following deformations to the original Lagrangian
\begin{equation}
\label{susy8}
\begin{split}
    \mathcal{L}^{^{(1)}}  =iM\alpha\overline{\psi}\Gamma^{123}\psi +
    N\alpha\phi_3F_{12},\ \ \ \ \
    \mathcal{L}^{^{(2)}} = V\alpha^2\phi_m^2 + W\alpha^2\phi_3^2,
\end{split}
\end{equation}
and by adding new terms to the supersymmetry transformations of the
fermions
\begin{equation}
\label{susy9}
\begin{split}
\delta^{^{(1)}} \psi = P\alpha \Gamma^m\Gamma^{123} \phi_m
\epsilon + G\alpha \Gamma^3\Gamma^{123} \phi_3 \epsilon,
\end{split}
\end{equation}
where $M,N,V,W,P,G$ are arbitrary parameters to be fixed by imposing the invariance of the complete
action. The size of the deformations is tuned by the natural mass scale $\alpha=1/(2R)$ provided by
the radius of the sphere.

Some comments on the form of (\ref{susy8}) and (\ref{susy9}) are
in order. The addition of mass terms for the scalars
($\mathcal{L}^{^{(2)}} $) is a common and well-known property for
supersymmetric theories in a background admitting Killing spinors.
Some of the mass terms can also be justified with the requirement
that the conformal invariance originally present in flat space is
preserved. In four dimensions, for $\mathcal{N}=4$ super
Yang-Mills, this is the only required modification of the
Lagrangian because of an accidental cancellation.  Since we are in
three dimensions, we are also forced to introduce a non-standard
mass term  for the fermions (the first term in
$\mathcal{L}^{^{(1)}} $). The natural supersymmetric companion for
a fermionic mass in $D=3$ is then a Chern-Simons-like term (the
second term in $\mathcal{L}^{^{(1)}} $). Its unusual form, $\phi_3
F_{12}$, mixes the scalar $\phi_3$ with the gauge-fields and is
inherited from the particular choice of the monomial
$\Gamma^{123}$ in (\ref{susy3}). Then the  modifications
(\ref{susy9}) in the supersymmetry transformations are the only
possible ones with the right dimensions and compatible with the
symmetries of the theory.

The most convenient and simple way to analyze the effect of the
additional terms in the Lagrangian (\ref{susy8}) and in the
supersymmetry transformations (\ref{susy9}) is to single out, in the
variation of the Lagrangian, different powers of the deformation
parameter $\alpha$. We start with the linear order in $\alpha$, the
zeroth order being automatically absent since our theory is
supersymmetric in flat space-time. At this order we have three
contributions: the original variation (\ref{susy7}), the variation
of the new Lagrangian $\mathcal{L}^{^{(1)}}$ with respect to the old
transformations (\ref{susy6}) \beq \label{susy11} \!\!
\begin{split}
\! \delta^{^{^{(0)}}}\!\!\!\mathcal{L}^{^{^{(1)}}}\!\!\!\!\!\!=&
2M\alpha\Real\{i\overline{\psi}(F_{ij}\Gamma^{ij}\!\!-2F_{0i}\Gamma^{0i}\!\!\!
-2D_0\phi_3\Gamma^{03}\!\!
+\!\!2D_i\phi_3\Gamma^{i3}\!+\!2D_0\phi_{\overline{m}}\Gamma^{0\overline{m}}\!\!\!
-2D_i\phi_{\overline{m}}\Gamma^{i\overline{m}}
+\\
&+2i[\phi_3,\phi_{\overline{m}}]\Gamma^{3\overline{m}}-i[\phi_{\overline{m}},
\phi_{\overline{n}}]\Gamma^{\overline{m}\,\overline{n}})\Gamma^{123}\epsilon\}+
iN\alpha(F_{ij}\overline{\psi}\Gamma^{ij}+2D_i\phi_3\overline{\psi}\Gamma^{i3})\Gamma^{123}\epsilon\,
\end{split}
\eeq and finally the variation of $\mathcal{L}^{^{(0)}}$ with
respect to (\ref{susy9})
\begin{equation}
\label{susy10}
\delta^{^{^{(1)}}}\!\!\mathcal{L}^{^{^{(0)}}}\!\!\!\!\! =\!
2\Real\{i\alpha\overline{\psi}( P \Gamma^{\mu m}D_\mu \phi_m \! -
\!i g P\Gamma^{mn}[\phi_m,\phi_n]\!+\!G \Gamma^{\mu 3}D_\mu \phi_3
\!-\!i g G\Gamma^{m3}[\phi_m,\phi_3] )\Gamma^{123}\epsilon \}.
\end{equation}
 See appendix A for all the different index conventions. It is
quite straightforward to derive (\ref{susy11}) and (\ref{susy10})
since at this order in $\alpha$ we can consider $\epsilon$ as a
constant spinor, namely $\nabla_\mu\epsilon=0$.
 Imposing that
$\delta^{^{^{(0)}}}\mathcal{L}^{^{^{(0)}}}+\delta^{^{^{(0)}}}\mathcal{L}^{^{^{(1)}}}+\delta^{^{^{(1)}}}
\mathcal{L}^{^{^{(0)}}}=\mathcal{O}(\alpha^2)$  gives a linear
system of eight equations in the five unknowns $M,~N~,P~,G~$ and
$\mathcal{B}$. The details are given in appendix \ref{Susyvar}.
Quite surprisingly, this system is still solvable and it fixes the
value of the above constants as
\begin{equation}
\label{susy12} M=-\frac{1}{2},\ \ \ \ N=4,\ P= -2,\ \ \ \ G= -2,\
\ \ \
 \mathcal{B} = \frac{1}{2}.
\end{equation}
The next and final step is to consider the order $\alpha^2$ in our supersymmetry variation. The
situation is much simpler now since we need to evaluate only few terms. We have in fact to consider
the effects of the corrected transformation (\ref{susy9}) on $\mathcal{L}^{^{(1)}}$
\begin{equation}
\label{susy13}
\begin{split}
    \delta^{^{(1)}}\mathcal{L}^{^{(1)}} &= iM\alpha \delta^{^{(1)}}(\overline{\psi}\Gamma^{123}\psi)
    = 2\Real\{i\alpha^2\overline{\psi}
    (\Gamma^{\overline{m}}\phi_{\overline{m}} - 2\Gamma^3
    \phi_3)\epsilon\}
\end{split}
\end{equation}
and we have to take care of  the terms coming from $\delta^{^{(1)}}\mathcal{L}^{^{(0)}}$ originated
from the covariant derivative of the Killing spinor $\epsilon$. We obtain
\begin{equation}
\label{susy14}
\begin{split}
    \delta^{^{(1)}}\mathcal{L}^{^{(0)}}=-
    2\Real\{i\alpha^2\overline{\psi}[3\Gamma^{\overline{m}}\phi_{\overline{m}} + 6\Gamma^3\phi_3]\epsilon\}.
\end{split}
\end{equation}
These two contributions are easily compensated by the variation of $\mathcal{L}^{^{(2)}}$,
\begin{equation}
\label{susy15}
\begin{split}
    \delta^{(0)}\mathcal{L}^{^{(2)}} &=
    -4i\alpha^2(V\phi_{\overline{m}}\overline{\psi}\Gamma^{\overline{m}}\psi+(V+W)\phi_3
    \overline{\psi}\Gamma^3\psi)\ ,
\end{split}
\end{equation}
By setting $V=-1$ and $W=-3$ no surviving term is left! We remark that there is no
$\mathcal{O}(\alpha^3)$ term, because there is neither an $\alpha$-dependent term in the variation of
bosons (which might produce a $\mathcal{O}(\alpha^3)$ term in the variation of
$\mathcal{L}^{^{(2)}}$) nor $\alpha^2$ term in the variation of fermions.

We have thus reached our original goal: to promote the ${\mathcal
N}=8$ theory in flat space in three dimensions to an ${\mathcal
N}=8$ theory in the curved background $\mathds{R}\times S^2$. Its
Lagrangian in a ten-dimensional language is thus given by
\begin{equation}
\label{susy16}
\begin{split}
    \mathcal{L} & = -\frac{1}{2}F_{MN}F^{MN} + i\overline{\psi}\Gamma^M
    D_M \psi -i\frac{\mu}{4}\overline{\psi}\Gamma^{123}\psi +
    2\mu\phi_3F_{12} -\frac{\mu^2}{4}\phi_{\overline{m}}^2
    -\mu^2\phi_3^2,
\end{split}
\end{equation}
and it is invariant under the supersymmetry transformations
\begin{equation}
\label{susy17}
\begin{split}
\delta A_M &= -2i\overline{\psi}\Gamma_M\epsilon,\\
\delta \psi &= F_{MN} \Gamma^{MN} \epsilon -\mu \Gamma^m\Gamma^{123}
\phi_m \epsilon -\mu \Gamma^3\Gamma^{123} \phi_3 \epsilon,
\end{split}
\end{equation}
where $\mu$ is the mass-scale $\mu=1/R$. Notice that the mass for
the scalars $\phi_{\overline{m}}\ (\mathrm{with}\
\overline{m}=4,5,\dots,9)$ in (\ref{susy16}) is that required by
conformal invariance on $\mathds{R}\times S^2$:
$m^2_{conf.}=\frac{\mathcal{R}}{8}=\frac{2}{8 R^2}=\frac{\mu^2}{4}.$
The mass of the scalar $\phi_3$ is, instead, different because
$\phi_3$ mixes with the gauge fields. This mixing also breaks the
original $SO(7)$ $R$-symmetry present in flat space to the smaller
group $SO(6)_R~(\simeq SU(4)_R)$: the bosonic symmetries
$\mathds{R}\times SO(3)\times SO(6)_R$ combine with the
supersymmetries into the supergroup $SU(2|4)$. We have to mention
that our presentation heavily relies on  the general analysis of
\cite{Blau:2000xg}, where the problem of the existence of globally
supersymmetric Yang-Mills theory on a curved space was addressed and
some general recipes on how to construct these models were given.
However, the Lagrangian (\ref{susy16}) does not directly belong to
the families of theories discussed in \cite{Blau:2000xg}, it
realizes nevertheless a straightforward generalization of them. We
have in fact allowed for a more general Killing spinor equation both
by including the additional matrix factor $K_{\mu}^{\ \nu}$ and by
considering a monomial factor $\Gamma^{123}$ mixing one of the
transverse compact directions with the two spatial directions of the
actual space-time of the theory.

The Lagrangian (\ref{susy16}) written in terms of the
three-dimensional fields becomes
\begin{equation}
\label{susy18}
\!\!\!\!\!\!
\begin{split}
    \mathcal{L} =&-\frac{1}{2}F_{\mu\nu} F^{\mu\nu}
    +2i\overline{\lambda}_i\gamma^\mu D_\mu\lambda^i
    -\frac{1}{2}D_\mu \phi_{ij}D^\mu \phi^{ij}
    - D_\mu  \phi_3 D^\mu \phi_3-2ig\overline{\lambda}_i [\phi_3, \lambda^i]+\\
    & -g\sqrt{2} \left(\lambda^{iT}[\phi_{ij},\varepsilon
    \lambda^j]\! -\! \overline{\lambda}_i [\phi^{ij},\varepsilon \overline{\lambda}^T_j] \right)
    \!\!+\!\frac{1}{8}g^2[\phi_{ij},\phi_{kl}][\phi^{ij},\phi^{kl}]
    \!\!+\! \frac{1}{2}g^2[\phi_3, \phi_{ij}][\phi_3,\phi^{ij}]+\\
    &-\frac{\mu}{2}\overline{\lambda}_i\gamma^0 \lambda^i
    -\frac{\mu^2}{8}\phi_{ij} \phi^{ij}-\mu^2\phi_3^2 + 2\mu\phi_3F_{12}\ .
\end{split}
\end{equation}
This is the $\mathcal{N}=8$ SYM Lagrangian on $\mathds{R}\times
S^2$ that will be used in computing the thermodynamic partition
function of the model. We have cast the contribution of the scalar
fields $(\phi_4,\dots,\phi_9)$ in an $SU(4)_R$ manifestly
covariant form, by rewriting their Lagrangian in terms of the
\textbf{6} representation of $SU(4)_R$, $\phi_{ij}$. The spinor
fields $\lambda_i$ are four Dirac spinors in $D=3$ originating
from the dimensional reduction of $\psi$.

Since we will be mainly interested in the finite temperature
features of the model, the Euclidean version of (\ref{susy18}) will
be more relevant. It is given by
\begin{equation}
\label{LEucl}
\begin{split}
    \mathcal{L} &=\frac{1}{2}F_{\mu\nu} F^{\mu\nu}
    -2i\overline{\lambda}_i\gamma^\mu D_\mu\lambda^i
    +\frac{1}{2}D_\mu \phi_{ij}D^\mu \phi^{ij}
    +D_\mu  \phi_3 D^\mu \phi^3+\\
    &+g\sqrt{2} \left(\lambda^{iT}[\phi_{ij}, \varepsilon
    \lambda^j] - \overline{\lambda}_i [\phi^{ij}, \varepsilon \overline{\lambda}^T_j] \right)
    +2ig\overline{\lambda}_i [\phi_3, \lambda^i]+\\
    &-\frac{1}{8}g^2[\phi_{ij},\phi_{kl}][\phi^{ij},\phi^{kl}]
    - \frac{1}{2}g^2[\phi_3, \phi_{ij}][\phi_3,\phi^{ij}]+\\
    &+\frac{i\mu}{2}\overline{\lambda}_i\gamma^0 \lambda^i
    +\frac{\mu^2}{8}\phi_{ij} \phi^{ij}+\mu^2\phi_3^2 -2\mu\phi_3F_{12}.
\end{split}
\end{equation}

We conclude by noting that, in the above analysis, we have made a
particular choice in considering the form of the Killing spinor
equation. A careful reader  might wonder if there are other
possibilities. Unfortunately, different choices in (\ref{susy3})
generally lead to inconsistencies: the Killing equation is not
integrable or no consistent supersymmetric deformation exists. For
example, the second type of inconsistency would occur if we had
simply chosen $K^{\ \nu}_\mu=\delta_{\mu}^\nu$. It is however
intriguing to note that the choice $K^{\ \nu}_\mu=\delta_{\mu}^\nu$
becomes consistent if we alter the background geometry from
$\mathds{R}\times S^2$ to AdS$_3$ and substitute $\Gamma^{123}$ with
$\Gamma^{012}$ or $\Gamma^{456}$. In the former case, we would have
found a maximally supersymmetric version of the topologically
massive theory, with bosonic symmetry group $SO(1,3)\times SO(7)$.
In the latter we would  have instead reached a massive deformation
of the maximally supersymmetric Yang-Mills with the peculiar
interaction $ \mathrm{Tr}(\phi_3[\phi_4,\phi_5])$ and symmetry group
$SO(1,3)\times SO(3)\times SO(4)$. This case was already considered
in \cite{Blau:2000xg}. It would be nice to understand better their
relations with higher dimensional theories and to explore the
possible existence of gravitational duals.

\sectiono{BPS vacua and their gravitational duals}
\label{VACUA}

In this section we shall briefly review the structure of the BPS
vacua of the $\mathcal{N}=8$ theory on $\mathds{R}\times S^2$
\cite{Lin:2005nh} that will be the main ingredients of the
thermodynamical investigation of section \ref{Thermobps1} and
\ref{Thermobps2}. More specifically, we shall be interested in those
vacua that maintain both the $R$-invariance and the geometrical
symmetries.

In order to have an $SU(4)_R$ invariant vacuum, we have to choose
$\phi_{ij}=0$. Moreover, to preserve the invariance under time
translations and the $SO(3)$ rotations of the background geometry,
we require that all the fields are time-independent and that the
chromo-electric field $E_i=F_{0i}$ vanishes, respectively. The BPS
condition can be derived from the requirement that on the
supersymmetric invariant vacuum the supersymmetry variations should
vanish. Fermions must be set to zero to saturate the BPS bound and
consequently the supersymmetry variations of bosons automatically
vanish on the vacuum. The supersymmetry variation of fermions,
instead, must be set to zero and with the above assumptions it reads
 \beq \label{vacua1}
0=\delta\psi=[2(F_{\theta\varphi}
-\frac{1}{\mu}\sin\theta\phi_3)\Gamma^{\theta\varphi}+ 2 D_\mu
\phi_3\Gamma^{\mu 3} ]\epsilon\ , \eeq ($\theta$ and $\varphi$ are
coordinates on $S^2$) which translates into two simple equations
\beq \label{vacua2} F_{\theta\varphi}
-\frac{1}{\mu}\sin\theta\phi_3=0,\ \ \ \ \ \ \ D_\mu \phi_3=0. \eeq
The reader familiar with YM$_2$ will immediately recognize in these
equations, those of Yang-Mills theory on the sphere $S^2$, for which
a complete classification of the solutions exists
\cite{Atiyah:1982fa,Gross:1994mr}. The general solution for a $U(N)$
theory is given by  a stack of $N$ independent $U(1)$ Dirac
monopoles of arbitrary charges. In detail, we have \beq
\label{vacua3}
 {\phi}_3=\frac{ \mu {\mathfrak{f}}}{2}\ \ \ \
 {F}_{\theta\varphi}=\frac{ {\mathfrak{f}}}{2}\sin\theta\ \ \ \
 {A}=
\frac{
{\mathfrak{f}}}{2}\frac{(1-\cos\theta)}{\sin\theta}(\sin\theta
d\varphi)\equiv {\frac{\mathfrak{f}}{2 }} \mathcal{A}, \eeq where
$\mathfrak{f}$ is a diagonal matrix with integer entries, for which
we shall use the short-hand notation
\begin{equation}
\label{vacuabreak}
 \mathfrak{f}=(n_1,N_1;n_2,N_2;\dots;n_k,N_k).
\end{equation}
Each $n_I$ represents the Chern-class of the corresponding Dirac
monopole and it assumes values in $\mathds{Z}$, while $N_I$ is the
number of times that this charge appears on the diagonal. The vacuum
\eqref{vacuabreak} then breaks the original $U(N)$ gauge symmetry to
a direct product $U(N_1)\times U(N_2)\times\dots U(N_k)$. However,
since all fields in \eqref{susy18} are in the adjoint
representation, this breaking will affect the dynamics only through
the relative charge ($n_I-n_J$) between different sectors, while the
global charge $Q=\sum_{I=1}^k N_I n_I$ will play no role.

The gravitational backgrounds dual to the vacua of these theories
were derived in~\cite{Lin:2005nh} and further discussed in
\cite{Ishiki:2006yr} (where also the relations between vacua of
theories with $SU(2|\,4)$ symmetry group are studied): they have an
$SO(3)$ and an $SO(6)$ symmetry and thereby the geometry contains
$S^2$ and $S^5$ factors, the remaining coordinates being time, a
non-compact variable $\eta$, $-\infty\le\eta\le\infty$, and a radial
coordinate $\rho$. These backgrounds are non-singular because the
dual theories have a mass gap. The relevant supergravity equations
can be reduced to a three-dimensional electrostatic problem where
$\rho$ is the radius of a charged disk. The ten dimensional metric
and the other supergravity fields are completely specified in terms
of the solution $V$ of the related Laplace equation\footnote{This
problem has been recently tackled in \cite{Ling:2006up} and
\cite{Ling:2006xi}, searching for a dual description of Little
String theory on $S^5$ }. The regularity condition requires that the
location where the $S^2$ shrinks are disks at constant $\eta_i$ (in
the $\rho,\eta$ space) while $S^5$ shrinks along the segment of the
$\rho=0$ line between two nearby disks. The geometry therefore
contains three-cycles connecting the shrinking $S^2$ and six-cycles
connecting the shrinking $S^5$, supporting respectively non-trivial
$H_3$ and $*F_4$ fluxes. There is a precise relation between these
quantized fluxes and the data of the electrostatic problem, namely
the electric charges $Q_i$ of the disks are related to the RR fluxes
while the distance (in the $\eta$ direction) between two disks
bounding a three cycle is proportional to the NS flux. To be more
specific, this {\it electrostatic} description of a non-trivial
vacuum  generically contains $k$ disks, whose positions are
parameterized by $k$ integers $n_I$ through the relations
\begin{equation}
\eta_I=\frac{\pi n_I}{2}.
\end{equation}
These integers are identified with the monopole charges $n_I$  in
\eqref{vacuabreak}. Moreover each disk carries a charge $Q_I$ given
by
\begin{equation}
Q_I=\frac{\pi^2 N_I}{8},
\end{equation}
where $N_I$ are the same integer numbers counting the degeneracy of
each monopole charge in the gauge theory. At the level of
supergravity data, the above picture realizes $k$ groups of $D2$
branes, each of $N_I$ elements, wrapping different two-spheres. This
is the geometric manifestation of the breaking of the gauge symmetry
to a direct product $U(N_1)\times U(N_2)\times\dots\times U(N_k)$.
The charges $n_I$ instead combine into NS$5$-fluxes given by
$n_I-n_J$. Again the total charge seems to play no role.

 In our field theoretical analysis we have neglected the
time component of the gauge field $A_0$, which disappears from
(\ref{vacua2}) when considering the solutions (\ref{vacua3}). Its
dynamics is implicitly governed by the requirement that $E_i=0$,
which, for a time-independent background,  becomes $D_i A_0=0$. It
is a trivial exercise  to show that the most general solution of
this equation is provided by $A_0=0$ when the topology of the time
direction is $\mathds{R}$. In the finite temperature case where time
is compactified to a circle $S^1$, the most general solution is,
instead, given by $A_0=a$, where $a$ is a constant diagonal matrix,
namely a flat-connection living on $S^1$. This will play a
fundamental role in studying the thermodynamical properties of the
theory.

It is instructive to look at the BPS vacua also at the level of the
Euclidean Lagrangian: this will elucidate the emerging of an
interesting class of instanton solutions thoroughly studied in
\cite{Lin:2006tr}. If we focus on the bosonic sector of our model
and we set $\phi_{ij}=0$ to preserve the $SU(4)_R$ symmetry, we can
write \beq \label{vacua4} \sqrt{g}\mathcal{L}=\frac{\sqrt{g}}{2}
F_{\alpha\beta} F^{\alpha\beta}+\sqrt{g} D_\alpha\phi_3D^\alpha
\phi_3+ \sqrt{g}\mu^2\phi_3^2-2\mu\phi_3  F_{\theta\varphi}. \eeq
This Lagrangian can be easily arranged in a \textsl{BPS-form},
\textit{i.e.} as a sum of squares and total divergences. In fact,
after some algebraic manipulation, the Euclidean Lagrangian can be
cast in the following form \beq \label{vacua5}
\begin{split}
\sqrt{g}\mathcal{L} &=\pm \frac{1}{\mu} \sin\theta D_t (\phi_3^2)\mp
D_{\alpha}(\phi_3  F_{\beta\rho}\epsilon^{\alpha\beta\rho})+
\sin\theta\left(F_{t\theta}\pm\frac{1}{\sin\theta}
D_\varphi\phi_3\right)^2+\\ &+\frac{1}{\sin\theta}
\left(F_{t\varphi}\mp\sin\theta
D_\theta\phi_3\right)^2+\frac{\mu^2}{\sin\theta}
\left(F_{\theta\varphi}- \frac{1}{\mu^2}\sin\theta(\mu\phi_3\mp
D_t\phi_3)\right)^2 .
\end{split}
\eeq Consequently, the minimum of the action is reached when the
fields satisfy the following BPS-equations \beq \label{vacua6} (a):\
F_{t\theta}\pm\frac{D_\varphi\phi_3}{\sin\theta} =0\ \ (b):\
F_{t\varphi}\mp\sin\theta D_\theta\phi_3=0\ \ (c):\
F_{\theta\varphi}- \frac{1}{\mu^2}\sin\theta(\mu\phi_3\mp
D_t\phi_3)=0, \eeq or in a compact and covariant notation \beq
\label{vacua7} \sqrt{g} \epsilon_{\rho\nu\lambda} F^{\nu\lambda}=\mp
2 D_\rho \phi_3  +2\mu {k}_\rho \phi_3, \eeq where ${k}_\rho$ is the
Euclidean version of the time-like Killing vector of the metric on
$\mathds{R}\times S^2$. The vacuum equations (\ref{vacua2}) are just
a particular case of (\ref{vacua6}) or equivalently (\ref{vacua7}).
They emerge when we add the requirement of time-independence and
vanishing of the chromo-electric field $E_i$. From (\ref{vacua5}) it
is manifest that all our vacua (\ref{vacua3}) possess  a vanishing
action and they are all equivalent from an energetic point of view.

It is natural to ask now what is the meaning of the Euclidean
time-dependent solutions of (\ref{vacua6}). The action on these
solutions reduces to \beq S_{class}=\mp\frac{1}{\mu}\int_{S^2}
d\theta d\varphi\sin\theta \int_{-\infty}^\infty dt \partial_t
\mathrm{Tr}(\phi_3^2)~, \eeq which is finite,  and thus relevant
for a semiclassical analysis of the theory, if and only if
$\phi_3(t=-\infty)=\frac{\mathfrak{f}_{-\infty}}{2\mu R^2}$ and
$\phi_3(t=\infty)=\frac{\mathfrak{f}_{\infty}}{2\mu R^2}$. In
other words, these solutions are interesting if and only if they
interpolate between two vacua: one at $t=-\infty$ and the other at
$t=+\infty$. Their finite action is then given by \beq
S_{class}=\mp\frac{1}{\mu}\int_{S^2}\sin\theta d\theta d\varphi
\int_{-\infty}^\infty dt \partial_t \mathrm{Tr}(\phi_3^2)=
\mp\frac{\pi}{ g^2_{YM}
R}(\mathrm{Tr}(\mathfrak{f}_{\infty}^2)-\mathrm{Tr}(\mathfrak{f}_{-\infty}^2))~,
\eeq where we have reintroduced the relevant coupling constant
factors. We recognize the characteristics of \textsl{instantons}
in these (Euclidean) time-dependent solutions. At the quantum
level, they will possibly induce a tunneling  process between the
different vacua. At zero temperature Lin \cite{Lin:2006tr}
discussed the effect of these instantons from the gauge
theoretical side, at weak coupling, and from the gravity side,
that should describe the strong-coupling limit of the theory (see
also \cite{vanAnders:2007ky}), finding precise agreement in both
regimes. Moreover he argued, in analogy with the plane-wave matrix
model, that because of the presence of fermionic
zero-modes\footnote{The instantons are 1/2 BPS solutions and
therefore we expect 8 fermionic zero-modes associated to the
broken supersymmetries} around these instanton solutions, the
path-integral for the tunneling amplitude is zero. The vacuum
energies would not be  corrected and the vacua are exactly
protected at the quantum mechanical level: in particular they
should remain degenerate.  This kind of instantons has also been
recently considered in \cite{Hikida:2007pr}.

In the rest of the paper, in any case, we shall neglect the effect
of these solutions since we shall work at zero-coupling and in
this limit the probability of tunneling is exponentially
suppressed anyway.

\sectiono{Free SYM partition functions in monopole vacua}

In this section we shall derive the finite temperature partition
function in the BPS vacua (\ref{vacua3}), taking the limit
$g^2_{_{YM}} R\to 0$. We follow a path-integral approach where the
computation is reduced to the evaluation of one-loop functional
determinants in the monopole backgrounds. Since at finite
temperature the Euclidean time is a circle $S^1$ of length
$\beta=1/T$, we can also allow for a flat-connection $a$ wrapping
this $S^1$. The mode $a$ will play a very special role because it is
the only zero-mode in the decomposition into Kaluza-Klein modes on
$S^2\times S^1$. Consequently, as stressed in \cite{Aharony:2003sx},
the fluctuations described by $a$ are always strongly coupled,
including in the limit $g^2_{_{YM}} R\to 0$.

When the vacuum is trivial, there is no breaking of the $U(N)$ gauge
symmetry and the final result for the partition function is given by
a matrix integral over the unitary matrix $U=\exp\bigl[i\beta
a\bigr]$
\begin{equation}\label{Zx1}
\mathcal{Z}(\beta)=\int[dU]\exp\left\{\sum_{n=1}^\infty\frac{1}{n}
\left[z_B(x^n)+(-1)^{n+1}z_F(x^n)\right]{\rm Tr}(U^n){\rm
Tr}(U^{-n})\right\}.
\end{equation}
The functions $z_{B,F}(x)$ are respectively the bosonic and
fermionic single-particle partition functions (here $x=e^{-\beta}$),
counting the one-particle states of the theory without the
degeneracy coming from the dimension of the representation (the
adjoint representation $Adj$ in our case) and without any gauge
invariant constraint
\begin{equation}
z_{B,F}(x)=\sum_i e^{-\beta E_i^{(B,F)}}.
\end{equation}
The explicit form of the thermal partition function is obtained by
integrating over the matrix $U$
\cite{Sundborg:1999ue,Aharony:2003sx}
\begin{eqnarray}
\mathcal{Z}(\beta)&=&\sum_{n_1=0}^{\infty}x^{n_1
E_1^{B}}\sum_{n_2=0}^{\infty}x^{n_2 E_2^{B}}..
\sum_{m_1=0}^{\infty}x^{m_1 E_1^{F}}\sum_{m_2=0}^{\infty}x^{m_2
E_2^{F}}...\times\nonumber\\
 &\,&\#\,{\rm of\, singlets\, in \,\,\,\{sym}^{n_1}(Adj)\otimes {\rm
sym}^{n_2}(Adj)\otimes\cdots\nonumber\\
 &\,&\otimes\, {\rm antisym}^{m_1}(Adj)\otimes {\rm
antisym}^{m_2}(Adj)\otimes\cdots\}:
\end{eqnarray}
the partition function is expressed as a sum over the occupation
numbers of all modes, with a Boltzmann factor corresponding to the
total energy, and a numerical factor that counts the number of
singlets in the corresponding product of representations. Particle
statistics requires to symmetrize (antysimmetrize) the
representations corresponding to identical bosonic (fermionic)
modes.

The same result can also be obtained starting from
\begin{equation}\label{ZH}
\mathcal{Z}(\beta)={\rm Tr}\left[e^{-\beta H}\right]\equiv{\rm
Tr}\left[x^{ H}\right],
\end{equation}
where $H$ is the Hamiltonian of the theory. To calculate (\ref{ZH})
at zero coupling we need a complete basis of states of the free
theory or, thanks to the state-operator correspondence, of
gauge-invariant operators and we should count them weighted by $x$
to the power of their energy. A complete basis for arbitrary
gauge-invariant operators follows naturally after we specify a
complete basis of single-trace operators.
At the end,  one can write  (\ref{ZH}) in terms of single-particle
partition functions $z^R_{B,F}(x)$ \cite{Aharony:2003sx} as
\begin{equation}\label{Zx2}
\mathcal{Z}(\beta)=\int[dU]\exp\left\{\sum_R\sum_{n=1}^\infty\frac{1}{n}
\left[z^R_B(x^n)+(-1)^{n+1}z^R_F(x^n)\right]\chi_R(U^n)\right\},
\end{equation}
where the sum is taken over the  representations $R$ of the $U(N)$
gauge group\footnote{We consider the possibility to have fields in
an arbitrary representation.} and $\chi_R(U)$ is the character for
the representation $R$. The result (\ref{Zx1}) is reproduced when
all fields are in the adjoint representation: the variable $U$ has
to be identified as the holonomy matrix along the thermal circle,
$i.e.$ the Polyakov loop. The path-integral approach provides
therefore a physical interpretation for the unitary matrix $U$,
otherwise missing in the Hamiltonian formalism. On the other hand
the Hamiltonian construction explains how the group integration
forces the projection into color singlets and how it emerges the
structure of the full Hilbert space.

From the previous results we learn that once the representation content is specified, the full
partition function is completely encoded into the single-particle partition functions $z^R_{B,F}$.
However, the structure of the gauge group is more complicated on monopole backgrounds, consisting
into a direct product of $U(N_I)$ factors: consequently our constituents fields transform also under
bifundamental representations, producing additional complications for the explicit expression of the
matrix model. We also remark that bifundamental fields can transform non-trivially under $U(1)$
rotations and implementing the Gauss's law hides some subtleties in three dimensions, when background
monopole fluxes are present \cite{Borokhov:2002ib}: this potential additional freedom could affect
non-trivially the spectrum of physical operators in our theory. For the theory we are investigating,
however, the free-field spectrum is simply obtained by truncating the four-dimensional parent theory,
suggesting that the ${\cal N}=8$ counting  is conveniently performed through the relevant $U(1)$
projection on the ${\cal N}=4$ single-particle partition functions. This is what we do in appendix C,
where we construct the projector that eliminates all the fields which are not invariant under the
$U(1)$ and we derive, even in the non-trivial vacuum, the single-particle partition functions for
bosons and fermions. While this is certainly the quickest way to obtain these quantities, we prefer
to adopt here a path integral approach which in turn provides also the contributions of fermions and
bosons to the Casimir energy and allows for a careful treatment of the fermion zero modes. In the
path-integral computation all the subtleties will be treated in the well-defined framework of the
$\zeta$-function regularization procedure and in this section we present only the final results,
referring for the technical details to appendix B.

\subsection{Scalars}

Let us first describe the contribution of the six $SU(4)_R$ scalars
$\phi_{ij}$ to the partition function in the background
(\ref{vacua3}) and in presence of the flat-connection $a$: it
amounts to the evaluation of the determinant of the scalar kinetic
operator. We have to solve the associated eigenvalue problem,
\textit{i.e.} \beq \label{scalar1} -\hat\square
\phi_{ij}+\frac{\mu^2}{4}
\phi_{ij}+[\hat{\phi}_3,[\hat{\phi}_3,\phi_{ij}]]=\lambda \phi_{ij},
\eeq where the hatted quantities are computed in the relevant
background. In the following we shall drop the subscript $_{ij}$ and
we shall consider just one field denoted by $\phi$. The total result
at the level of free energy is then obtained by multiplying by six
the single-component contributions. Since $\phi$ is a matrix-valued
field, we shall expand it in the Weyl-basis, whose elements are the
generators $H_i$ of the Cartan subalgebra and the ladder operators
$E^\alpha$ \beq \label{scalar2} \phi=\sum_{i=1}^{N-1} \phi_{i} H^i
+\sum_{\alpha\in \mathrm{roots}} \phi_{\alpha}E^\alpha. \eeq We
shall also expand the background fields in this basis and define the
following two accessory quantities \beq \label{scalar3}
a_\alpha=\langle \alpha|a\rangle\ \ \ \ \textrm{and} \ \ \ q_\alpha=
\frac{\langle \alpha|\mathfrak{f}\rangle}{2 }. \eeq Here $a_\alpha$
denotes  the projection of the flat-connection $a$ along the root
$\alpha$ and $q_\alpha$ is the effective monopole charge measured
along the same root. Once the time-dependence is factored out, the
original eigenvalue problem splits into two subfamilies: $N(N-1)$
independent eigenvalues coming from each direction along the ladder
generator and $N-1$ independent eigenvalues coming from the
directions along the Cartan subalgebra. We can simply focus our
attention on the first family, since the latter can be obtained as a
limiting case for $a_\alpha,\ q_\alpha\to 0$. The relevant
eigenvalue equation can be solved algebraically if we introduce the
angular momentum operator in the presence of a $U(1)$ monopole of
charge $q_\alpha$, as explained in appendix B, and the resulting
spectrum does not depend on the sign of $q_\alpha$. By using
$\zeta$-function regularization, the scalar contribution to the
effective action can be easily computed as \beq \label{scalar12}
\Gamma^{Sc.}\!=\!\!\!\sum_{\alpha\in
\mathrm{roots}}\!\left(\frac{|q_\alpha|}{12}\left( 4
|q^2_\alpha|-1\right)\beta\mu+
\sum_{n=1}^\infty\frac{z^{scal.}_{q_\alpha}(x^n)}{n}e^{i n \beta
a_\alpha}\right)\!\!+\!(N\!-\!1)\sum_{n=1}^\infty
\frac{z^{scal.}_0(x^n)}{n}, \eeq where the scalar single-particle
partition function is given by \beq \label{scalar13}
{z^{scal.}_{q_\alpha}(x)}=x^{|q_\alpha|+1/2}\left(\frac{1+x}{(1-x)^2}+\frac{2|q_\alpha|}{1-x}\right).
\eeq

\subsection{Vectors}

Evaluating the contribution of the system $(A_\mu,\phi_3)$ is more
subtle and involved: the fields are coupled through the
Chern-Simons term and  the  Lagrangian for $A_\mu$ requires a
gauge-fixing procedure, with the consequent addition of a ghost
sector. A convenient choice for such a gauge-fixing appears  to be
\beq \label{vector1} \mathcal{L}_{g.f.}=(\hat{D}_\nu
A^\nu-i[\hat\phi_3,\phi_3])^2, \eeq where $\hat\phi_3=\frac{ \mu
{\mathfrak{f}}}{2}$ and the hatted derivative is defined in
(\ref{scalar5}). With this choice some of the mixing-terms in the
Euclidean quadratic Lagrangian cancel and we obtain the relevant
eigenvalue-problem  for computing the vector-scalar contribution
to the partition function: it is defined by the system of coupled
equations, written explicitly in (\ref{vector2}). Since both the
geometrical and the gauge background are static, the
time-component of the vector field $A_0$ decouples completely from
the eigenvalue system and satisfies the massless version of the
scalar equation previously studied. For the moment we shall forget
about $A_0$ since its contribution will be cancelled by the ghost
determinant. We are left with a purely two-dimensional system
where all the indices run only over space: the spectrum is again
conveniently determined by factoring out the time-dependence and
projecting the eigenvalue equations on the Weyl basis. We remark
that the equations involve also the Laplacian on vectors in the
background of a monopole of charge $q_\alpha$, besides the
Laplacian on scalars. The full computation of the spectrum is
reported in appendix B: we obtained three families of eigenvalues,
denoted by $\lambda_+$, $\lambda_-$ and $\lambda_3$. The
contribution of $\lambda_3$ will be cancelled by the ghost
determinant and we just consider, at the moment, the first two
families $\lambda_\pm$, which instead yield the actual vector
determinant in the roots sector \beq \label{vector181}
\Gamma^V_{r}= \sum_{\alpha\in \mathrm{roots}}\left(-\frac{1}{3}
 \left(4 q_{\alpha }^3+5 q_{\alpha }\right)\beta\mu
-2\sum_{n=1}^\infty\frac{z^{vec.}_{q_\alpha}(x^n)}{n}\,e^{i n
\beta a_\alpha}\right), \eeq where \beq \label{vector191}
z^{vec.}_{q_\alpha}(x)= x^{q_\alpha}\left[\frac{4x}{(1-x)^2}-1+2
q_\alpha\frac{1+x}{1-x}\right]. \eeq We remark that the results
(\ref{vector181}) and (\ref{vector191})  were shown to hold under
the initial assumption $q_\alpha\ge 1$. The extra-cases to be
considered are $q_\alpha=\frac{1}{2},\,0$. By recomputing the
spectrum for $q_\alpha=1/2$ we get the same results: quite
surprisingly this does not happen, instead, for $q_\alpha=0$ and
we get \beq \label{vector211} \Gamma^V_r(q_\alpha=0)=
-2\sum_{n=1}^\infty\frac{z_{0}^{vec.}(x^n)}{n}\, e^{i n \beta
a_\alpha}\ \ \ \  \mathrm{with}\ \ \
z^{vec}_0(x)=\frac{4x}{(1-x)^2}, \eeq a factor $-1$ missing in the
limit. To complete the discussion, we notice that, when multiplied
by $(N-1)$, (\ref{vector211}) is the contribution of the Cartan
components; the results (\ref{vector181}) and (\ref{vector191})
extends also to negative charges $q_\alpha$ by simply replacing
$q_\alpha$ with $|q_\alpha|$.

\subsection{Ghosts and $A_0$}

Let us discuss now the contributions to the partition function of
the eigenvalues $\lambda_{3}$, of the field $A_0$ and of the
determinant of ghost operator \beq \label{measure1}
-\hat\square\cdot+[\hat\phi_3,[\hat\phi_3,\cdot]]: \eeq they do
not cancel completely but, importantly, they give a measure of
integration for the flat-connection. It is possible to show that
when $q_\alpha\neq 0$ we have a complete cancellation of the
different contributions: crucially for $q_\alpha=0$ this does not
happen and a modification of the measure for the flat-connection
is induced \beq \label{measure4} \prod_{\alpha\in
\mathrm{roots}\atop \mathrm{with}\ q_\alpha=0} 2 i e^{-i
\frac{\beta a_\alpha}{2}} \sin \left(\frac{\beta a_\alpha
}{2}\right)= \prod_{\alpha\in \mathrm{positive\ roots}\atop
\mathrm{with}\ q_\alpha=0} 4 \sin^2 \left(\frac{\beta a_\alpha
}{2}\right). \eeq The meaning of this measure is quite
transparent: the monopole background breaks the original $U(N)$
invariance to the subgroup $\prod_{I=1}^kU(N_I)$, (\ref{measure4})
being the product of the Haar measure of each $U(N_I)$ component,
as can be easily checked by recalling the explicit form of the
roots and the definition of $q_\alpha.$ As a matter of fact, in
non-trivial monopole backgrounds, when we shall write the integral
over the flat-connections we will be naturally led to consider a
unitary {\it multi-matrix} model instead of an ordinary one.

\subsection{Fermions}
\label{fermions}

The contribution of the fermions to the total partition function
needs a careful analysis. At first sight, apart from having
antiperiodic boundary conditions along the time circle, the
computation of the fermion determinants seems to follow closely
the bosonic cases. We have again $N(N-1)$ independent eigenvalues
coming from each direction along the ladder generators and $N-1$
independent eigenvalues coming from the directions along the
Cartan subalgebra, that can obtained as limit of vanishing flux.
The computation of the spectrum is quite technical as in the
vector case and boils down in solving the eigenvalue problem for a
family of effective massless Dirac operators
$\mathfrak{D}^{(\alpha)}$ (see app. \ref{SPINOR}) on the
two-sphere, in the effective monopole backgrounds provided by
$q_\alpha$. The spectrum of $\mathfrak{D}^{(\alpha)}$, as expected
in two dimensions, consists in a set non-vanishing eigenvalues,
symmetric with respect the zero, and in a finite kernel, as
predicted by the Atiyah-Singer theorem. These zero-modes are
chiral and can be classified by using the eigenvalues of the
operator $(\sigma\cdot\hat{r})$, playing the role of $\gamma_5$:
we shall denote $\nu_\pm$ the number of zero modes with eigenvalue
$\pm 1$. A simple application of the index theorem shows that
$\nu_+=|q_\alpha|-q_\alpha\ \ $ and $ \ \ \
\nu_-=|q_\alpha|+q_\alpha$, namely for positive $q_\alpha$ we have
only zero modes with negative chirality and viceversa. As shown in
appendix \ref{SPINOR}, the contribution of the first set of
eigenvalues to the effective action can be easily evaluated \beq
\label{spinor12} \Gamma^S_{1}=\!\!\!\!\sum_{\alpha\in
\mathrm{roots}}\!\!\!\left(\!\!-\frac{\beta \mu}{3}
\left(2|q_{\alpha}|^3\!\!
   +3|q_{\alpha }|^2\!\!+ |q_{\alpha }|\right)\!\!
-\!\!\sum_{n=1}^\infty \frac{(-1)^n}{n} {z^{spin.}_{q_\alpha
1}(x^n)} e^{i\beta n a_\alpha} \right)\!\!, \eeq with \beq
\label{spinor13} z^{spin.}_{q_\alpha 1}(x)=2
x^{|q_\alpha|+1}\left(\frac{1}{(1-x)^2}+
\frac{|q_\alpha|}{1-x}\right)(x^{\frac{1}{4}}+ x^{-\frac{1}{4}}).
\eeq Next we consider the contribution of the zero-modes of the
effective Dirac operators: in a monopole background, this
subsector originates the spectral asymmetry
\cite{Alvarez-Gaume:1984nf}
 of the three dimensional fermionic
operator and therefore the potential appearance of a parity
violating part in the effective action. In particular, we could
expect the generation of the Chern-Simons anomalous term (we refer
to \cite{Deser:1997nv,Deser:1997gp} for a complete discussion of
this issue). Concretely, in our case, the explicit computation of
the zero-mode contribution amounts to evaluate a family of
one-dimensional massive fermion determinants, in a flat-connection
background (see appendix \ref{SPINOR}). It is well-known that the
$\zeta$-function regularization scheme carries an intrinsic
regularization ambiguity\footnote{This ambiguity is not something
peculiar of the $\zeta$-function regularization, but it appears in
different forms also in other regularizations: in the usual
Pauli-Villars approach, for example, this ambiguity translates into
a dependence of the local terms in the effective action on the sign
of the mass of the regulator.} in this case, depending on the choice
of some branch-cuts in the $s$-plane, affecting the local terms in
the effective action \cite{Deser:1997nv,Deser:1997gp}. For us all
the different possibilities boil down to two alternatives: we can
regularize the contributions associated to the \textit{zero-modes}
of negative and positive chirality by choosing opposite cuts in
defining the complex power of the eigenvalues (one on the real
positive axis and the other on the real negative axis) or by
choosing the same cut. We find quite natural to use the same
procedure for ${\it all}$ the four fermions present in the theory:
we surely preserve the $R$-symmetry and the global non-abelian
symmetry in this way. Within this choice, the following results hold
from our one-dimensional fermion determinants: taking opposite cuts
we get \beq \label{spinor151}
\Gamma^{S}_{0,A}=\sum_{\alpha\in\mathrm{roots}} (1-r)\beta \mu
\left( q_\alpha^2+\frac{|q_\alpha| }{4} \right)
-\sum_{\alpha\in\mathrm{roots}} \sum_{n=1}^\infty\frac{(-1)^n}{n}
2|q_\alpha|x^{n |q_\alpha|} e^{i\beta n a_\alpha} x^{\frac{n}{4}}.
\eeq Here $r=\pm 1$ and its specific value depends on the cut
selected  for the zero-modes of positive chirality. Choosing instead
the same cuts we obtain \beq \label{spinor152}
\Gamma^{S}_{0,B}=\sum_{\alpha\in\mathrm{roots}}\Bigl[\beta
\mu\left(|q_\alpha|^2+\frac{|q_\alpha|}{4}\right)+  i r \beta
a_\alpha q_\alpha\Bigr]-\sum_{\alpha\in\mathrm{roots}}
\sum_{n=1}^\infty\frac{(-1)^n}{n} 2|q_\alpha|x^{n |q_\alpha|}
e^{i\beta n a_\alpha} x^{\frac{n}{4}}. \eeq Again $r=\pm 1$
according to the specific choice of the cut (real positive or
negative axis): we must stress, however, that this ambiguity will
become irrelevant when we shall perform the integration over the
flat-connections.

We remark that there is an important difference between the two
expressions: in the second case we have a new term in the
effective action, depending explicitly on the flat connection. To
understand its nature, it can be equivalently written as
\begin{equation}
i r\sum_{\alpha}\beta q_\alpha a_\alpha=i r\beta (N \mathrm{Tr}(a
\mathfrak{f})-\mathrm{Tr}(a) \mathrm{Tr}(\mathfrak{f})).
\end{equation}
We immediately recognize the $SU(N)$ part of the usual Chern-Simons
term, calculated in our particular background. The related
regularization choice is therefore consistent with the intrinsic
parity anomaly of three dimensional gauge theories. We stress that
the above contribution arises just in the monopole vacua and it is
related to non-perturbative properties of the fermion determinants.
We also observe that the two results differ in the charge-dependent
contribution linear in $\beta$, and we will see this to modify
crucially the Casimir energy.

Summing now, in both cases, the kernel contribution to
$\Gamma^S_1$ we get
\beq \label{spinor161} \Gamma^{S}_{A}=\!\!\!\!\sum_{\alpha\in
\mathrm{roots}}\!\!\! \left(\!\!-\frac{\beta \mu}{12}
\left(8|q_{\alpha}|^3\!\!
   +12 r |q_{\alpha }|^2\!\!+(3 r+1) |q_{\alpha }|\right)
-\!\!\sum_{n=1}^\infty \frac{(-1)^n}{n} {z^{spin.}_{q_\alpha}(x^n)}
e^{i\beta n a_\alpha} \right)\!\!, \eeq with the first choice  and
\beq \label{spinor162} \Gamma^{S}_{B}=\!\!\!\!\sum_{\alpha\in
\mathrm{roots}}\!\!\!\left(\!\!-\frac{\beta \mu}{3}
\left(2|q_{\alpha}|^3\!\!
   + \frac{|q_{\alpha }|}{4}\right)\!\!+i r q_\alpha a_\alpha
-\!\!\sum_{n=1}^\infty \frac{(-1)^n}{n}
{z^{spin.}_{q_\alpha}(x^n)} e^{i\beta n a_\alpha} \right)\!\!,
\eeq in the latter. Happily the single-particle partition function
is the same for both the regularization choices \beq
\label{spinor17} z^{spin.}_{q_\alpha }(x)= x^{\left|q_{\alpha
}\right|} \left(\frac{2
   x}{(1-x)^2}+\frac{2 |q_\alpha|\sqrt{x}}{1-x}\right)(x^{\frac{1}{4}}+
x^{-\frac{1}{4}}). \eeq \noindent The contribution of the Cartan
components is of course obtained from the above results by simply
setting $q_\alpha=0$.

\subsection{Partition functions}

The next step is to collect the different contributions, coming from
the functional determinants, and write down the total result as a
compact integral over unitary matrices. According to the previous
discussion, we must distinguish two cases, depending on the form of
the spinor determinant (\ref{spinor161}) or (\ref{spinor162}). We
shall first consider the choice (\ref{spinor161}). The complete
effective action, obtained by including roots and Cartan
contributions with the appropriate multiplicities, can be expressed
as \beq \label{partition11}
\begin{split}
S_{eff.}=&-\beta V_0+
\sum_{\alpha\in\mathrm{roots}}\sum_{n=1}^\infty\frac{1 }{n}(6
z^{scal.}_{q_\alpha}(x^n)+z^{vec.}_{q_\alpha}(x^n)+(-1)^{n+1} 4
  z^{spin.}_{q_\alpha}(x^n))e^{i n \beta
a_\alpha}+\\
&+(N-1)\sum_{n=1}^\infty\frac{1 }{n}(6
z^{scal.}_0(x^n)+z^{vec.}_0(x^n)+(-1)^{n+1} 4
z^{spin.}_{0}(x^n))\equiv\\
\equiv&-\beta
V_0+\sum_{\alpha\in\mathrm{roots}}\sum_{n=1}^\infty\frac{1
}{n}z^{tot.}_{q_\alpha}(x^n) e^{i n \beta
a_\alpha}+(N-1)\sum_{n=1}^\infty\frac{1 }{n}z^{tot.}_0(x^n),
\end{split}
\eeq where we have introduced the total single-particle partition
functions and the Casimir  energy $V_0$ of the configuration \beq
V_0=r \sum_{\alpha\in\mathrm{roots}}(4 |q_\alpha|^2 +|q_\alpha|).
\eeq The matrix structure hidden in (\ref{partition11}) appears
manifest when writing the original Polyakov loop $U=\exp(i \beta
a)$, associated to the  diagonal flat-connection $a$, through $k$
sub-matrices $U_I$ acting on the invariant subspaces implicitly
defined by the monopole background $(\ref{vacuabreak})$. The
$N_I\times N_I$ unitary matrices $U_I$ have the form
$U_{I}=\mathrm{diag}(e^{i\beta a^I_{1}},\dots,e^{ i\beta
a^I_{N_I}})$, where we have parameterized the original flat
connection $a$ as follows: \beq \label{flat} {a}=
\mathrm{diag}(\underbrace{a^1_1,\dots, a^1_{N_1}}_{N_1},
\underbrace{a^2_1,\dots, a^2_{N_2}}_{N_2},\dots\dots,
\underbrace{a^I_1,\dots, a^I_{N_I}}_{N_I},\cdots). \eeq Let us
consider now the subset $\mathcal{A}_{IJ}$ of the positive
roots\footnote{ The roots of $SU(N)$ are all the $N(N-1)$
permutations of the $N-$vector $(1,-1,0,\cdots,0)$ and they can be
separated in \textsl{positive} and \textsl{negative} according to
the sign of the first non zero entry.} of $SU(N)$ whose first  and
second non vanishing entries belong respectively to the $I^{th}$ and
$J^{th} $ invariant subspace of $\mathfrak{f}$. The effective
charges $q_\alpha= \frac{\langle \alpha
|\mathfrak{f}\rangle}{2}=\frac{n_I-n_J}{2}$ and, consequently, the
$z^{tot.}_{q_\alpha}$ take always the same value for this class of
roots. The sum over roots on this subset reduces to \beq
\label{matrix11} \sum_{\alpha\in\mathcal{ A}_{IJ}}e^{i n \beta
a_\alpha}=\sum_{i=1}^{N_I}\sum_{j=1}^{N_J} e^{i n \beta
(a^I_{i}-a^J_{j})}=\mathrm{Tr}(U_I^n)\mathrm{Tr}(U_J^{\dagger n});
\eeq the analogous subsector $\bar{\mathcal{A}}_{IJ}$ given by the
negative roots yields $\mathrm{Tr}(U_I^{\dagger
n})\mathrm{Tr}(U_J^{n})$. We remark that the pre-factor
$z^{tot.}_{q_\alpha}$ is however the same for both cases since it
depends just on the modulus of the effective monopole charge. The
subset of roots $\mathcal{B}_{I}$ whose first and second non
vanishing entries live in the same $I^{th}$ invariant subspace of
$\mathfrak{f}$ have instead effective monopole charge zero. Then the
contribution of this subsector is simply given by \beq
\label{matrix21}
\begin{split}
\sum_{n=1}^\infty\frac{1 }{n}z^{tot.}_0(x^n)\sum_{\alpha\in
\mathcal{B}_I} e^{i n \beta a_\alpha}&=\sum_{n=1}^\infty\frac{1
}{n}z^{tot.}_0(x^n) \sum_{i\ne j=1}^{N_I} e^{i n \beta
(a^I_{i}-a^I_{j})}=\\
&=\sum_{n=1}^\infty\frac{z^{tot.}_0(x^n)
}{n}(\mathrm{Tr}(U^{\dagger n}_I)\mathrm{Tr}(U^n_I) -N_I).
\end{split}
\eeq Because of the results (\ref{matrix11}) and (\ref{matrix21}),
it is convenient to change our notation and to define the $k\times
k$ matrix-valued single-particle partition function
$z^{tot.}_{IJ}$: the diagonal elements are
$z^{tot.}_{II}=z^{tot.}_0$, the off-diagonal ones are instead
identified with the function $z^{tot.}_{q_\alpha}$, associated to
the charge $\frac{n_I-n_J}{2}$. The matrix $z^{tot.}_{IJ}$ is
symmetric since everything depends just on the modulus of the
charge. The complete effective acton takes the elegant form
 \beq
\label{partition21} S_{eff.} =-\beta
V_0+\sum_{IJ}\sum_{n=1}^\infty\frac{1
}{n}z^{tot.}_{IJ}(x^n)\mathrm{Tr}(U_I^n)\mathrm{Tr}(U_J^{\dagger
n})
 -\sum_{n=1}^\infty\frac{1 }{n}z^{tot.}_{II}(x^n).
\eeq The last term drops if we consider $U(N)$ instead of $SU(N)$.
Remarkably the structure of the matrix action is perfectly
consistent with the measure found in (\ref{measure4}), which is
exactly the Haar measure for this multi-matrix model.

The above analysis is practically unaltered when considering the
fermionic contribution (\ref{spinor162}) in the effective action,
except on a couple of points. It changes the value of the Casimir
energy $V_0$, which now vanishes identically, and we have a new
important addition to (\ref{partition21}), that can expressed in
terms of the determinants of the unitary matrices $U_I$ \beq
\label{Penner11} i r\beta \sum_{\alpha\in \mathrm{roots}} q_\alpha
a_\alpha= \log\left(\prod_{I=1}^k\det(U_I)^{r (N n_I
-Q)})\right)=r\sum_{I=1}^k (N n_I -Q) \log(\det(U_I)), \eeq where
$Q=\sum_{I=1}^k N_I n_I$. As a first remark, we notice that new
contributions depends still on the differences $n_I-n_J$,
consistently with the decoupling of the total $U(1)$ charge of the
monopole configuration. Then we observe that the two different
values $r=\pm 1$, related to our regularization choice, produce the
same result when integrating over the unitary group: the difference
can be reabsorbed just changing integration variable $U_I\mapsto
(U_I)^{-1}$, which leaves the measure and (\ref{partition21})
unaltered. From now on, we shall set $r=1$.

In the trivial vacuum we obtain a partition function that is a
straightforward generalization of the unitary matrix model
discussed in \cite{Aharony:2003sx} \beq \label{matrix-model1aa}
\mathcal{Z}=\int dU \exp\left(
\sum_{n=1}^\infty\frac{1}{n}z_0^{tot.}(x^n)
\mathrm{Tr}(U^n)\mathrm{Tr}(U^{\dagger n})\right) \eeq where the
function $z_0^{tot.}(x^n)$ encodes the dynamical content of the
three-dimensional supersymmetric theory. Notice that the Casimir
energy is identically zero, since it vanishes for each
contribution both bosonic and fermionic.

The situation changes in non-trivial monopole vacua: we get
respectively \beq \label{matrix-model12}
\mathcal{Z}_A=\int\prod_{I=1}^k [dU_I] \exp\left(-\beta
V_0+\sum_{IJ}\sum_{n=1}^\infty\frac{1 }{n}z^{tot.}_{IJ}(x^n)
\mathrm{Tr}(U_I^n)\mathrm{Tr}(U_J^{\dagger n})\right) \eeq and
\beq \label{matrix-model21b} \mathcal{Z}_B=\int\prod_{I=1}^k
[dU_I] \exp\left(\sum_{IJ}\sum_{n=1}^\infty\frac{1
}{n}z^{tot.}_{IJ}(x^n) \mathrm{Tr}(U_I^n)\mathrm{Tr}(U_J^{\dagger
n})\right)\prod_{I=1}^k\det(U_I)^{(N n_I -Q)}, \eeq depending on
our regularization choice. First of all we see that the partition
function is related to a unitary multi-matrix model: the gauge
group is broken in factors and states in the bifundamental
representation are present, with energies clearly encoded into the
off-diagonal entries of the single-particle partition function
$z^{tot.}_{IJ}$. Let us discuss on general grounds the effects of
the different choices for the fermion determinants. A first mild
diversity arises in the Casimir energies: from (\ref{spinor161})
we have a non-vanishing $V_0$, with arbitrary sign, while
(\ref{spinor162}) leads to a vanishing result. We recall that the
Casimir energy is supposed to correspond to the mass of the dual
geometry \cite{Hikida:2006qb}: in the first case it seems that
different backgrounds supports different, monopole dependent,
masses, suggesting a possible lifting of the vacua degeneracy at
quantum level. The second choice is instead consistent with the
believed degeneracy: unfortunately no computation from the
gravitational side seems to be available up to now and we do not
have further insights on the meaning of the different results.

The presence of the new terms (\ref{Penner11}) in the matrix model
(\ref{matrix-model21b}) can be, instead, better understood at the
level of partition functions. First of all we notice that the matrix
integral implementing the Gauss's law is actually over $unitary$
matrices $U_I$: the $U(1)$ phases contained into the the $U_I$'s
play a non-trivial role in the monopole background. This has to be
contrasted with the trivial vacuum: there the effective action is
invariant under $U(1)$ transformations and we can simply forget the
integration over the center. In the non-trivial vacuum the resulting
effective action (\ref{partition21}) is not invariant under phase
rotations, as an effect of the off-diagonal terms in the
single-particle partition function, and the $U(1)$ integrations
precisely correspond to selection rules in the bifundamental sector.
It is not difficult to realize that within the first regularization
the matrix integrals select states having vanishing $U(1)$ charge,
with respect to all $U(N_I)$ group factors. To understand the effect
of the new terms in (\ref{matrix-model21b}) instead, we simply
observe that the determinants depend just on the $U(1)$ phases and
modify non-trivially the selection rules of the bifundamental
sectors, according to the charges of the monopole background. We
shall say in this case that our regularization procedure correspond
to the choice of a charged vacuum, as discussed in
\cite{Jackiw:1975fn}, while we will refer to the first possibility
as to the uncharged vacuum. Since at the quantum field theory level
both choices seems to be allowed, we think it is instructive to
investigate the thermodynamics in both cases, deferring a deeper
understanding of the different possibilities to future studies, in
the context of supersymmetry and gravitational duals.

We end this section introducing the simple modification to the
effective action due to chemical potentials for the $SU(4)$
$R$-charge. In the path integral approach their effect amounts to
simply adding an imaginary $SU(4)$ flat connection
$\mathbb{A^{\mathbf{R}}}=i(\Omega_1 Q_1^\mathbf{R} +\Omega_2
Q_2^\mathbf{R}+\Omega_3 Q_3^\mathbf{R})$ in the Euclidean time
direction. Here $Q_i^\mathbf{R}$ are the Cartan generators of
$SU(4)$ and $\mathbf{R}$ denotes the relevant representation:
$\mathbf{4}$ for the spinors and $\mathbf{6}$ for the scalars. One
finds the new partition functions
\begin{equation}
\label{partition3}
\begin{split}
4 z^{spin.}_{q_{\alpha}}\mapsto    z_{IJ}^{spin.} &=\,x^{
    |q_\alpha|}\left(\frac{2x}{(1-x)^2}\sum_{p=1}^4\left(x^{\frac{1}{4}}y^{-\widetilde{\Omega}_p}+x^{-\frac{1}{4}}y^{\widetilde{\Omega}_p}\right)
    +\right.\\
    &\ \ \ \ \ \ \ \ \ \ \ \
    \left.+2|q_\alpha|\frac{x^\frac{1}{4}}{1-x} \sum_{p=1}^4\left(y^{-\widetilde{\Omega}_p}+x^{\frac{1}{2}}y^{\widetilde{\Omega}_p}\right)\right),\\
6 z^{scal.}_{q_{\alpha}}\mapsto     z_{IJ}^{scal.} &= \,x^{
|q_\alpha|+1/2}\left(\frac{x +
    1}{(1-x)^2}+2|q_\alpha|\frac{1}{1-x}\right)\sum_{p=1}^3\left(y^{\Omega_p}+y^{-\Omega_p}\right),
\end{split}
\end{equation}
with $y=e^{-\beta}$ and
\begin{eqnarray}
\tilde{\Omega}_1&=\displaystyle{\frac{1}{2}}\left(\Omega_1+\Omega_2+\Omega_3\right)\
\ \ \ \ \
\tilde{\Omega}_2&=\displaystyle{\frac{1}{2}}\left(\Omega_1-\Omega_2-\Omega_3\right)\cr
\tilde{\Omega}_3&=\displaystyle{\frac{1}{2}}\left(-\Omega_1+\Omega_2-\Omega_3\right)\
\ \ \
\tilde{\Omega}_4&=\frac{1}{2}\left(-\Omega_1-\Omega_2+\Omega_3\right).
\end{eqnarray}
\sectiono{Thermodynamics in the trivial vacuum}
\label{Thermodynamics}

We have seen in the previous section that the thermodynamics in
the trivial vacuum is governed, in the zero-coupling
approximation, by the {\it one-component} unitary matrix model
\beq \label{matrix-model1a} \mathcal{Z}=\int dU \exp\left(
\sum_{n=1}^\infty\frac{1}{n}z_0^{tot.}(x^n)
\mathrm{Tr}(U^n)\mathrm{Tr}(U^{\dagger n})\right) \eeq where the
function $z_0^{tot.}(x^n)$ encodes the dynamical content of the
three-dimensional supersymmetric theory. Notice that the Casimir
energy is identically zero, since it vanishes for each
contribution both bosonic and fermionic.

When $N$ is large we can trade the integration in
(\ref{matrix-model1a}) over the unitary group for an integration
over the normalized distribution function $\rho(\theta)$ of the
continuous eigenvalues $e^{i \theta}$ of $U$, with
$-\pi<\theta\le\pi$. More precisely we can write the integral over
the unitary matrices in terms of the Fourier-modes
$(\rho_n,\bar\rho_n)$ defined as \beq \label{matrix-model2a}
\rho(\theta)=\frac{1}{2\pi}+\sum_{n=1}^\infty(\rho_{ n} e^{i
n\theta}+\bar{\rho}_{ n} e^{-i n\theta})\ . \eeq Following
\cite{Sundborg:1999ue,Aharony:2003sx}, we can then reduce the
integral to the standard form \beq \label{matrix-model3a}
\begin{split}
\mathcal{Z}=\int D\rho_{n} D\bar{\rho}_{n}\exp\left(- N^2 \sum_{n=1}^\infty{\rho_{n} \bar{\rho}_{n}
}V(x^n)\right) \ \ \ \ \ \ \mathrm{with}\ \ \ V(x^n)=\frac{1}{n}(1-z_0^{tot.}(x^n))\ .
\end{split}
\eeq In the large $N$ limit, (\ref{matrix-model3a}) is dominated
by the absolute minimum of the quadratic action $S= \sum\rho_{n}
\bar{\rho}_{n}V(x^n)$ which is reached  for $\rho_{n}=0$ for every
$n$ if  $V(x^n)$ is positive definite. For small temperatures,
namely small $x$, the function $V(x^n)$ is positive for any $n$
and close to $1/n$ since $V(x^n) \sim\frac{1}{n}$ for $x\ll 1$.
(Recall that $z_0^{tot.}(x^n)$ vanishes as $x$ approaches zero.)
Therefore the partition function is $1$ at the leading order and
it is simply given by the small fluctuation around the minimum at
the subleading order: \beq
\mathcal{Z}\propto\prod_{n=1}^{\infty}\frac{1}{(1-z_0^{tot.}(x^n))}.
\eeq When we increase the temperature, $x$ approaches $1$ and the
above description is reliable up to the smallest value $x_c$ where
$V(x^n)$ becomes negative. Since $z_0^{tot.}(x)$ is a monotonic
function ranging from $0$ to infinity, this value always exists
and it is reached for $n=1$, namely \beq
V(x_c)=1-z_0^{tot}(x_c)=0. \eeq This algebraic condition, whose
explicit form is \beq \label{criticaltemp1}
\begin{split}
V(x_c)= 1-z_{0}^{tot.}&= 1-(4
z_0^{spin.}+6z_0^{scal.}+z_0^{vec.})=\\
&=\frac{\left(\sqrt[4]{{x_c}}+1\right)^4 \left({x_c}-4 {x_c}^{3/4}+4
   \sqrt{{x_c}}-4 \sqrt[4]{{x_c}}+1\right)}{(1-{x_c})^2}=0,
   \end{split}
\eeq can be exactly solved, since it can be reduced to an equation
of fourth degree. It possesses just one solution in the interval
$[0,1]$ given by \beq \label{criticaltemp2} x_c=\left(2+2
\sqrt{2}-\sqrt{11+8 \sqrt{2}}\right)^2\simeq(0.104688)^2. \eeq It
is interesting  to compare this value with the critical
temperature computed in \cite{Hikida:2006qb} for $\mathcal{N}=4$
on $S^3/\mathds{Z}_k$. This theory should in fact reproduce our
model when $k$ goes to infinity. However, the three-dimensional
theory obtained in this limit lives on a $S^2$ sphere whose radius
is half of the radius of the original $S^3$: this means that
$x_c=\lim_{k\to\infty}x^2_c(k)$. To facilitate the comparison with
the four dimensional literature and in particular with the results
of~\cite{Harmark:2006di,Hikida:2006qb} in what follows we shall
replace the basic variable $x$ with $x^2$. In \cite{Hikida:2006qb}
the $x_c(k)$ for $k=10$ is $0.104689$ which is already very close
to (\ref{criticaltemp2}).

 Above this critical value the integral
\eqref{matrix-model3a} is no longer dominated by the trivial
minimum $\rho_n=\bar\rho_n=0$ and one has to look for other
saddle-points \cite{Sundborg:1999ue,Aharony:2003sx}. Following
\cite{Aharony:2003sx}, one can easily show that above $x_c$ the
dynamics is governed by a distribution different from zero only in
the interval $[-\theta_0,\theta_0]$ and given, in first
approximation \footnote{We are assuming that the relevant features
are completely captured by the first mode $n=1$.}, by \beq
\rho(\theta) = {\cos\left({\theta \over 2} \right)\over \pi \sin^2
\left({\theta_0 \over 2} \right)} \sqrt{\sin^2 \left({\theta_0
\over 2} \right) - \sin^2 \left({\theta \over 2} \right)} \ \ \
\mathrm{with} \ \ \ \cos^2\left({\theta_0\over 2}\right) =
\sqrt{1-\frac{1}{z_0^{tot.}(x)}}. \eeq This behavior  at $x_c$
produces   a first-order transition with the same qualitative
characteristics of the four-dimensional  model.

\subsection{Chemical potentials}
A natural and intriguing  generalization is to add chemical
potentials for the $R$-charges, while  maintaining the trivial
vacuum as a gauge background.

The critical equation has still the form (\ref{criticaltemp1}) but $4 z_0^{spin.}$ and $6
z_0^{scal.}$ are substituted by
\begin{equation}
\begin{split}
4 z_0^{spin.}\mapsto &\frac{2x}{(1-x)^2}\sum_{p=1}^4\left(x^{\frac{1}{4}}y^{-\widetilde{\Omega}_p}+
x^{-\frac{1}{4}}y^{\widetilde{\Omega}_p}\right)\\
6 z_0^{scal.}\mapsto     & \,x^{1/2}\frac{x +
    1}{(1-x)^2}\sum_{p=1}^3\left(y^{\Omega_p}+y^{-\Omega_p}\right),
\end{split}
\end{equation}
which is (\ref{partition3}) for $q_\alpha=0$.
 The effect of small
chemical potentials can be easily computed by treating them as a perturbation and expanding around
$(\Omega_1,\Omega_2,\Omega_3)=(0,0,0)$. This yields the following result
\begin{equation}
\begin{split}
T_H(\Omega)&=T_H(0)-0.113946\sum_{i=1}^3\Omega_i^2-0.054438\prod_{i=1}^3\Omega_i\\&-0.036442
\sum_{i=1}^3\Omega_i^4-0.014059\sum_{i<j}\Omega_i^2\Omega_j^2+ O(\Omega^5),
\end{split}
\end{equation}
where all the numerical coefficients are actually known exactly, but their explicit expression is
long and irrelevant. The presence
of small chemical potentials \emph{decreases} the Hagedorn temperature. 

\begin{figure}[h]
\begin{center}
\includegraphics[width=10cm]{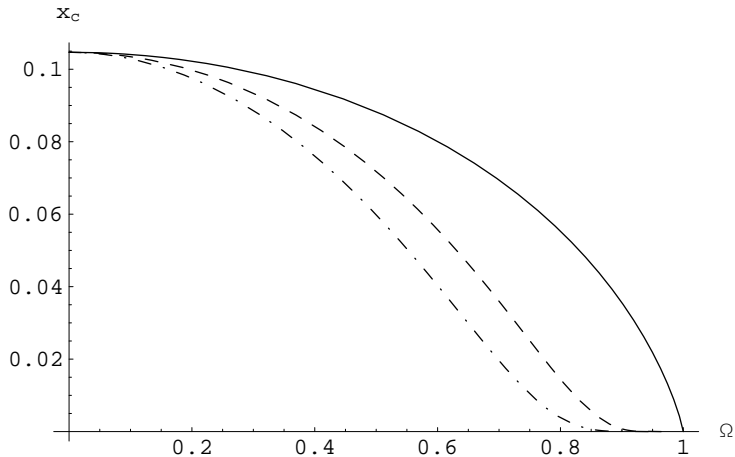}\\
\end{center}
\caption{\footnotesize The continuous, dashed and dot-dashed lines correspond to \label{chempotfig1}
$(\Omega_1,\Omega_2,\Omega_3) = (\Omega, 0, 0)$, $(\Omega_1,\Omega_2,\Omega_3) =(\Omega,\Omega,0)$
and $(\Omega_1,\Omega_2,\Omega_3) =(\Omega,\Omega,\Omega)$ respectively. All the curves reach
$\Omega=1$ when $x_c$ approaches zero.}
\end{figure}

In fig.\ref{chempotfig1} we display the dependence of the critical
temperature $T_{H}$ for the three particular choices of critical
potential $(\Omega_1,\Omega_2,\Omega_3) = (\Omega, 0, 0)$,
$(\Omega_1,\Omega_2,\Omega_3) =(\Omega,\Omega,0)$ and
$(\Omega_1,\Omega_2,\Omega_3) =(\Omega,\Omega,\Omega)$.
In all three cases, the behavior around $\Omega=1$, in a trivial
vacuum background, is similar to that of the $\mathcal{N}=4$
theory in four dimensions discussed
in~\cite{Yamada:2006rx,Harmark:2006di}. We find, in fact: \beq
\begin{split}
&(\Omega_1,\Omega_2,\Omega_3) = (\Omega, 0, 0):\ \ \
T_H=-\frac{1}{\log(1-\Omega)}\left[1-\frac{\log\left(-
\log[1-\Omega]\right)}{\log(1-\Omega)}+\dots\right],\\
&(\Omega_1,\Omega_2,\Omega_3) = (\Omega, \Omega, 0):\ \ \ T_H=\frac{1-\Omega}{\log
2}\left[1-\frac{1}{\log 2}e^{-\frac{\log 2}{2(1-\Omega)}}+\mathcal{O}(e^{- \frac{\log
2}{(1-\Omega)}})\right],\\
&(\Omega_1,\Omega_2,\Omega_3) = (\Omega, \Omega, \Omega):\ \ \ T_H=\frac{1-\Omega}{\log
4}\left[1-\frac{30}{\log 4}e^{-2\frac{\log 4}{(1-\Omega)}}+\mathcal{O}(e^{-3\frac{\log
4}{(1-\Omega)}})\right]\ ,
\end{split}
\end{equation}
which have the same qualitative behavior of the analogous equations found in~\cite{Harmark:2006di}
for the $\mathcal{N}=4$ theory. This similarity suggests the possibility to consider decoupling
limits analogous to those performed in~\cite{Harmark:2006di} for the $\mathcal{N}=4$ theory. This
might help to single out some subsectors of the present model with simple properties at the (full)
quantum level~\cite{Harmark:2006ta,Harmark:2006ie,Harmark:2007px}. However, this analysis is left for
future research.

\subsection{High temperatures}
In the high temperature regime the eigenvalue distribution becomes
almost like a delta-function \cite{Aharony:2003sx}. Therefore
$\rho_n=1$ and the free energy can be evaluated by looking at the
expression of the functional determinants in the background of
vanishing flat-connections. When the chemical potentials are
strictly zero the leading contribution to the free energy
$F=-T\log{\cal Z}$ is (see (\ref{Appdet9}), (\ref{Appdet14}) and
(\ref{Appdet15})) \beq
F=-\frac{7}{\pi}\,\zeta(3)\,V(S^2)N^2T^3+{\cal
O}(T^2).\label{alta}\eeq We see that the limiting free energy
density here coincides precisely with that of the ${\cal N}=8$
super Yang-Mills theory in flat three-dimensional space. Taking
the dimensionless parameter $TR$ to infinity is equivalent to
taking the limit of large volume at fixed temperature, loosing in
this way any memory of the original deformed supersymmetry. We can
also notice that no dependence appears, at the leading order, on
the particular monopole vacuum on which the expansion has been
performed and the result (\ref{alta}) is actually general.

It is interesting to consider the corrections to this result when
chemical potentials are taken into account. The first non-trivial
contribution is easily evaluated by using the expansions of ${\rm
Li}_3(z)$ presented in (\ref{Appdet15}): we simply notice that
chemical potentials appear as imaginary parts of the
flat-connections and are contained in the variable $z$ introduced
in the appendix \ref{determinants}. Summing carefully the
contributions coming from bosons and fermions, we obtain the free
energy \beq \label{HighT1}
F=-V(S^2)N^2T^3\left[\frac{7}{\pi}\,\zeta(3)+\sum_{i=1}^{3}\frac{y_i^2}{4\pi}\left(3-\log\frac{y_i^2}{4}\right)\right]+{\cal
O}(T^2), \eeq where we introduced the relevant combination
$y_i=\Omega_i/T$. This result is perfectly consistent with the
computation performed in \cite{Harmark:1999xt}, for a system of
$N$ free D2 branes in the presence of chemical potentials.

\sectiono{Thermodynamics in non-trivial vacua I}
\label{Thermobps1}

We shall first consider the multi-matrix model,
\eqref{matrix-model12}, which originates from the \textit{uncharged
vacuum.} We recall that in this case the partition function is
defined by the matrix integral \beq \label{matrix-model1}
\mathcal{Z}_A=\int\prod_{I=1}^k [dU_I] \exp\left(-\beta
V_0+\sum_{IJ}\sum_{n=1}^\infty\frac{1 }{n}z^{tot.}_{IJ}(x^n)
\mathrm{Tr}(U_I^n)\mathrm{Tr}(U_J^{\dagger n})\right), \eeq where
\beq V_0=r \displaystyle{\sum_{\alpha\in\mathrm{roots}}(4
|q_\alpha|^2 +|q_\alpha|)}\eeq is the Casimir energy. The value of
the Casimir energy is puzzling not only for the $r$ dependence,
making its sign ambiguous, but also because it depends on the charge
of the vacuum $q_\alpha$ so that it is different for different
vacua. At the supergravity level we expect instead these vacua to be
degenerate. This last feature is reproduced within our second
regularization choice, giving a vanishing Casimir energy and
consequently degenerate vacua: the price we pay is the introduction
of the logarithmic interactions (\ref{matrix-model21b}) that will be
studied in the next section.

The large $N$-limit of the matrix-model (\ref{matrix-model1}) is
investigated by generalizing to a multidimensional case the
technique presented in the previous section: we introduce the
density functions $\rho_I(\theta_I)$ associated to the matrices
$U_I$ and in terms of the Fourier-modes $\rho_{In}$ \beq
\label{matrix-model3}
\rho_I(\theta_I)=\frac{1}{2\pi}+\sum_{n=1}^\infty(\rho_{I n} e^{i
n\theta_I}+\bar{\rho}_{I n} e^{-i n\theta_I}), \eeq the matrix
integral (\ref{matrix-model3}) reduces as well to an infinite set
of independent gaussian integrals \beq \label{matrix-model5}
\begin{split}
\!\! \mathcal{Z}_A\!&=\!\!\!\int\prod_{I=1}^k D\rho_{In}
D\bar{\rho}_{In}\exp\!\!\left(-\beta V_0-N^2\sum_{IJ}
\sum_{n=1}^\infty\rho_{In} \bar{\rho}_{Jn}\underbrace{\frac{1 }{n}
(\delta_{I J}-z^{tot.}_{IJ}(x^n))
 s_I s_J}_{V_{IJ}(x^n)}\right),
\end{split}
\eeq where we have introduced the filling fractions $s_I=N_I/N$.
In the large $N$ limit (\ref{matrix-model5}) is dominated by the
absolute minimum of the quadratic action \beq
\label{matrix-model6} S=\sum_{IJ} \sum_{n=1}^\infty\rho_{In}
\bar{\rho}_{Jn}V_{IJ}(x^n), \eeq which is given by $\rho_{In}=0$
for every $I$ and $n$ if the quadratic form $V_{IJ}(x^n)$ is
positive definite. For small temperatures, namely small $x$, the
eigenvalues of the matrix $V_{IJ}(x^n)$ are all positive and close
to $1/n$ since $V_{IJ}(x^n) \sim\frac{1}{n}\delta_{IJ}$ for $x\ll
1$ (we recall that $z_{IJ}^{tot.}(x^n)$ vanishes as $x$ approaches
zero). Therefore the partition function is simply given by the
Casimir contribution at the leading order and by the small
fluctuation around the minimum at the subleading order \beq {\cal
Z}_A \propto e^{-\beta V_0}
\prod_{n=1}^{\infty}\frac{1}{\det\left(V_{IJ}(x^n)\right)}. \eeq
When we increase the temperature, $x$ approaches $1$ and the above
description is reliable until the quadratic form $V_{IJ}(x^n)$
develops the first  negative eigenvalue. This occurs at the
smallest $x_c$ for which one of the eigenvalues of $V_{IJ}(x^n)$
vanishes, or equivalently for which \beq \label{matrix-model8}
\det\left(V_{IJ}(x^n_c)\right)=0. \eeq The smallest
$x_c=e^{-1/T_c}$, namely the smallest critical temperature, is
obviously obtained for $n=1$ which provides the strongest
condition. Moreover this critical value always exists since
$z_{IJ}^{tot.}(x)$ is a monotonic function ranging from $0$ to
infinity when $x\in[0,1]$.

We are now ready to investigate the dependence of the critical
temperature on the non trivial monopole background. We start by
considering a configuration $\mathfrak{f}$ with just two sectors of
equal length. It is given by \beq \label{back1}
\mathfrak{f}=(n_1,\dots,n_1,n_2,\dots,n_2)\ . \eeq The $z_{IJ}$ and
thus the critical temperature depend only on the absolute effective
charge, namely $ q={|n_1-n_2|}/{2}$. This property reflects the fact
that the global $U(1)$ sector of charge $(n_1+n_2)/2$ does not
affect the thermodynamics in the large $N$ limit, since there are no
degree of freedom which couples to it. We also observe that the
critical equation is independent of the filling fractions $s_I$ and
it is obtained by requiring the vanishing of the determinant \beq
\det\begin{pmatrix}1-z^{tot.}_{11}(x) & -z_{12}^{tot.}(x)\\
-z_{21}^{tot.}  &
1-z^{tot.}_{22}(x)\end{pmatrix}=(1-z_{0}^{tot.})^2-(z_{12}^{tot.})^2
=0,\eeq where we have used that the matrix $V_{IJ}$ is symmetric
($z_{12}^{tot.}=z_{21}^{tot.}$) and that
$z_{11}^{tot.}=z_{22}^{tot.}=z_{0}^{tot.}$ is the partition
function in the trivial vacuum. This equation naturally splits
into two simpler equations \begin{eqnarray} &(a)&:\ \ \ \ \
\lambda_-(x)=1-z_{0}^{tot.}(x)-
z^{tot.}_{12}(x)=0\label{Hag2}\\&(b)&:\ \ \ \ \
\lambda_+(x)=1-z_{0}^{tot.}(x)+z^{tot.}_{12}(x)=0.
\end{eqnarray} The critical temperature is determined by the
lowest zero of these two equations. Since $\lambda_+-\lambda_-=2
z_{12}\ge0$ and $\lambda_+(0)=\lambda_-(0)=1$, $\lambda_-(x)$
reaches its zero at a smaller temperature: in determining $x_c$ we
can then neglect $\lambda_+(x)$.

From the structure of the critical equation, $\lambda_-(x)=0$, we can deduce two general properties
of the critical temperature. First, the positivity of $z^{tot}_{12}$ also ensures that
$\lambda_-(x)\le \lambda_0(x)=(1-z_0^{tot})$. This means that the critical temperature in a
non-trivial monopole background will always be smaller than the corresponding one in the trivial
vacuum. Second, the function $z_{12}^{tot.}$ decreases with the monopole charge $q$ (in the interval
$x\in[0,1]$): this implies that the critical temperature increases with the monopole charge. When $q$
approaches infinity the value of the critical temperature becomes that of the trivial vacuum. Below
we present a table for the critical temperature, where the behaviors described above are manifest
\begin{center}
\begin{tabular}{|c|c|c|}
  \hline
  q & $x_c$ & $T_c$ \\
  \hline
 ${1/2}$   &$0.085786$   &$0.407183$  \\
  $1$ &  $0.099771$ & $0.433863$  \\
 ${3}/{2}$ & $0.103842$  &$0.441523$   \\
  $2$& $0.104567$  & $0.442884$  \\
  ${5}/{2}$ &$0.104672$   &$0.443081$   \\
  $3$ & $0.104686$  &  $0.443107$ \\
  ${7}/{2}$ & $0.104688$  &  $0.443111$ \\
  \hline
\end{tabular}

\medskip
{\small Table 1: $x_c$ and $T_c$ in the two sectors situation as a function of the relative monopole
charge $q$.}
\end{center}

When the number $k$ of sectors grows, the dependence of the critical temperature $T_c$ on the
relative monopole charges becomes quite  intricate. However, some general behaviors can be
anticipated. Consider, for example, a generic background of the form \beq
\mathfrak{f}=(n_1,\dots,n_1,n_2,\dots,n_2,\dots\dots,n_k,\dots,n_k), \eeq where the induced relative
monopole charges \beq q_{IJ}=\frac{|n_{I}-n_{J}|}{2} \eeq are large, namely $n_I$ and $n_J$ are very
different from each other. Then the Hagedorn temperature is dominated by the smallest charge and the
off-diagonal terms associated  to the other charges can be considered as small perturbations. The
determinant is approximately given by \beq \det(V_{IJ})\approx
(1-z_{0})^{k-2}((1-z^{tot}_{0})^2-(z^{tot}_{q_{min}})^2). \eeq Exploiting what we have learned for
the $k=2$ system, the lowest transition temperature is an approximate solution of the equation
$1-z^{tot}_{0}-z^{tot}_{q_{min}}=0$.

Another interesting family of configurations is built by
considering long sequences of sectors with equal length and
monopole charge increasing by a fixed value $\mathfrak{q}$, namely
\beq
\mathfrak{f}=(n_0,\dots,n_0,n_0+\mathfrak{q},\dots,n_0+\mathfrak{q},n_0+2\mathfrak{q},\dots,n_0+2\mathfrak{q},\dots\dots,
n_0+k \mathfrak{q},\dots,n_0+k\mathfrak{q}). \eeq When  the number
of sectors $k$ goes to infinity, the Hagedorn temperature in these
vacua approaches that of $\mathcal{N}=4$ on the Lens space
$S^3/\mathbb{Z}_\mathfrak{q}$ in the sector described by a
vanishing flat-connection. For example for $\mathfrak{q}=1$, a
simple numerical analysis shows that $T_c$ goes to that of pure
$\mathcal{N}=4$, $T^{D=4}_c=-1/\log(7-4\sqrt{3})\simeq
0.379663$~\cite{Aharony:2003sx}, (see table below).
\begin{center}
\begin{tabular}{|c|c|c|}
  \hline
  $k$ & $x_c$ & $T_c$ \\
  \hline
  $2$ & $0.085786$ & $0.407184$  \\
  $3$ & $0.079653$ & $0.395245$  \\
  $10$& $0.072873$ & $0.381820$  \\
  $15$& $0.072312$ & $0.380697$  \\
  $20$& $0.072098$ & $0.380267$  \\
  $30$& $0.071936$ & $0.379942$  \\
  $60$& $0.071833$ & $0.379736$  \\
  \hline
\end{tabular}

\medskip
{\small Table 2: $x_c$ and $T_c$ in the $k$ sectors situation at
$\Omega=0$. The vacua are labelled by $\mathfrak{f}_k =
\mathrm{diag}(k-1,...,k-2,...,0)$.}
\end{center}
Analytically, this result can be argued by noting that the matrix
$V_{IJ}$, of which we have to compute the determinant, is of
Toeplitz type, namely a matrix in which each descending diagonal
from left to right is constant. Consequently its entries do not
depend on $I$ and $J$ separately, but only on the difference
$I-J$. For this kind of matrices, when the dimension is large, the
determinant is approximated by that of their circulant
version~\cite{mcgray}. This means that the smallest zero of the
determinant can be found as a solution of \beq \label{circulant}
1-\sum_{k=-\infty}^\infty z^{tot.}(k \mathfrak{q},x)=0, \eeq which
is the smallest eigenvalue of the corresponding circulant matrix.
In (\ref{circulant}) $z^{tot.}(k \mathfrak{q},x)$ is the
single-particle partition function in the sector of charge $k
\mathfrak{q}$. It is now possible to show that this infinite sum
produces the single-particle partition function  of the
$\mathcal{N}=4$ SYM theory in the trivial vacuum of
$S^3/\mathbb{Z}_\mathfrak{q}$ (see \cite{Hikida:2006qb} for
comparison). In other words (\ref{circulant}) coincides with the
critical equation for the $\mathcal{N}=4$ SYM theory in the
trivial vacuum of $S^3/\mathbb{Z}_\mathfrak{q}$.

\begin{figure}[h]
\begin{center}
\includegraphics[width=10cm]{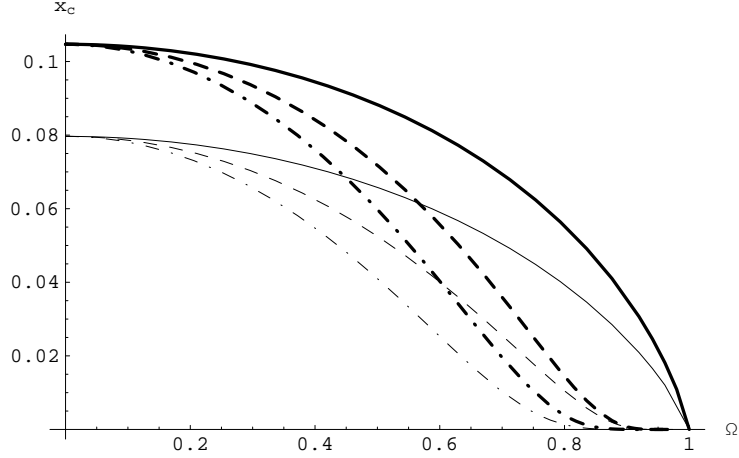}\\
\end{center}
\caption{\footnotesize Transition lines for three sectors vacuum:
$\mathfrak{f}=\mathrm{diag}(i,...,0,...,-i)$.  Narrow lines corresponds to $i=1$, thick lines to
$i=10$. The convention for continuous, dashed and dot-dashed are those of fig. \ref{chempotfig1}. The
qualitative behavior is the same for every number of sectors.} \label{chempotN=3}
\end{figure}

Finally we consider the addition of chemical potentials to a
monopole configurations. Their introduction does not alter
significantly the picture and a numerical analysis is given in
fig.~\ref{chempotN=3}.

\subsection{Just above the critical temperature}

To understand what happens when we cross the critical temperature,
we shall now focus our attention on the two-sectors configuration
(\ref{back1}). In this case, if we introduce the combination \beq
\rho_\pm=\frac{1}{2}(\rho_1\pm\rho_2)~, \eeq the action takes a
diagonal form \beq S=2 \sum_{n=1}^\infty\left(
\frac{1}{n}\lambda_-(x^n)\bar{\rho}_{+n}\rho_{+n}+
\frac{1}{n}\lambda_+(x^n)\bar{\rho}_{-n}\rho_{-n}\right). \eeq
Above the critical temperature, $\lambda_-(x)$ is negative and the
dominant saddle-point is no longer realized by a flat distribution
$\rho_{1n}=\rho_{2n}=0$ ($\rho_{+n}=\rho_{-n}=0$). In fact, as
the temperature is increased, the attractive term in the pairwise
potential continues to increase in strength, so the eigenvalues
become increasingly bunched together, occupying, at the end, only
a finite interval $I = [-\theta_0, \theta_0]$ on the circle (we
arbitrarily choose the middle of this interval to be at $\theta=
0$ for convenience). However, since $\lambda_+(x)$ is still
positive, we can safely assume that the new dominant saddle point
satisfies \beq \label{r1=r2}
\rho_{-n}=\frac{1}{2}(\rho_{1n}-\rho_{2n})=0,\ \ \ \
\mathrm{i.e.}\ \ \rho_1=\rho_2=\rho_+\ . \eeq In other words, the
problem reduces to an effective one matrix model governed by the
action \beq
\begin{split}
S=&2\int d\theta d\theta^\prime \rho_+(\theta)\rho_+(\theta^\prime)
\sum_{n=1}^\infty \left[\frac{(\lambda_{-}(x^n) -1)}{n}
\cos( n(\theta-\theta^\prime))\right]+\\
&+ 2\int d\theta d\theta^\prime \rho_+(\theta)\rho_+(\theta^\prime)
\log\left|\sin\frac{\theta-\theta^\prime}{2}\right|,
\end{split}
\eeq where the distribution function has support in the interval
$[-\theta_0, \theta_0]$. In complete analogy with what we found in
trivial vacuum case (\ref{above}), we have \beq\label{above}
\rho(\theta) = {\cos\left({\theta \over 2} \right)\over \pi \sin^2
\left({\theta_0 \over 2} \right)} \sqrt{\sin^2 \left({\theta_0
\over 2} \right) - \sin^2 \left({\theta \over 2} \right)} \ \ \
\mathrm{with} \ \ \ \cos^2\left({\theta_0\over 2}\right) =
\sqrt{{\lambda_-(x) \over \lambda_-(x)-1}}. \eeq Near the critical
temperature, for $T>T_{H}$ we have the following expansion for the
partition function \beq {F\over N^2} = {T_H \over 2} \lambda_-(x)
+ {\cal O}((\lambda_-)^2) = {T_H\over 2} (T - T_H) {{\partial
\lambda_-} \over {\partial T}} \bigg|_{T=T_H} + {\cal
O}((T-T_H)^2), \eeq which gives the characteristic first-order
transition, already found   in the four-dimensional
 model.

\sectiono{Thermodynamics in non-trivial vacua II}
\label{Thermobps2} We discuss now the thermodynamical behavior
arising when the second regularization scheme, considered for the
fermions in section~4, is adopted. As previously derived, a
non-trivial logarithmic deformation of the multi-matrix model
\eqref{matrix-model1} has to be considered \beq
\label{matrix-model21} \mathcal{Z}_B=\int\prod_{I=1}^k [dU_I]
\exp\left(\sum_{IJ}\sum_{n=1}^\infty\frac{1 }{n}z^{tot.}_{IJ}(x^n)
\mathrm{Tr}(U_I^n)\mathrm{Tr}(U_J^{\dagger
n})\right)\prod_{I=1}^k\det(U_I)^{(N n_I -Q)}. \eeq To illustrate
the effect of the new interactions on the large $N$ dynamics, we
shall make a very drastic assumption and we shall focus our
attention just on the first winding, $n=1$. With this choice the
original matrix integral reduces to \beq \int\prod_{I=1}^k DU_I
\exp\left(\sum_{IJ} z^{tot.}_{IJ}(x)
\mathrm{Tr}(U_I)\mathrm{Tr}(U_J^{\dagger
})\right)\prod_{I=1}^k\det(U_I)^{N n_I -Q}. \eeq It is useful, as a
first step, to introduce a set of $k$ complex Lagrange multipliers
$\lambda_I$ and the partition function can be written as\beq
\begin{split}
&\frac{\prod_{I=1}^k N^{2}_I}{(\det(z_{IJ}))^k}\int \prod_{J=1}^k d\lambda_J
d\bar{\lambda}_J\exp(-\sum_{IJ}N_I N_J\bar \lambda_I z^{-1}_{IJ}(x)\lambda_J)\times\\
&\prod_{I=1}^k \int DU_I \exp\left(\bar \lambda_I N_I \mathrm{Tr}(U_I)+\lambda_I
N_I\mathrm{Tr}(U_I^{\dagger })\right)\det(U_I)^{ N n_I -Q}.
\end{split}
\eeq Next we use the polar decomposition $\lambda_I=\gamma_I
e^{i\alpha_I}$ for each Lagrange multipliers. The phases
$e^{i\alpha_I}$ are  then decoupled from the matrix integration by
means of the change of variables $U_I\mapsto U_I e^{i\alpha_I}$.
This procedure yields the following integral
 \beq
 \label{saddle1}
\begin{split}
\frac{\prod_{I=1}^k N^{2}_I}{(\det(z_{IJ}))^k}\int &\prod_{J=1}^k d\gamma_I
d\alpha_I\exp\left(-\sum_{IJ}N_I N_J\gamma_I z^{-1}_{IJ}(x)
e^{i(\alpha_J-\alpha_I)}\gamma_J+ i\sum_{I=1}^k N_I (N n_I -Q)\alpha_I\right)\\
&\times\prod_{I=1}^k \int DU_I \exp\left(\gamma_I N_I( \mathrm{Tr}(U_I)+\mathrm{Tr}(U_I^{\dagger
}))\right)\det(U_I)^{N n_I -Q}.
\end{split}
\eeq In \eqref{saddle1} the group integrations over the unitary matrices $U_I$ are completely
decoupled. Each matrix integration corresponds to a Gross-Witten model~\cite{Gross:1980he} with a
coupling $\gamma_I$ and an additional logarithmic potential proportional to $\log(\det(U_I))$. We
remark that these kinds of deformations for unitary matrix models were widely considered in the early
eighties (see e.g. \cite{Green:1981mx,Rossi:1982vw}). The determinant operator was expected to act as
an order parameter for the large $N$ phase transitions characterizing this class of models
\cite{Rossi:1996hs}: unfortunately, we cannot simply borrow the old results. In \eqref{saddle1} in
fact we have a new and decisive ingredient with respect to the original investigations: the power of
the determinant is not a fixed number, but it grows linearly with $N$. This last feature dramatically
alters the usual large $N$ dynamics since the integral \eqref{saddle1} is not dominated anymore by
the same family of saddle-points of the familiar Gross-Witten model, as we will see in the following.

\subsection{Solution of unitary matrix model with logarithmic potential}

\label{Pennersec} The phase structure of \eqref{saddle1} can be
naturally studied along the lines proposed in
\cite{Alvarez-Gaume:2006jg}.  We will first perform the integration
over the unitary matrices and then the integration over the Lagrange
multipliers. We will then start by studying the large $N$ properties
of the reduced model \beq \label{Penner} \mathcal{Z}(\gamma,p)=\int
DU \exp\left(\gamma N( \mathrm{Tr}(U)+\mathrm{Tr}(U^{\dagger
}))\right)\det(U)^{N p}, \eeq where $N p$ is an integer, whose sign
is irrelevant because we can transform $N p$ into $-N p$ by
performing the change of variable $U\mapsto U^\dagger$. For this
reason, from now on, we shall take $p$ to be positive. The first
important effect of the new logarithmic interaction concerns the
small $\gamma$ behavior of (\ref{Penner}): differently from the
Gross-Witten model ($p=0$), where $\mathcal{Z}(\gamma,p)$ is finite
as $\gamma$ approaches zero, here $\mathcal{Z}(\gamma,p)$  vanishes
as $\gamma^{N^2 p}$. This leading behavior is determined by
expanding the exponential around $\gamma=0$ and performing the
integral term by term. The first non-vanishing contribution is fixed
by the selection rule imposed by the $U(1)$ factor present in $U(N)$
and it is given by
 \beq
\begin{split}
\mathcal{Z}(\gamma,p)\approx& \frac{(\gamma N)^{N^2 p}}{(N^2 p)!}\!\!\int\!\! DU~
\mathrm{Tr}(U^{\dagger })^{N^2p}\det(U)^{N p} \!=\!{(\gamma N)^{N^2 p}}\!\prod_{i=0}^{N-1}\!\!\frac{
i!}{(i+N p)!}\!=\!{(2\gamma)^{N^2 p}}e^{N^2 C}\!,
\end{split}
\eeq where the constant $C$ in the large $N$ limit is given by \beq
C =-\frac{1}{2} \left( (\log (4)-3) p+(p+1)^2 \log (p+1)-p^2\log
(p)\right). \eeq
 In other words, the free
energy ${\cal F}(\gamma,p)=\log{\cal Z}(\gamma,p)=N^2{\cal
F}_0(\gamma,p)+..$ of the present unitary matrix model starts, at
leading $N^2$ order, with a logarithmic singularity similar to the
one of the usual Penner model \cite{Distler:1990mt,Ambjorn:1994bp}
\beq \label{boundary0} {\cal
F}_0(\gamma,p)=p\log(2\gamma)+C+O(\gamma^2). \eeq This new
behavior suggests that the usual strong-coupling expansion of
\eqref{Penner} might be radically different from that of the
Gross-Witten model, which is simply given by $e^{N^2\gamma^2}$. To
explore this idea,  one could perform a full strong-coupling
expansion and to resum the resulting series in the large $N$
limit; however the presence of the determinant factor much
complicates this approach.  Here, we shall choose a simpler path
and consider a different expansion, peculiar of the present model,
namely $p$ very large. In this limit we can perform a
semiclassical analysis on the integral (\ref{Penner}): the
relevant classical potential is, in this case, \beq p
V(\theta_i)=p N\left(2\frac{\gamma}{p}\sum_{i=1}^N \cos\theta_i+ i
\sum_{i=1}^N\theta_i\right). \eeq The equations for the critical
point are easily derived and solved (we will denote from now on
$4\gamma^2=t$)\beq -\frac{\sqrt{t}}{p}\sin\theta_i+ i=0\ \ \
\Rightarrow \ \ \ \theta_i = i \sinh ^{-1}\left(\frac{p}{\sqrt{t}
}\right). \eeq The semiclassical approximation is then obtained by
expanding the classical action around the critical point up to the
quadratic order \beq N^2 \left(\sqrt{{p^2}+t}-p \sinh
^{-1}\left(\frac{p}{\sqrt{t} }\right)\right)-\frac{N}{2} \left(
\sqrt{p^2+t} \right)\sum_{i=1}^N
\hat{\theta}_i^2+O\left(\hat{\theta}_i^3\right), \eeq with \beq
\hat{\theta}_i\equiv\theta_i -i \sinh ^{-1}\left(\frac{p}{\sqrt{t}
}\right). \eeq We remark that this is a good approximation as long
as $\sqrt{p^2+t}\gg1$: in this limit the gaussian integration
covers the whole real line and the Haar measure over the unitary
matrices becomes the usual measure over the hermitian matrices. We
can easily perform the integration over the angles $\theta_i$ and
up to a constant independent of $p$ we get \beq
\begin{split}
\label{poppa} {\cal F}_0(t,p)&= \left(\sqrt{{p^2}+t}-p \sinh
^{-1}\left(\frac{p}{\sqrt{t}
}\right)- 1/2\log( \sqrt{p^2+t})\right)=\\
&= \left(\sqrt{{p^2}+t}-p \log \left(\frac{p}{\sqrt{t}}+\sqrt{\frac{p^2}{t}+1}\right)- 1/2 \log(
\sqrt{p^2+t})\right).
\end{split}
\eeq For $p$ large and $t$ finite or small, we finally arrive to the following expansion
\beq\label{plarge}
\begin{split}
{\cal F}_0(t,p)=&p\left(\frac{\log (t)}{2}-\log
\left({p}\right)-\log (2)+1\right)-\frac{1}{2} \log (p)+\frac{t}{4
p}-\frac{1}{4} t
\left(\frac{1}{p}\right)^2-\\
&-\frac{1}{32} t^2 \left(\frac{1}{p}\right)^3+\frac{1}{8} t^2
\left(\frac{1}{p}\right)^4+O\left(\left(\frac{1}{p}\right)^5\right).
\end{split}
\eeq This result is quite remarkable: we see that the above
expansion reproduces exactly the large $p$ limit of
(\ref{boundary0}) and contains a infinite series of corrections in
powers of $t$. Since \eqref{plarge} holds also for small $t$, we
must conclude that the strong-coupling expansion of our deformed
Gross-Witten model leads to a non-trivial function of $t$ and $p$,
eventually encoding an intriguing modification of the $p=0$
result.

It is quite easy to repeat the same analysis taking $t$ large,
exploring in this way the deformation of the weak-coupling phase of
the familiar unitary model. In this limit we should obtain, at
leading order in $t$, the very same result for the free energy as in
the Gross-Witten case: again we could expect a non-trivial
deformation due to the presence of $p$. Actually, performing the
same steps as before, we get again
\eqref{poppa}\footnote{\label{ciro1001} The semiclassical
computation really holds for $\sqrt{p^2+t}>>1$ and this condition is
realized by taking either $p$ or $t$ large. Therefore we have to
obtain the same free energy \eqref{poppa} in the large $t$ case as
well.}, which expanded for large $t$ gives
\beq\label{boundaryinfinity}
\begin{split}
{\cal F}_0(t,p)=&-\frac{3}{4}+\sqrt{t}-\frac{1}{4}
\log(t)-\frac{1}{2} p^2 \sqrt{\frac{1}{t}}-\frac{p^2}{4
t}+\frac{1}{24} p^4 \left(\frac{1}{t}\right)^{3/2}+\frac{1}{8} p^4
\left(\frac{1}{t}\right)^2-\\
&-\frac{1}{80} p^6 \left(\frac{1}{t}\right)^{5/2}-\frac{1}{12} p^6
\left(\frac{1}{t}\right)^3+O\left(\left(\frac{1}{t}\right)^{7/2}\right).
\end{split}
\eeq We recognize in the first three terms the exact large $N$ result of the Gross-Witten
weak-coupling phase: it does not come as a surprise, being the semiclassical approximation exact in
this phase. As expected, we also observe an infinite series of corrections, depending on $p$, that
modify non-trivially the usual spherical free energy of the weak-coupling phase.

We do not expect, of course, that the above expansions yield the
exact large $N$ free energy: these results are semiclasssical, in
the sense that we missed the contribution of the Vandermonde
determinants associated to the measure over ${\it unitary}$
matrices, that is essential in recovering the correct spherical
free energy. Nevertheless they should capture the leading order
behavior at large $p$ or $t$ of the complete large $N$ answer, and
also a certain series of subleading terms (as we will explicitly
check in the following).

These computations suggest an intriguing possibility: we observe
non-trivial deformations of both strong and weak-coupling expansion
of the Gross-Witten model, involving complicated functions of $p$
and $t$. It is quite natural to conjecture, at this point, that a
unique non-trivial analytic function ${\cal F}_0(t,p)$ exists,
reproducing for $p\neq 0$ both behaviors and being the large $N$
free energy of the model. This is also suggested by the fact that
the same free energy \eqref{poppa} describes smoothly either the
large $p$ or the large $t$ region (see footnote \ref{ciro1001}). If
this is the case, the presence of the logarithmic interaction would
smooth out the third-order phase transition of the Gross-Witten
model, the parameter $p$ providing an analytic interpolation between
the strong and the weak-coupling phase.

In order to prove this idea, we have to solve exactly the large
$N$ dynamics: we shall exploit the beautiful relation between our
model and the Painlev\'e $\mathrm{III}$ system illustrated in
\cite{FW2002}. In that paper the authors have shown that it is
possible to construct an auxiliary function, \beq \label{rock}
\sigma(t)=-t\frac{d}{d t}\log\left((t N^2)^{N^2 p^2/2}e^{-N^2
t/4}\mathcal{Z}(t,p)\right), \eeq that satisfies, at finite $N$,
the following non-linear differential equation \beq \label{Pain1}
-\frac{1}{16} p^2 N^6+\left(p^2-1\right) \sigma '(t)^2 N^2+\sigma
'(t) \left(4 \sigma '(t)-N^2\right) \left(\sigma (t)-t \sigma
'(t)\right)+t^2 \sigma ''(t)^2=0. \eeq  In the large $N$ limit,
the spherical ansatz for the partition function
$\mathcal{Z}(t,p)=e^{N^2 {\cal F}_0(t,p)}$ dictates the following
scaling for the auxiliary $\sigma(t)$ \beq\sigma(t)=N^2
\rho(t).\eeq Thus, at the leading order in $N^2$, we obtain a nice
first-order differential equation for the reduced function
$\rho(t)$ \beq \label{Pain2} -4 t \rho '(t)^3+\left(p^2+t+4 \rho
(t)-1\right) \rho '(t)^2-\rho (t) \rho '(t)-\frac{p^2}{16}=0. \eeq
The analysis for small and large $t$  given in \eqref{boundary0}
and \eqref{boundaryinfinity} provides two possible boundary
conditions for the above equation:
\begin{itemize}
\item[(s):]\,\,  $\rho(t)|_{t=0}=-\frac{1}{2}(p^2+p)$;
\item[(w):]\,\,
$\rho(t)|_{t\to\infty}=\frac{t}{4}-\frac{1}{2}\sqrt{t}$.
\end{itemize}
Since \eqref{Pain2}  is a first-order differential equation, these
boundary values will correspond, in general, to two different
solutions: the former, which satisfies $(s)$, is denoted with
$\rho_s(t)$ and it is supposed to describe the strong-coupling
regime\footnote{In the matrix model language, $\gamma$ is
conventionally identified with the inverse of the fundamental
coupling constant. Thus small values of $t=4\gamma^2$ are in the
strong-coupling region.}; the latter, $\rho_w(t)$, obeys $(w)$ and
it is expected to  hold in the weak-coupling regime. The two
corresponding free energies ${\cal F}_0^{s,w}(t,p)$ are then
constructed by integrating the simple relation \beq \label{FEdef}
\frac{d{\cal F}_0^{s,w}(t,p)}{dt}=\left(\frac{1}{4}-\frac{p^2}{2
t}-\frac{\rho_{s,w}(t)}{t}\right) \eeq which follows from
\eqref{rock} once we have used the spherical ansatz
$\mathcal{Z}(t,p)=e^{N^2 {\cal F}_0^{s,w}(t,p)}$.

The above simple picture works very well  at $p=0$, where our
model reduces to the usual Gross-Witten model. In this case the
differential equation becomes extremely tractable, factorizing
into two simple first-order equations: the solution ${\cal
F}_0^s(t,0)$ and ${\cal F}_0^w(t,0)$ can be obtained explicitly
and they exactly coincides with the well-known free energies of
the model at strong and weak coupling. The condition ${\cal
F}_0^s(t,0)={\cal F}_0^w(t,0)$ defines the correct critical value
for the coupling constant $(t_c=1)$. When $p\not=0$, the situation
reserves some surprises as we shall illustrate below.

As thoroughly described in appendix \ref{D}, the general case can
be solved exactly, in spite of the apparent difficult
non-linearity of the differential equation. In particular there
are two relevant solutions, describing respectively the
deformations of $\rho_s(t)$ and $\rho_w(t)$ found in the
Gross-Witten case. Integrating (\ref{FEdef}) we get a candidate
${\cal F}_0^s(t,p)$ given by
 \beq
\label{kkksec} {\cal F}_0^s(t,p)= -\frac{1}{2} \left( (\log (4)-3)
p+(p+1)^2 \log (p+1)-p^2\log
(p)\right)+\frac{t}{4(1+p)}-\frac{p}{2}\log(t)~ ,\eeq while ${\cal
F}_0^w(t,p)$ has the form \beq \label{caro2sec}
\begin{split}
{\cal F}_0^w(t,p)=&f_w\!\!+\!\!
\left(\frac{p^2}{4\rho'_{w}}-\frac{p^2}{64
\left(\rho'_{w}\right)^2}+\frac{1}{2} \left(\log
\left(\rho'_{w}\right) p^2- 2 p\tanh
^{-1}\left(p+4\left(\frac{1}{p}-p\right)
   \rho'_{w}\right)+\right.\right.\\
   &\left.\left. +\log \left(1-4 \rho'_{w}\right)+\frac{2}{1-4 \rho'_{w}}\right)\right),
\end{split}
\eeq where the constant $f_w$ is given by \beq \label{caro3sec}
f_w=-\frac{3}{4}+\frac{1}{4} p ((-3+\log (16)) p-2 \log (p-1)+2
\log (p+1)). \eeq Here $\rho'_w(t)$ is the solution of the fourth
order algebraic equation \eqref{sol2}, which respects the large
$t$ behavior implied by the boundary condition $(w)$. One can
easily check that ${\cal F}_0^w(t,p)$ smoothly reduces, as $p$
goes to zero, to the free energy of the Gross-Witten model in the
weak-coupling phase, and accurately reproduces the semiclassical
expansion (\ref{boundaryinfinity}), up to higher order terms in
$p^{2n}/t^{n+m/2}$, coming from the exact large $N$ solution
encoded into the differential equation. It is also evident from
(\ref{kkksec}) that ${\cal F}_0^s(t,p)$ reproduces, in the limit
of vanishing $p$, the Gross-Witten strong-coupling result.

On the other hand, we already know that ${\cal F}_0^s(t,p)$, as
given by (\ref{kkksec}), $cannot$ provide the right solution
describing the small $t$ regime! The large $p$ expansion of
(\ref{kkksec}) is quite boring and does not reproduce the
non-trivial series (\ref{plarge}), obtained from the semiclassical
approximation. On the other hand it is possible to show that
${\cal F}_0^w(t,p)$, for $p\neq 0$, also satisfies the right
boundary condition to describe the strong-coupling region (see
appendix \ref{D}) and, more importantly, correctly reproduces
(\ref{plarge}) in the large $p$ limit (up to higher order
corrections in $t^n/p^{2n+m}$, coming from the exact large $N$
solution of the model).

We arrive therefore to the conclusion that the critical behavior of
the standard unitary matrix model is completely modified by the
addition of our logarithmic interaction. As long as $p\neq 0$ the
system is always in a ``weak-coupling" phase, described by the free
energy ${\cal F}_0^w(t,p)$: this solution has the correct boundary
condition both at small and at large $t$ and smoothly interpolates
between them. We also identify  ${\cal F}_0^s(t,p)$ with an {\it
unphysical}  solution of the differential equation (\ref{Pain2}) and
therefore we neglect it. The situation drastically changes for
$p=0$: it is possible to show that, starting from ${\cal
F}_0^w(t,p)$, the limiting behavior changes discontinuously at
$t=1$. On the other hand, taking $p=0$ at level of the differential
equation (\ref{Pain2}), the strong-coupling phase is instead encoded
into the solution $\rho_s(t)$.

\subsection{Phase-structure in non-trivial vacua}
In this subsection, we shall explore the consequences of the
previous results on the phase structure of the theory. After having
performed the integration over the unitary matrices in the deformed
Gross-Witten models, we are left with the integration over the
Lagrange multipliers
 \beq
 \label{saddle2}
\begin{split}
\int &\prod_{J=1}^k d\gamma_I  d\alpha_I\exp\left( N^2 \sum_{I=1}^k
(i s_I (n_I -q)\alpha_I+ s^2_I {\cal
F}_0(\gamma_I,p))-N^2\sum_{IJ}s_I s_J\gamma_I z^{-1}_{IJ}(x)
e^{i(\alpha_J-\alpha_I)}\gamma_J \right)
\end{split}
\eeq with $q=Q/N$. Since $N$ is  large, we can perform this integral
in the  saddle-point approximation as well. The saddle-points which
dominate this integration are determined by \beq \label{saddle3}
\begin{split}
2&\sum_{I=1}^k s_I \gamma_I z^{-1}_{IJ} \gamma_J \sin(\alpha_J-\alpha_I)+i (n_J-q)=0\\
-2&\sum_{I=1}^k s_I \gamma_I z^{-1}_{IJ}  \cos(\alpha_J-\alpha_I)+
s_J {\cal F}'_0(\gamma_J,p)=0.
\end{split}
\eeq To be concrete, we shall consider only the  case $k=2$: here
the relevant combinations of the parameters are given by $n_1-q=s_2
(n_1-n_2) \equiv s_2 n$ and $n_2-q=-s_1(n_1-n_2)\equiv-s_1 n $
($n=n_1-n_2>0$). The first equation  in \eqref{saddle3} then
produces two conditions \beq 2 \gamma_2 z^{-1}_{21} \gamma_1
\sin(\alpha_1-\alpha_2)+i n=0\ \ \ \mathrm{and}\ \ \ 2 \gamma_1
z^{-1}_{12} \gamma_2 \sin(\alpha_2-\alpha_1)-i  n=0.
\label{saddlea}\eeq These two equations are obviously equivalent and
they are solved by \beq \sin(\alpha_1-\alpha_2)=-i \frac{n}{2
\gamma_2 z^{-1}_{21} \gamma_1}\ \ \ \Rightarrow \ \ \ \
\cos(\alpha_1-\alpha_2)=\pm\sqrt{1+\frac{n^2}{4(\gamma_2 z^{-1}_{21}
\gamma_1)^2}}. \eeq Substituting this result into \eqref{saddle3},
the second equation provides two relations, which determines
$\gamma_1,\gamma_2$
 \beq
\begin{split}
&-2 s_1 \gamma_1^2 z^{-1}_{11} \mp 2 s_2   \sqrt{(\gamma_2
z^{-1}_{21} \gamma_1)^2+\frac{n^2}{4}}+s_1\gamma_1 {\cal
F}'_0(\gamma_1,p) =0,
\\
&\mp 2 s_1 \sqrt{(\gamma_2 z^{-1}_{21} \gamma_1)^2+\frac{n^2}{4}} -2
s_2 \gamma_2^2 z^{-1}_{11}  +s_2 \gamma_2 {\cal F}'_0(\gamma_2,p)=0.
\end{split}
\eeq In the following, we shall further simplify our example  and we
shall choose two sectors of equal length, namely we shall set
$s_1=s_2=1/2$. Then, by taking the difference of the two equations
and using the fact that ${\cal F}_0(t,p)$ is a monotonic function,
one can immediately show that $\gamma_1=\gamma_2$. We remain with
just one equation, which determines $t_1=4\gamma_1^2$ \beq
\mp\sqrt{\frac{1}{4}(z^{-1}_{12}t_1)^2 +{n^2}} -\frac{t_1}{2}
z^{-1}_{11}  +2 t_1 {\cal F}'_0(t_1,p) =0, \eeq which is
conveniently rewritten in terms of $\rho_w(t_1)$ as follows \beq
\label{pio} f_{\pm}(t_1)\equiv\pm
\sqrt{\frac{1}{4}(z^{-1}_{12}t_1)^2 +{n^2}}+ \frac{t_1}{2}
(1-z^{-1}_{11})  +n=2 \left(\rho_w(t_1)+\frac{n}{2}(n+1)\right).
\eeq When $t_1$ runs from zero to infinity, the r.h.s of (\ref{pio})
spans the same region. Thus a necessary condition for having a
non-trivial  solution is that the l.h.s. of (\ref{pio}) is not
negative definite. Let us discuss the first equation: $f_+(t)$ has
the following properties
\begin{eqnarray}
f_+(0)&=&2n,\,\,\,\,\,\,\,\,\,f'_+(0)=\frac{1}{2}\frac{\det
z-z_{11}}{\det z}\nonumber\\
f'_+(t)&=&0\,\,\,\Rightarrow\,\,
\,\,\,\,\,\,\,\,t^2=-4n^2\frac{(\det
z-z_{11})^2}{(z_{12})^2}\frac{\det z}{\det(1-z)}.
\end{eqnarray}
Moreover we have that for large $t$\beq
f_+(t)\to\frac{t}{2}\frac{z_{11}+z_{12}-1}{z_{11}+z_{12}}+n+{\cal
O}(1/t).\eeq We immediately conclude that for temperatures near zero
$(x\ll1)$, $f_+(t)$ is always decreasing: for $T<T_H$, where $T_H$
is the Hagedorn temperature defined by the equation \beq
z_{11}+z_{12}=1\eeq as in (\ref{Hag2}), we have that $\det(1-z)\geq
0$, implying that $f'_+(t)$ never vanishes for $t>0$. Therefore in
this range of temperature there is always one solution to the
saddle-point equation at $t\neq 0$. At the Hagedorn temperature
$T_H$ the function $f_+(t)$ is still decreasing but becomes positive
definite, asymptotically approaching the value $n$. Above $T_H$ we
see that $f_+(t)$ develops a minimum at finite $t$ and then becomes
monotonically increasing. The minimum disappears at the temperature
$T_2$ defined by \beq \det z=z_{11}\eeq and the function becomes
monotonically increasing for any $T>T_2$.
\begin{figure}[htbp]
\begin{center}
\includegraphics[height=8cm,width=12cm]{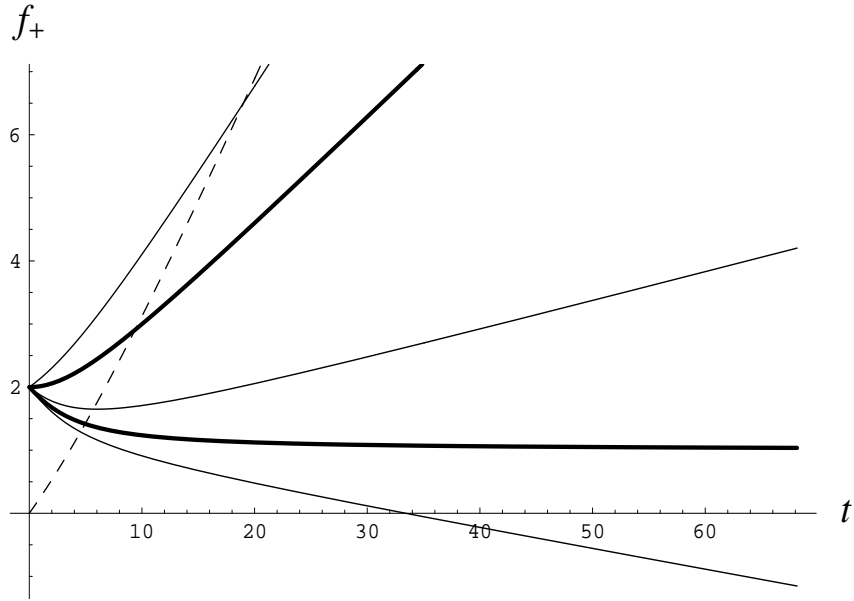}\\
\end{center}
\caption{\footnotesize Plot of $f_+(t)$ for different values of $T$
and $n=1$, $q=1/2$. Going bottom-up, the solid lines illustrate the
behavior for $T<T_{H}$, $T=T_{H}$ (lower thick line), $T_{H}<T<T_2$,
$T=T_2$ (upper thick line), $T>T_2$. The dashed line is the r.h.s.
of (\ref{pio}) as a function of $t$.}\label{graph:effepiu}
\end{figure}

\begin{figure}[htbp]
\begin{center}
\includegraphics[height=8cm,width=12cm]{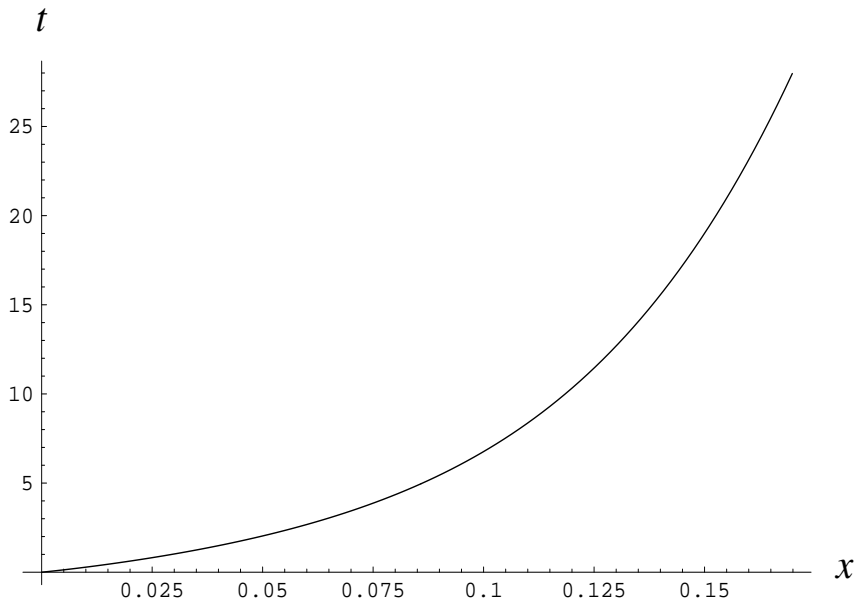}\\
\end{center}
\caption{\footnotesize Saddle point as a smooth function of the
temperature $x$ for $n=1$, $q=1/2$. The graph covers all the
different regimes, the Hagedorn temperature being $x_{H} =
0.0857864$ and $T_2$ corresponding to $x_2 =
0.115493$.}\label{graph:saddle}
\end{figure}

In spite of these changes of behavior, one can check that there is
always one solution to the saddle-point equation as shown, in
different regimes, in figure \ref{graph:effepiu}. Moreover the
position of this saddle-point changes smoothly as function of the
temperature (fig. \ref{graph:saddle}).

\begin{figure}[htbp]
\begin{center}
\includegraphics[height=5.5cm,width=7.7cm]{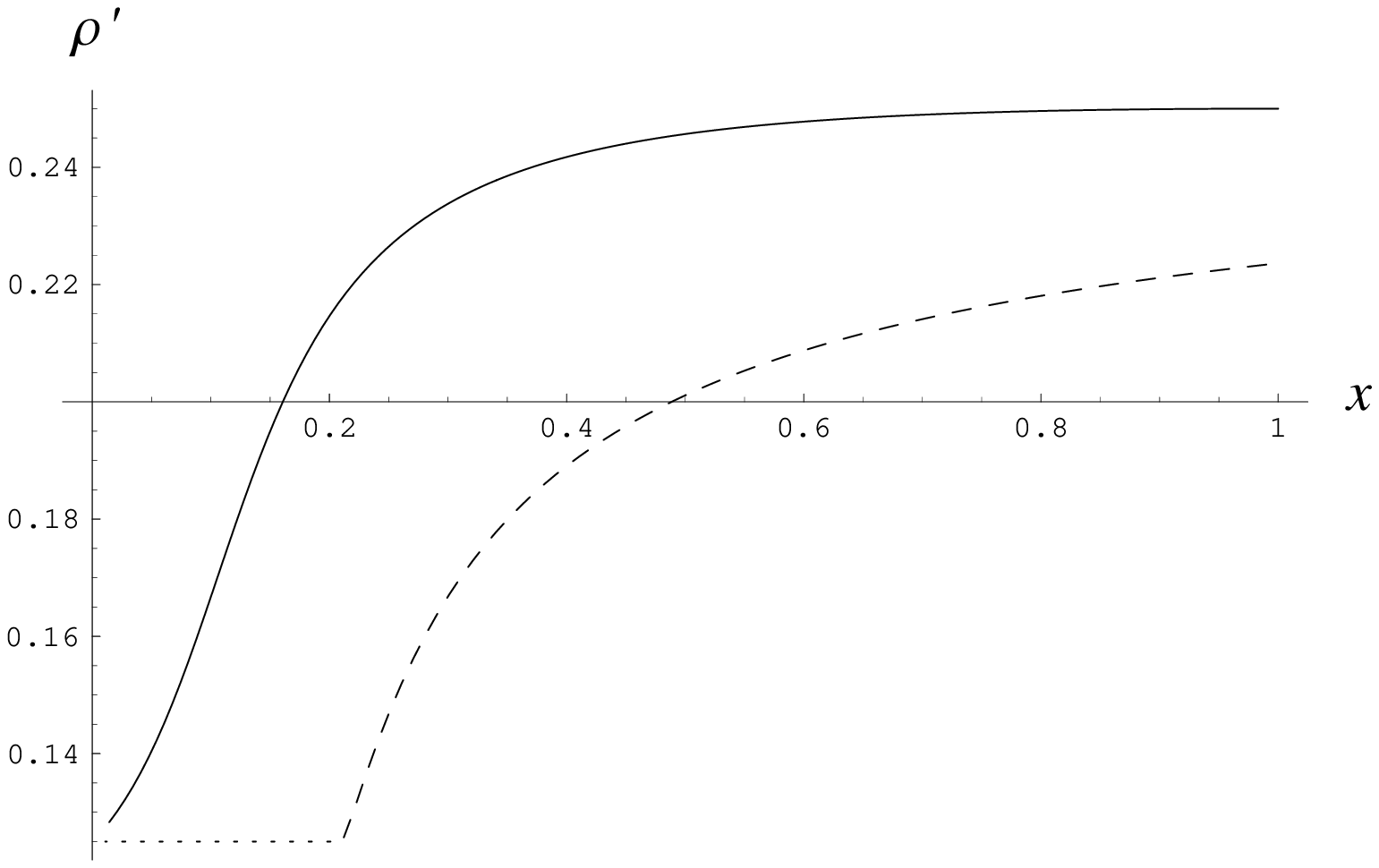}
\includegraphics[height=5.5cm,width=7.7cm]{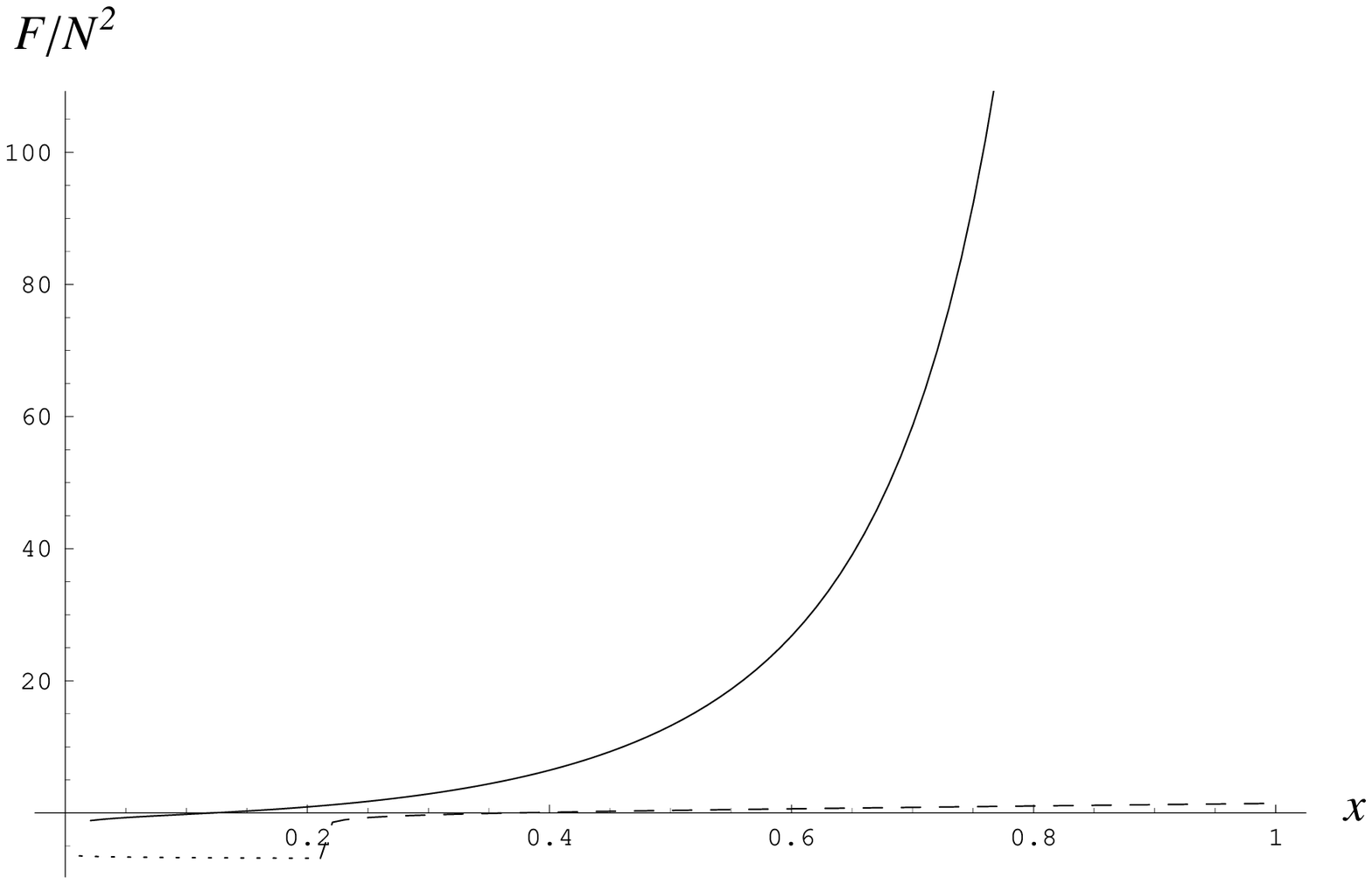}\\
\end{center}
\caption{\footnotesize On the left side the saddle-points (for
$n=1$, $q=1/2$) in terms of $\rho '(x)$ associated to $f_+$
(continuous line) and $f_-$ (dashed line) are shown. At $x=0.212352$
the $f_-$ solution intersects with the $t=0$ unphysical solution
(dotted line). On the right the free energies for both
cases.}\label{graph:rhoprime}
\end{figure}

Let us examine the second saddle-point equation, the one involving
$f_-(t)$. We can repeat the same analysis: the main conclusion is
that for temperature $T<T_2$ we see $f_-(t)$ being monotonically
decreasing and therefore, because $f_-(0)=0$, there is no solution
for $t\neq 0$ to the saddle-point equation. We notice that $t=0$ is
not acceptable because of (\ref{saddlea}). For $T>T_2$ it  is not
easy to see analytically if $f_-(t)$ provides new solutions to the
saddle-point equation: we have done a numerical study, showing that
a new solution appears for $x\geq 0.212352$. However, the resulting
free energy is always subdominant with respect to the other one as
illustrated in fig. \ref{graph:rhoprime}.
So the solution associated to $f_+$ is the only relevant
saddle-point in the large $N$ limit.

We conclude therefore that within our approximation, that consisted
in taking just the first winding in the matrix model action ($n=1$),
we have always a non-trivial saddle-point giving a free energy
$F_B=\log{\cal Z}_B$ of order $N^2$. Moreover this saddle-point
varies continuously with the temperature: in particular at the
Hagedorn temperature $T_H$, representing the point of the
first-order phase transition in our first regularization scheme, the
free energy remains smooth and no discontinuous behavior appears in
this second scheme.

\sectiono{Conclusions and future directions}
In this paper we have studied the maximal supersymmetric gauge
theory on $\mathds{R}\times S^2$, with particular attention to its
thermodynamical properties in the limit of zero 't Hooft coupling.
In the case of the trivial vacuum, we found a behavior similar to
the parent four-dimensional theory, with a first-order Hagedorn
transition separating a ``confining'' phase from a ``deconfined"
one, with non-trivial expectation value for the Polyakov loop. We
have repeated the analysis for monopole vacua and we have apparently
different behaviors, depending on the regularization procedure: this
actually reflects the particular choice of the fermionic
three-dimensional vacuum, that is related to generation of
Chern-Simons terms when monopole are present on the sphere. We have
presented two opposite choices, both allowed at quantum field theory
level, generating different unitary multi-matrix models describing
the thermal partition function. The critical behaviors we found,
under suitable assumptions on the relevant contributions at small
temperature, are very different: in particular we have observed that
no Hagedorn transition seems to be present within our second
regularization choice.
 Further studies are surely necessary to elucidate the situation: first of all
we expect that supersymmetry should play a role in order to distinguish between the different
regularization choices and consistency with the SUSY algebra could probably select a preferred
``vacuum charge". On the other hand the relation with the gravitational duals should also be
investigated to provide a physical interpretation of the Casimir energies and of the Chern-Simons
contributions. Apart from solving the puzzles arisen in this paper, there are a lot of potential
interesting developments involving the study of the ${\cal N}=8$ three-dimensional supersymmetric
theory considered here. It would be important of course to determine the nature of the phase
transition beyond zero 't Hooft coupling and to discuss the issue of exact decoupling limit using
chemical potentials, in the spirit of \cite{Harmark:2006di,Harmark:2006ie,Harmark:2007px}. We also
plan to consider the phase diagram in the presence of background scalars as in
\cite{Hollowood:2006xb,Gursoy:2007np}.
 More generally one could try to explore
if some remnant of four-dimensional integrability persists in three
dimensions and to make some quantitative connection, in the
strong-coupling limit, between the gauge theory and its gravity
dual. It would also be  very interesting to study BPS Wilson loops
on $S^2$: in four dimensions there have been exact results for
particular classes of loops, the computations reducing to matrix
integrals \cite{Erickson:2000af,Drukker:2000rr} . It is natural to
ask if a similar phenomenon takes place in three dimensions too.

\section*{Acknowledgments}
We thank Valentina Forini for a fruitful collaboration at the
early stages of this work. We also thank Leonardo Brizi, Gianni
Cicuta, Filippo Colomo, Troels Harmark, Carlo Meneghelli, Marta
Orselli, Ettore Vicari for useful discussions. We wish to thank
the Galileo Galilei Institute for Theoretical Physics for
hospitality during the last stages of this work.

\appendix
\addcontentsline{toc}{section}{Appendices}

\sectiono{Conventions and supersymmetry variations}
\label{ConvSusy} Before discussing in more details the
supersymmetry variations considered in section$\,$\ref{sectD10D3},
we shall briefly summarize our conventions and identities on
$\Gamma$-matrices.
\paragraph{Metric and gauge conventions:}
The metric is taken diagonal and with Minkowskian signature:
$\eta_{MN} = \{-,+,...,+\}$. The capital letters $M,N,\dots$ will
span the ten dimensional spacetime indices $(0,1,\dots,9)$, while
the Greek letters $\mu,\nu\dots$  will denote the  three dimensional
spacetime indices $(0,1,2)$. The indices $i,j,k$ are associated to
the directions $(1,2)$  along the sphere $S^2$, while the directions
$(3,\dots,9)$ transverse  to $S^2$ are  indicated with $m,n,\dots$.
Finally a special index notation is also reserved to the set of
directions $(4,\dots,9)$ for which we shall use the overlined
letters $\bar{m},\bar{n},\dots$.

\medskip
 The gauge fields  $A=A^a t^a$ are taken to be hermitian
and the generator $t^a$ are normalized so that  $ \mathrm{Tr}(t^a
t^b) = \frac{1}{2}\delta^{ab}$. The covariant derivatives are then
defined as follows $D_\mu = \nabla_\mu -ig[A_\mu, \cdot]$, where
$\nabla_\mu$ is the \textsl{geometrical} covariant derivative. In
general we shall omit the trace over the gauge generators in our
expressions, unless it is source of confusion.

\paragraph{Some useful $\Gamma$-identities:} For convenience, here we have collected some
$\Gamma$-identities, which are useful in checking the
supersymmetry invariance of the Lagrangian of our model:
\begin{equation}
\!
\begin{array}{llllllll}
\Gamma^i \Gamma^{jk} \Gamma^i =-2 \Gamma^{jk}, &\
&\Gamma^0\Gamma^{jk} \Gamma^0 = -\Gamma^{jk},&\ &\Gamma^i
\Gamma^{0j} \Gamma^i=0,
&\ &\Gamma^0\Gamma^{0j} \Gamma^0 = \Gamma^{0j},\\
\Gamma^i \Gamma^{jm} \Gamma^i = 0, &\ &\Gamma^0\Gamma^{jm} \Gamma^0
= -\Gamma^{jm}, &\ &\Gamma^i \Gamma^{0m} \Gamma^i = 2 \Gamma^{0m},
&\ &\Gamma^0\Gamma^{0m} \Gamma^0 = \Gamma^{0m},\\
\Gamma^i \Gamma^{mn} \Gamma^i = 2 \Gamma^{mn}, &\
&\Gamma^0\Gamma^{mn} \Gamma^0 = -\Gamma^{mn}. &\ & &\ &
\end{array}
\end{equation}
Summation over repeated index is understood. Here $\Gamma^M$ denotes
the ten dimensional matrices, while the symbol $\gamma^\mu$ is used
for the three dimensional Dirac matrices. The symbol $\Gamma^{M_1
M_2\dots M_N}$ defines the completely antisymmetrized product of the
matrices $\Gamma^{M_1}$, $\Gamma^{M_2}$,\dots ,$\Gamma^{M_N}$.
\paragraph{Three-dimensional fields:} The scalar field $\phi^{ij}$ is antisymmetric in $i$, $j$,
which are $SU(4)_R$ indices and it satisfies reality condition:
\begin{equation}
    \phi^{ij} \equiv (\phi_{ij})^{\dagger} =
    \frac{1}{2}\epsilon^{ijkl}\phi_{kl}.
    \end{equation}
It is defined in terms of the old fields $\phi_{\overline{m}}$ by
the relations:
\begin{equation}
\begin{split}
\phi_4&= \frac{\phi_{14} + \phi_{23}}{\sqrt{2}}, \ \ \ \ \
\phi_5=\frac{-\phi_{13} + \phi_{24}}{\sqrt{2}}, \ \ \ \ \
\phi_6= \frac{\phi_{12} + \phi_{34}}{\sqrt{2}}, \\
\phi_7&= i\frac{\phi_{14} - \phi_{23}}{\sqrt{2}}, \ \ \ \ \
\phi_8=i\frac{\phi_{13} + \phi_{24}}{\sqrt{2}}, \ \ \ \ \
\phi_9=i\frac{-\phi_{12} + \phi_{34}}{\sqrt{2}}.
\end{split}
\end{equation}
The spinor fields $\lambda_i$ (again, $i$ is an $SU(4)_R$ index)
denote the Dirac spinors in $D=3$ originating from the dimensional
reduction of $\psi_M$, while $A_\mu$ describes the three-dimensional
gauge field.
\subsection{Supersymmetry variations}
\label{Susyvar} In this appendix, for completeness, we shall write
the conditions for the vanishing of the variation at the order
$\alpha$ and at the order $\alpha^2$. At the linear order the
complete variation can be summarized by the following table:
\begin{equation}
\begin{array}{c|c}
\mathrm{\textbf{Term}} & \mathrm{\textbf{Condition}}\\
\hline
 & \\2\Real\{\alpha g[\phi_{\overline{m}},\phi_{\overline{n}}]\overline{\psi}\Gamma^{\overline{m}\,\overline{n}}\Gamma^{123}\epsilon\}&\mathcal{B}+2+
P + M = 0\\
 & \\2\Real\{\alpha g[\phi_3,\phi_{\overline{m}}]\overline{\psi}\Gamma^{3\overline{m}}\Gamma^{123}\epsilon\}&
2\mathcal{B} + 4 +2 P +  G -2 M=0\\
 & \\2\Real\{\alpha i D_0
 \phi_3\overline{\psi}\Gamma^{03}\Gamma^{123}\epsilon\}& 4
 -2\mathcal{B}+ P + G -2 M=0\\
 & \\2\Real\{\alpha i D_0
 \phi_{\overline{m}}\overline{\psi}\Gamma^{0{\overline{m}}}\Gamma^{123}\epsilon\}&
 4 -2\mathcal{B}+ P +2 M = 0\\
 & \\2\Real\{\alpha i D_i
 \phi_3\overline{\psi}\Gamma^{i3}\Gamma^{123}\epsilon\}& 2\mathcal{B}
 + P +  G +2 M +  N = 0\\
 & \\2\Real\{\alpha i D_i
 \phi_{\overline{m}}\overline{\psi}\Gamma^{i{\overline{m}}}\Gamma^{123}\epsilon\}&
 2\mathcal{B} + P -2 M = 0\\
 & \\2\Real\{\alpha iF_{0i}\overline{\psi}\Gamma^{0i}\Gamma^{123}\epsilon\} &
 -2\mathcal{B} - 2 M =0\\
 & \\2\Real\{\alpha iF_{ij}\overline{\psi}\Gamma^{ij}\Gamma^{123}\epsilon\}
 & \mathcal{B} -2 + M + \frac{ N}{2} = 0
\end{array}
\end{equation}
There are eight different kind of terms, listed in the first column,
and they must vanish separately: this leads to the conditions in the
second column.

\medskip
 At the quadratic order in $\alpha$ we have simply
\begin{equation}
\begin{array}{c|c}
\mathrm{\textbf{Term}} & \mathrm{\textbf{Condition}}\\
\hline
 &
 \\2\Real\{i\alpha^2\phi_{\overline{m}}\overline{\psi}\Gamma^{\overline{m}}\psi\}
& -2V + (2 - \frac{\beta}{\alpha})P + MP = 0\\
& \\2\Real\{i\alpha^2\phi_3\overline{\psi}\Gamma^3\psi\} & -2(V+W) +
(2 - \frac{\beta}{\alpha})(P+G) - M(P+G) = 0
\end{array}
\end{equation}
\sectiono{Computing the one loop partition function}
Here we give all the details of the calculation of the partition
function in a monopole background. For the free model the one-loop
contribution of each field is a functional determinant, giving the
single-particle partition function.

\subsection{Computing determinants: the master-formula}
\label{determinants} We illustrate our regularization scheme:
readers who are not interested in these details can take
(\ref{Appdet8}) and (\ref{Appdet17}) as main results, and skip to
next subsection.\newline All the determinants appearing in the
evaluation of the \textit{free} partition function contains, as a
key ingredient, the evaluation of the following infinite product
\beq \label{Appdet1} \Sigma(\eta,\rho,\beta,w):=\prod_{j=0}^\infty
\prod_{n=-\infty}^\infty \left[(j+\eta)^2+ \frac{4\pi^2}{\beta^2}
\left(n+ w\right)^2\right]^{2 j+\rho}. \eeq This quantity is
divergent and it must be regularized. Here, we shall adopt the
standard $\zeta-$function regularization and we shall define \beq
\label{Appdet2} \Sigma(\eta,\rho,\beta,w):=\e^{-\zeta^\prime(0)},
\eeq where \beq \label{Appdet3} \zeta(s)=\sum_{j=0}^\infty
\sum_{n=-\infty}^{\infty}\frac{2j+\rho}{\left[(j+\eta)^2+
\frac{4\pi^2}{\beta^2} \left(n+ w\right)^2\right]^s}\, . \eeq Notice
that (\ref{Appdet3}) defines the function $\zeta(s)$ only for
$|s|>1$. In order to compute $\zeta^\prime(0)$, we have to consider
its analytical continuation to a neighborhood of the origin in the
$s$-plane.  This is achieved through a standard technique:  firstly,
we shall use the Mellin-Barnes representation and subsequently we
shall  perform  a Poisson-resummation in $n$ \beq \label{Appdet4}
\begin{split}
\!\!\!\zeta(s)&=\frac{1}{\Gamma(s)}\sum_{j=0}^\infty(2j+\rho)
\sum_{n=-\infty}^\infty\int_0^\infty\!\!\!\!\! \!dt~ t^{s-1} e^{-t
(j+\eta)^2-t \frac{4\pi^2}{\beta^2} \left(n+ w\right)^2}
\!=\\
&= \frac{\beta}{2 \sqrt{\pi}\Gamma(s)} \sum_{j=0}^\infty(2j+\rho)
\sum_{n=-\infty}^\infty\int_0^\infty\!\!\!\!\! \!dt~
t^{s-\frac{3}{2}}  e^{-t (j+\eta)^2}
 e^{-\frac{\beta^2 n^2}{4 t }-2\pi i w n}=\\
 &=\frac{\beta \,{\Gamma}(s- \frac{1}{2}  )}{2\,{\sqrt{\pi }}\,{\Gamma}(s)}\sum_{j=0}^\infty\frac{\left( 2\,j+\rho \right) \,\,}
  {{\left( j + \eta \right) }^{2\,s-1}}+\\
  &\ \ \ \ \ \  \ \ +\frac{2^{\frac{3}{2}-s} \beta ^{s+\frac{1}{2}}}{\sqrt{\pi }
   \Gamma (s)}\sum_{j=0}^\infty\sum_{n=1}^\infty
  {\frac{(2 j+\rho)}{n^{\frac{1}{2}-s}(j+\eta)^{s-\frac{1}{2}}}    K_{\frac{1}{2}-s}(n (j+\eta) \beta)
   \cos (2 n \pi  w)}{}=\\
 &=\frac{\beta \,{\Gamma}(s- \frac{1}{2}  )}{2\,{\sqrt{\pi }}\,{\Gamma}(s)}\left(2 \zeta(2 s-2,\eta)-(2 \eta-\rho)\zeta(2 s-1,\eta)
 \right)\\
  &\ \ \ \ \ \  \ \ +\frac{2^{\frac{3}{2}-s} \beta ^{s+\frac{1}{2}}}{\sqrt{\pi }
   \Gamma (s)}\sum_{j=0}^\infty\sum_{n=1}^\infty
  {\frac{(2 j+\rho)}{n^{\frac{1}{2}-s}(j+\eta)^{s-\frac{1}{2}}}    K_{\frac{1}{2}-s}(n (j+\eta) \beta)
   \cos (2 n \pi  w)}{}.
\end{split}
\eeq The only contribution to  $\zeta^\prime(0)$ in
(\ref{Appdet4}) arises when the derivative acts on $1/\Gamma(s)$
since this quantity vanishes  as $s$ approaches $0$. We obtain
\beq \label{Appdet5} \zeta^\prime(0)=-\beta  (2 \zeta
(-2,\eta)+(\rho -2 \eta) \zeta
(-1,\eta))+\sum_{j=0}^\infty\sum_{n=1}^\infty \frac{2 e^{-n \beta
(j+\eta )} (2 j+\rho ) \cos (2 n \pi  w)}{n}. \eeq From the final
expression (\ref{Appdet5}) we can  deduce two equivalent
representations of this result, which are both useful for our
goals. Firstly we can perform the sum over $j$, which yields \beq
\label{Appdet6} \zeta^\prime(0)=\beta \left(\frac{2}{3}
B_3(\eta)+\frac{1}{2}(\rho -2 \eta) B_2 (\eta)\right)
+\sum_{n=1}^\infty\frac{2 x^{n \eta } \left(\rho -x^n (\rho
-2)\right) }{n \left(x^n-1\right)^2}\cos (2 n \pi  w), \eeq with
$x:=e^{-\beta}$ and $B_k(\eta)$ being the Bernoulli polynomial.
Next, we shall define the ``single-particle'' partition function
\beq \label{Appdet7} z_{single}(x):=\frac{ x^{ \eta } \left(\rho
-x (\rho -2)\right) }{ \left(1-x\right)^2}, \eeq and finally write
\beq \label{Appdet8} \log\left(\Sigma(\eta,\rho,\beta,w)\right)=-
\beta \left(\frac{2}{3} B_3(\eta)+\frac{1}{2}(\rho -2 \eta) B_2
(\eta)\right) -2\sum_{n=1}^\infty\frac{z_{single}(x^n)}{n}\cos (2
n \pi  w). \eeq This representation will be the most natural when
discussing the matrix model and the position of the Hagedorn
transition.

 Alternatively, in (\ref{Appdet5})  we can first sum
over $n$ \beq \label{Appdet9} \zeta^\prime(0)=\beta  (\frac{2}{3}
B_3(\eta)+\frac{1}{2}(\rho -2 \eta) B_2 (\eta))- \sum_{j=0}^\infty
(2 j+\rho ) \left(\log \left(1-\bar z x^j \right)+\log \left(1-z
x^{j} \right)\right) \eeq where $z:=e^{-\beta \eta+ 2 i \pi w}$. If
we define \beq \label{Appdet10} \eta(z,x):=\prod_{j=0}^\infty
\left(1- z x^j \right)\  \ \ \  \mathrm{and}\  \ \ \ \
\mathcal{M}(z,x):=\prod_{j=0}^\infty \left(1-z x^j \right)^j, \eeq
we can recast the above result in a very compact form \beq
\label{Appdet11} \Sigma(\eta,\rho,\beta,w)=e^{-\beta  (\frac{2}{3}
B_3(\eta)+\frac{1}{2}(\rho -2 \eta) B_2 (\eta))}
\left|\eta(z,q)\right|^{2\rho} \left|\mathcal{M}(z,q)\right|^{4}.
\eeq This second representation will be the most suitable when
discussing the high temperature behavior. In this limit the leading
contribution is encoded in the function $F_{\rho}(z,x)$ \beq
F_{\rho}(z,x)= \sum_{j=0}^\infty (2 j+\rho ) \log \left(1-z
x^{j}\right). \label{Appdet12} \eeq The $x\to 1$ behavior is
transparent by rewriting $F_{\rho}(z,x)$ as \beq F_{\rho}(z,x)=
-\sum_{m=1}^\infty \frac{z^m}{m}\,\left[(\rho-2)\frac{1}{1-
x^{m}}+\frac{2}{(1- x^{m})^2}\right], \label{Appdet13} \eeq and
expanding in $\beta$, at fixed $z$, we get \beq F_\rho(z,x)=
-\frac{2}{(\beta)^2}\,{\rm Li}_3(z)-\frac{\rho}{\beta}\,{\rm
Li}_2(z)+(\frac{\rho-2}{2}+\frac{5}{6})\log(1-z)+O(\beta)
.\label{Appdet14}\eeq To recover (\ref{HighT1}), where the
contribution of chemical potentials to the high-temperature limit
has been presented, we need further expand ${\rm Li}_3(z)$ for $z\to
1$: we are interested in the case when $w=0$ and $w=1/2$, appearing
respectively in the bosonic and fermionic case, and with
zero flat-connection ($z=e^{-y}$, $y\to 0$) \beq\begin{split} {\rm Li}_3(e^{-y})&=\zeta(3)-\frac{\pi^2}{6}\,y+\left(\frac{3}{4}-\frac{1}{4}\log y^2\right)\,y^2+O(y^3),\\
{\rm Li}_3(-e^{-y})&=
-\frac{3}{4}\zeta(3)+\frac{\pi^2}{12}\,y-\frac{1}{4}\log
(4)\,y^2+O(y^3).\end{split}\label{Appdet15}\eeq

\paragraph{Fermionic zero modes:} In order to compute the contribution of the fermion zero modes,
we need to compute the product
$\mathfrak{F}=\displaystyle{\prod_{n=-\infty}^{\infty}\left[\frac{2\pi}{\beta}(n+w)
\right]^\rho}$. If we adopt the zeta function regularization as
before, we are led to compute the following accessory sum \beq
\label{Appdet16}
G(s)=\frac{\beta^s}{(2\pi)^s}\sum^\infty_{n=-\infty}\frac{\rho}{(n+w)^s}=
\frac{\beta^s}{(2\pi)^s}\rho(\zeta(s,w)+e^{i\pi s}\zeta(s,1-w)).
\eeq Then \beq \label{Appdet17}
\log(\mathfrak{F})=-G^\prime(0)=-\rho\sum_{n=1}^{\infty}
\frac{e^{-2\pi i n w}}{n}. \eeq

\subsection{The scalar determinant}
\label{SCALAR}

Let us discuss the solution of the eigenvalue problem
(\ref{scalar1}). Since our background is static, we can factor out
the time-dependence in the eigenfunction by posing
$\phi(t,\theta,\phi)\sim \phi_n(\theta,\phi)~ e^{-\frac{2\pi i
n}{\beta} t}$. Then  the eigenvalue problem in the Weyl basis
(\ref{scalar2}) takes the form \beq \label{scalar4}
\begin{split}
&\sum_{\alpha\in \mathrm{roots}}\left[\frac{4\pi^2}{\beta^2}
\left(n+\frac{\beta a_\alpha}{2\pi}\right)^2\phi_{\alpha n}
-\hat\triangle\phi_{\alpha n} +\frac{\mu^2}{4}\phi_{\alpha n}+\mu^2
q^2_\alpha \phi_{\alpha n} \right]
E^\alpha +\\
&+\sum_{i=1}^{N-1} \left(\frac{4\pi^2 n^2}{\beta^2} \phi_{i
n}+\frac{\mu^2}{4}\phi_{i n}-\triangle \phi_{i n}\right) H^i=
\lambda\sum_{i=1}^r \phi_{i n} H^i +\lambda\sum_{\alpha\in
\mathrm{roots}} \phi_{\alpha n}E^\alpha,
\end{split}
\eeq where  $\triangle$ denotes the geometrical Laplacian for a
scalar on the sphere. The symbol $\hat\triangle$ instead represents
the geometrical Laplacian in the background of a $U(1)$ magnetic
monopole of charge $q_\alpha$. This Laplacian is constructed with
the covariant derivative \beq \label{scalar5} \hat {D}_\mu
=\nabla_\mu-i q_\alpha \mathcal{A}_\mu, \eeq where $\nabla_\mu$ is
the geometrical covariant derivative. In (\ref{scalar4}) the
components along the different directions in the Lie algebra do not
interfere and we can consider them as independent. This allows us to
split the original eigenvalue problem into two subfamilies, we have:
(a) $N(N-1)$ independent eigenvalues coming from each direction
along the ladder generator \beq \label{scalar6}
\frac{4\pi^2}{\beta^2}\left(n+\frac{\beta
a_\alpha}{2\pi}\right)^2\phi_{\alpha n} -\hat\triangle \phi_{\alpha
n} +\frac{\mu^2}{4}\phi_{\alpha n}+\mu^2 q^2_\alpha \phi_{\alpha
n}=\lambda_{\alpha n} \phi_{\alpha n}, \eeq and (b) $N-1$
independent eigenvalues coming from the directions along the Cartan
subalgebra \beq \label{scalar7} \frac{4\pi^2 n^2}{\beta^2} \phi_{i
n}+\frac{\mu^2}{4}\phi_{i n}-\triangle \phi_{i n}=\lambda_{i n}
\phi_{in}. \eeq To begin with, we shall focus our attention on the
family (a), since the family (b) can be obtained from (a) as a
limiting case for $a_\alpha,\ q_\alpha\to 0$. The solution of the
eigenvalue equation (\ref{scalar6}) can be translated into an
algebraic problem  if we introduce the angular momentum operator in
the presence of a $U(1)$ monopole of charge $q_\alpha$. Its form
\cite{Wu:1976ge} is \beq \label{scalar8}
L_i^{(\alpha)}=\epsilon_{ijk}x_j( -i
\partial_k-q_\alpha A_k)- q_\alpha
\frac{x^i}{|x|}\equiv\epsilon_{ijk}x_j P_k-q_\alpha \frac{x^i}{|x|}.
\eeq Here $x^i$ are the Cartesian coordinates of a flat
$\mathds{R}^3$ where our sphere $S^2$ is embedded. In terms of this
auxiliary operator, the  kinetic operator in (\ref{scalar6}) takes
the form \beq \label{scalar9} \mu^2 (L^{(\alpha)})^2 \phi_{\alpha
n}+ \left[\frac{4\pi^2}{\beta^2}\left(n+\frac{\beta
a_\alpha}{2\pi}\right)^2+\frac{\mu^2}{4}\right]\phi_{\alpha n}=
\lambda_{\alpha n} \phi_{\alpha n}. \eeq Thus our task is reduced to
finding the eigenvalues and the eigenfunctions of this
\textsl{dressed} angular momentum operator  $(L^{(\alpha)})^2.$ Its
spectrum\footnote{The eigenfunctions are also known and they are
given by the so-called \textsl{monopole harmonics}
$Y_{qjm}(\theta,\varphi)$. They are a straightforward generalization
of the usual spherical harmonics, but we shall not need their
explicit form here. We refer the reader to \cite{Wu:1976ge} for more
details.} was determined thirty years ago by Wu and Yang
\cite{Wu:1976ge} and it is formally equal to that of the usual
angular momentum: the eigenvalues are $j_\alpha(j_\alpha+1)$ and
their degeneracy is $2j_\alpha+1$. What changes is the range spanned
by the index
 $j_\alpha$, which now  is $|q_\alpha|,\ |q_\alpha| + 1,\ |q_\alpha| + 2,\ \cdots$.
Putting everything together the spectrum of the kinetic operator
(\ref{scalar6}) turns out to be\beq \label{scalar10} \lambda_{\alpha
n}= \mu^2\left(j_\alpha+\frac{1}{2}\right)^2+\frac{4\pi^2}{\beta^2}
\left(n+ \frac{\beta a_\alpha}{2\pi}\right)^2\ \ \mathrm{with}\ \
j_\alpha=|q_\alpha|,\ |q_\alpha| + 1,\ |q_\alpha| + 2 \cdots, \eeq
and each eigenvalue has degeneracy $2j_\alpha+1$. Notice that the
spectrum does not depend on the sign of $q_\alpha$. The contribution
of the family (a) to the effective action is given by the infinite
product \beq \label{scalar11}
{\Gamma^{Sc.}}_{(a)}=\log\left(\prod_{\alpha\in
\mathrm{roots}}\prod_{j_{\alpha}=|q_{\alpha}|}^\infty\prod_{n=-\infty}^\infty
\left[\mu^2\left(j_\alpha+\frac{1}{2}\right)^2+\frac{4\pi^2}{\beta^2}
\left(n+ \frac{\beta
a_\alpha}{2\pi}\right)^2\right]^{2j_\alpha+1}\right), \eeq which is
easily computed by using the results of appendix \ref{determinants}
(with $\rho=1+2|q_\alpha|$, $\eta=1/2+|q_\alpha|$, $w=\frac{\beta
a_\alpha}{2\pi}$). Setting $x=e^{-\beta\mu}$, we obtain
\eqref{scalar12} and \eqref{scalar13}
The contribution of the family (b) is then obtained from the
above results by setting $q_\alpha= a_\alpha=0$.

\subsection{The vector/scalar determinant}
\label{VECTOR}

The eigenvalue problem for the coupled system $(\phi_3, A)$ can be
simplified by choosing the gauge-fixing \eqref{vector1}.  This
choice allows us to cancel some of the mixed terms $(\phi_3 A)$ in
the Euclidean quadratic Lagrangian and to obtain \beq
\label{vector2}
\begin{split}
\mathcal{L}^{(2)}_{(A_\mu,\phi_3)}=& - A_\nu \hat{D}_\mu\hat D^\mu
A^\nu+ R_{\mu\nu} A^\mu A^\nu-i \hat{F}_{\nu\mu}  [ A^\nu,A^\mu]
-[A_\rho,\hat\phi_3][A^\rho,\hat\phi_3]+ \\
&+\hat D_\rho \phi_3 \hat D^\rho \phi_3+\mu^2\phi_3^2- [\hat
\phi_3,\phi_3]^2-2\frac{\mu}{\sqrt{g}}\phi_3
\epsilon^{\rho\nu\lambda}{k}_\rho \hat D_\nu A_\lambda.
\end{split}
\eeq Then, the following coupled eigenvalue problem
\begin{eqnarray}
\label{vector3} -\hat{\square }\phi_{3}+{\mu^2}
\phi_{3}+[\hat{\phi}_3,[\hat{\phi}_3,\phi_{3}]]-\mu
\sqrt{g} \epsilon_{\rho\nu\lambda} {k}^\rho \hat D^\nu A^\lambda=\lambda \phi_{3},\\
\label{vector4} -\hat\square A_\nu+R_{\mu\nu} A^\mu + i
[\hat{F}_{\nu\mu},A^\mu]+[\hat\phi_3,[\hat\phi_3,A_\nu]]+\mu
\sqrt{g} \epsilon_{\rho\lambda\nu} {k}^\rho \hat D^\lambda
\phi_3=\lambda A_\nu.
\end{eqnarray}
Since both the geometrical and the gauge background are static,
the time-component of the vector field $\omega= k^\rho A_\rho=A_0$
decouples completely  from the above system. It satisfies the
massless version of the scalar equation studied in~\ref{SCALAR},
namely the eigenvalue problem associated to this component is \beq
\label{vector5}
-\hat\square\,\omega+[\hat\phi_3,[\hat\phi_3,\omega]]=\lambda_0\omega.
\eeq We shall forget about $\omega$ since its contribution is
cancelled by the ghost determinant. We are left with the system
given by (\ref{vector3}) and (\ref{vector4}) where the indices run
only over space.

We expand the coupled system (\ref{vector3}) and (\ref{vector4}) in
the Weyl basis and we factor out the time-dependence of the
eigenfunctions: $A_{\mu }(t,\theta,\phi)\sim A_{n\mu} (\theta,\phi)~
e^{-\frac{2\pi i n}{\beta} t}$ and $\phi_{3}(t,\theta,\phi)\sim
\phi_{3 n} (\theta,\phi)~ e^{-\frac{2\pi i n}{\beta} t}$. Along the
directions associated to the ladder operators $E_\alpha$ we find
\beq \label{vector6}
\begin{split}
&-\hat\triangle A_{i \alpha n} +  m^2_n A_{i \alpha n} +
i\mu^2q_\alpha\sqrt{g}\epsilon_{ij} A^j_{\alpha n}+ \mu^2 q^2_\alpha
A_{i \alpha n}+\mu
\sqrt{g} \epsilon_{j i} \hat D^j \phi_{3\alpha n}=\lambda_{\alpha n} A_{i \alpha n},\\
&-\hat{\triangle}\phi_{3\alpha n}+ m^2_n \phi_{3\alpha n}+\mu^2
q^2_\alpha \phi_{3\alpha n}-\mu \sqrt{g} \epsilon_{ij}\hat D^i
A^j_{\alpha n}=\lambda_{\alpha n} \phi_{3\alpha n},
\end{split}
\eeq where $m^2_n=(2\pi n/\beta+ a_\alpha)^2+\mu^2$. In the first
equation, the symbol $\hat{\triangle}$ denotes the Laplacian on
vectors in the background of a monopole of charge $q_\alpha$, while
in the second represents the Laplacian on scalars. Along the Cartan
directions we shall again get the system (\ref{vector6}) but for
$q_\alpha=0$.

To find explicitly the spectrum of  system (\ref{vector6}), it is
convenient to decompose our vector $A_{i\alpha n}$ in its selfdual
part $A^{+}_{i\alpha n}$ and anti-selfdual part $A^{-}_{i\alpha
n}$. Consequently we shall introduce the differential operators
$O^{(\alpha)}_\pm$ mapping (anti-)selfdual vectors into scalars
and their adjoints, mapping scalars into (anti-)selfdual vectors.
They are defined by \beq \label{vector7} O^{(\alpha)}_\pm
V_{\pm}\equiv O^{i(\alpha)}_\pm V_{\pm i}
=\frac{1}{\sqrt{g}}\epsilon^{ij} \hat D_i V_{(\pm)j},\ \ \ \ \ \ \
O^{(\alpha)\dagger}_\pm\phi\equiv O^{i(\alpha)\dagger}_\pm\phi=\mp
\frac{i}{2}\left(g^{ij}\pm \frac{i}{\sqrt{g}} \epsilon^{ij}\right)
\hat{D}_j\phi, \eeq where $\hat{D}$ as in (\ref{vector6}) stands
for the covariant derivative in the background of a monopole of
charge $q_\alpha$. In terms of these operators, the system
(\ref{vector6}) takes the form
\begin{eqnarray}
\label{vector8} &&\!\!\!\!\!\!\!\!\!\!\! O^{(\alpha)\dagger}_{+}
O^{(\alpha)}_{+}A^{+}_{\alpha n}+\frac{q^2_\alpha\mu^2}{2}
A^{+}_{\alpha n} + \frac{\ell^2_n}{2}A^{+}_{\alpha n}
 -\frac{\mu}{2} O^{(\alpha)\dagger}_+\phi_{3\alpha n}=\frac{\lambda_{\alpha n}}{2} A^{+}_{\alpha n},\nonumber\\
&&\!\!\!\!\!\!\!\!\!\!\! O^{(\alpha)\dagger}_{-}
O^{(\alpha)}_{-}A^{-}_{\alpha
n}+\frac{q_\alpha^2\mu^2}{2}A^{-}_{\alpha n}  +
\frac{\ell^2_n}{2}A^{-}_{\alpha n}
  -\frac{\mu}{2} O^{(\alpha)\dagger}_-\phi_{3\alpha n}=\frac{\lambda_{\alpha n}}{2}A^{-}_{\alpha n},\\
&&\!\!\!\!\!\!\!\!\!\!\! \frac{1}{2}(O^{(\alpha)}_-
O^{(\alpha)\dagger}_-\!\!+O^{(\alpha)}_+
O^{(\alpha)\dagger}_+\!\!+{q^2_\alpha}\mu^2+ \ell^2_n+\mu^2)
 \phi_{3\alpha n}
-\frac{\mu}{2} O^{(\alpha)}_+ A^{+}_{\alpha n}  -\frac{\mu}{2}
O^{(\alpha)}_- A^{-}_{\alpha n} =\frac{\lambda_{\alpha
n}}{2}\phi_{3\alpha n}.\nonumber
\end{eqnarray}
Here we have dropped the index $i$ because it is immaterial and we
have set $m^2_n=\ell_n^2+\mu^2$. At first sight the eigenvalue
problem might appear cumbersome, but in this representation it is
quite simple to provide a basis where our problem reduces to
diagonalizing an infinite set of  three by three matrices. In
fact, let us take $q_\alpha \ge 1$\footnote{The case $q_\alpha
\leq -1$ is obtained by exchanging the role of self-dual vectors
with that of anti-self dual vectors. The value $q_\alpha=\pm 1/2$
and $q_\alpha=0$ will be discussed separately.}  and consider  the
following \textsl{basis} for scalars, selfdual and  anti-selfdual
vectors on the sphere \beq \label{vector9}
e^{+\alpha}_{jm}=O^{(\alpha)\dagger}_+ Y_{q_{\alpha}jm}\ \ \
\mathrm{for} \ \  j\ge q_\alpha+1,\ \ \
e^{-\alpha}_{jm}=O^{(\alpha)\dagger}_-Y_{q_{\alpha}jm} \ \
\mathrm{and}\ \ e^{3\alpha}_{jm}=Y_{q_{\alpha}jm} \ \ \
\mathrm{for} \ \  j\ge q_\alpha. \eeq Here $Y_{q_{\alpha}jm}$ are
the monopole harmonics, namely the eigenfunctions of the angular
momentum (\ref{scalar8}). For the anti-selfdual vector we have to
add also $2 (q_\alpha-1)+1$ elements coming from the zero modes of
$O^{(\alpha)\dagger}_{-}O^{(\alpha)}_{-}$. We shall denote them as
$e^{-\alpha}_{(q_\alpha-1) m}$. For a detailed proof that
(\ref{vector9}) with the addition of the zero modes is a basis, we
refer the reader to \cite{Weinberg:1993sg}, where the following
two useful identities are also shown to hold: \beq
\label{vector10} O^{(\alpha)}_{(\pm)}O^{(\alpha)\dagger}_{(\pm)}
e^{3\alpha}_{jm}= \frac{\mu^2}{2}((L^{(\alpha)})^2-q_\alpha^2\mp
q_\alpha)Y^{q_{\alpha}jm}= \frac{\mu^2}{2}(j(j+1)-q_\alpha^2\mp
q_\alpha))e^{3\alpha}_{jm}, \eeq and \beq \label{vector11}
O^{(\alpha)\dagger}_{(\pm)}O^{(\alpha)}_{(\pm)}e^{\pm\alpha}_{\pm
jm}=O^{(\alpha)\dagger}_{(\pm)}O^{(\alpha)}_{(\pm)}
O^{(\alpha)\dagger}_{(\pm)}e^{3\alpha}_{\pm
jm}=\frac{\mu^2}{2}(j(j+1)-q_\alpha^2\mp q_\alpha))
e^{\pm\alpha}_{jm}. \eeq Because of (\ref{vector10}) and
(\ref{vector11}) and the definitions (\ref{vector9}),
$e^{\pm\alpha}_{jm}$ and $e^{3}_{jm}$ for fixed $j\ge q_\alpha+1$
and fixed $m$ generate an invariant three-dimensional linear
subspace for the eigenvalue  problem (\ref{vector8}). The original
problem can be then separately solved in each subspace, where it
reduces to diagonalizing the following three by three matrix \beq
\label{vector12} \mbox{ \small $
\begin{pmatrix}
  m_n^2-\mu^2+ j (j+1) \mu^2-q_{\alpha} \mu ^2 & 0 & -\mu^2 \\
 0 & {m_n^2-\mu^2}+j (j+1) \mu ^2+{q_{\alpha} \mu ^2} &-\mu^2  \\
 -\frac{\mu ^2}{2} \left(-q_{\alpha}^2-q_{\alpha}+j (j+1)\right)  & -\frac{\mu ^2}{2}
 \left(-q_{\alpha}^2+q_{\alpha}+j (j+1)\right) &
 m_n^2+j (j+1) \mu ^2
\end{pmatrix}
$ }. \eeq The three distinct eigenvalues of this matrix are given by
\beq \label{vector13} \lambda_+=\ell_n ^2+j^2 \mu ^2,\ \ \ \
\lambda_-= \ell_n ^2+(j+1)^2 \mu ^2, \ \ \ \lambda_3=\ell_n ^2+j
(j+1) \mu ^2,\ \ \ \ \ \mathrm{with}\   j\ge q_\alpha+1. \eeq For
$j=q_\alpha$ self-dual vectors do not exist and the invariant
subspace is generated only by $e^{-\alpha}_{q_\alpha m}$ and
$e^{3\alpha}_{q_\alpha m}$. Instead of  (\ref{vector12}), we have
the two by two matrix that is obtained from (\ref{vector12}) by
dropping the first row and the first column. Its diagonalization
produces the following two eigenvalues $\lambda_-= \ell_n
^2+(q_\alpha+1)^2$ and $\lambda_3=\ell_n ^2+q_\alpha (q_\alpha+1)$.
Finally, we have to consider $j=q_\alpha-1$. In this case, we are
left with a one-dimensional invariant subspace generated by
$e^{-\alpha}_{(q_\alpha-1)m}$. The eigenvalue is simply $\lambda_-=
\ell_n ^2+q_\alpha^2$. Summarizing we have $ \lambda_-= \ell_n
^2+(j+1)^2 \mu ^2$ for $j\ge q_\alpha-1$ and $ \lambda_3= \ell_n ^2+
j(j+1)\mu ^2$ for $j\ge q_\alpha$ so that we have extended the range
of existence of the eigenvalues (\ref{vector13}). The degeneracy is
always $2j+1$.

In the following we shall neglect the family with eigenvalue
$\lambda_3$, since its contribution is cancelled by the ghosts. We
shall just consider the first two families $\lambda_\pm$, which
instead yield the actual vector determinant. The contribution of
$\lambda_+$ is obtained from the results of appendix
\ref{determinants} by setting $w=\beta a_\alpha/(2\pi)$,
$\eta=q_\alpha+1$ and $\rho=2 q_\alpha+3$ \beq \label{vector14}
\Gamma^V_+=\!\!\!\!\!\sum_{\alpha\in
\mathrm{roots}}\left(-\frac{1}{12} \beta \mu \left(8 q_{\alpha
}^3+18 q_{\alpha }^2+10 q_{\alpha }+1\right)
-2\sum_{n=1}^\infty\frac{z^{vect.}_{q_\alpha+}(x^n)}{n}\,e^{ i n
\beta a_\alpha}\right), \eeq with \beq \label{vector15}
z^{vect.}_{q_\alpha+}(x)= x^{q_\alpha+1}
\left[\frac{(3-x)}{(1-x)^2}+\frac{2 q_\alpha}{1-x}\right]. \eeq
 The contribution of $\lambda_-$
is instead obtained setting $w=\beta a_\alpha/(2\pi)$,
$\eta=q_\alpha$ and $\rho=2q_\alpha-1$: \beq \label{vector16}
\Gamma^V_-=\!\!\!\!\!\sum_{\alpha\in
\mathrm{roots}}\!\!\!\left(-\frac{1}{12} \beta \mu \left(8 q_{\alpha
}^3-18 q_{\alpha }^2+10 q_{\alpha }-1\right)
-2\sum_{n=1}^\infty\frac{z_{q_\alpha-}^{vect.}(x^n)}{n}\,e^{i n
\beta a_\alpha}\right)\ , \eeq with \beq \label{vector17}
z^{vect.}_{q_\alpha-}(x)=
x^{q_\alpha}\left[\frac{x(1+x)}{(1-x)^2}-1+\frac{2
q_\alpha}{1-x}\right]. \eeq When adding these two contributions, we
obtain \eqref{vector181} and \eqref{vector191}.
For what concerns the non-negative values of the monopole charge,
there are still two cases to be considered: $q_\alpha=1/2$ and
$q_\alpha=0$. In both cases, the elements of the basis coming from
the additional zero modes of the operator
$O^{(\alpha)\dagger}_{-}O^{(\alpha)}_{-}$ disappear
\cite{Weinberg:1993sg}.  For $q_\alpha=0$, in the basis
(\ref{vector9})   the element $e^{-\alpha}_{jm}$ with
$j=q_\alpha=0$ is absent. The net effect is to reduce  the range
of the existence of the eigenvalues
$\lambda_-=\ell_n^2+(j+1)^2\mu^2$ to $j\ge q_\alpha$ for
$q_\alpha=0,1/2$ and of the eigenvalues
$\lambda_3=\ell_n^2+j(j+1)\mu^2$ to $j\ge 1$ for $q_\alpha=0$. By
recomputing  the contribution of $\lambda_-$, for $q_\alpha=1/2$
we get the same results (\ref{vector16}) and (\ref{vector17}) when
we use the appropriate values for $\rho$ and $\eta$ in appendix
\ref{determinants}: $\eta=3/2$ and $\rho=2$. This does not happen,
instead, for $q_\alpha=0:$ by using $\eta=1$ and $\rho=1$, we get
\eqref{vector211}.


\subsection{The spinor determinant}
\label{SPINOR}

In  determining the contribution to the total partition function
of the spinors $\lambda_i$, we shall follow closely the steps of
the previous appendix. The fermion kinetic operator expanded
around the background (\ref{vacua3}) has the following eigenvalue
problem \beq \label{spinor1} -i\gamma^\mu \hat{D}_\mu \lambda+ i
[\hat \phi_3,\lambda]+i\frac{\mu}{4}\gamma^0\lambda=\rho\lambda,
\eeq where we dropped the $SU(4)_R$ index since all the components
give the same contribution.  Expanding the matrix-valued field
$\lambda$ in the Weyl basis and separating the time-dependence we
get \beq \label{spinor2} \lambda=\left(\sum_{\ell=1}^{N-1}
\lambda_{\ell n} H^\ell +\sum_{\alpha\in \mathrm{roots}}
\lambda_{\alpha n}E^\alpha\right) e^{-\frac{2\pi
i}{\beta}\left(n+\frac{1}{2}\right) t}. \eeq The only real
difference with the scalar and vector cases is that fermions have
antiperiodic boundary conditions along the time circle. The usual
procedure will, in turn, disentangle the different components
along the Lie algebra and it will divide the eigenvalue problem
(\ref{spinor1}) into two subfamilies. As in the scalar case, we
have: (a) $N(N-1)$ independent eigenvalues coming from each
direction along the ladder generator \beq \label{spinor3}
\not\hskip-3.5pt\mathcal{D}^{(\alpha n)}\lambda_{\alpha
n}\equiv-\gamma^0\left
 [\frac{2\pi}{\beta}\left(n+\frac{1}{2}\right)+a_\alpha-i\frac{\mu}{4}\right]\lambda_{\alpha n}-i
{\gamma^i\hat{D}_i}
 \lambda_{\alpha n}+ i \mu q_\alpha \lambda_{\alpha n}=\rho_{\alpha n} \lambda_{\alpha
 n},
\eeq and (b) $N-1$ independent eigenvalues coming from the
directions along the Cartan subalgebra \beq \label{spinor4}
\not\hskip-4pt\mathcal{D}^{(\ell n)}\lambda_{\ell n}\equiv
-\gamma^0\left
[\frac{2\pi}{\beta}\left(n+\frac{1}{2}\right)-i\frac{\mu}{4}\right]\lambda_{\ell
n}-i {\gamma^i\nabla_i}
 \lambda_{\ell n}=\rho_{\ell n} \lambda_{\ell n}.
\eeq In (\ref{spinor4}) the symbol $\nabla$ denotes the
geometrical covariant derivative on spinors while $\hat{D}_i$ in
(\ref{spinor3}) is the covariant derivative in the background of a
$U(1)$ magnetic monopole of charge $q_\alpha$, \textit{i.e.} \beq
\label{spinor5} \hat{D}_i=\partial_i +\frac{i}{2}\Gamma^{ab}_i
\Sigma_{ab}-i q_\alpha A_i. \eeq We shall first consider the
family $(a)$. The problem of diagonalizing the operator
(\ref{spinor3}) can be solved algebraically by exploiting the
unitary transformation $ U=e^{\frac{i}{2} \theta\sigma_2}
e^{\frac{i}{2} \varphi\sigma_3}$. In fact, after performing this
transformation, the operator (\ref{spinor3}) becomes directly
related to the total angular momentum
$J^{(\alpha)}=L^{(\alpha)}+\frac{\sigma}{2}$ in the monopole
background
 \beq \label{spinor6} \mathcal{S}\equiv U^{\dagger}
\not\hskip-4pt\mathcal{D}^{(\alpha n)} U= -\left
[\frac{2\pi}{\beta}\left(n+\frac{1}{2}\right)+a_\alpha-i\frac{\mu}{4}\right]
(\sigma\cdot \hat{r})+\mu\epsilon^{ijk}{\hat{r}}_i \sigma_j
J^{(\alpha)}_k+i\mu q_\alpha. \eeq Here $\hat r$ stands for the
usual radial unit vector in three dimensions while $\sigma_i$ are
the Pauli matrices. In (\ref{spinor6}), the operator $\mathcal{S}$
is the sum of three contributions. There is a reduced Dirac operator
\beq \label{spinor7}
\mathfrak{D}^{(\alpha)}\equiv\mu\epsilon^{ijk}\hat{r_i} \sigma_j
J^{(\alpha)}_k=i\mu\hat r\cdot\sigma+ \mu\epsilon^{ijk}{\hat r_i}
\sigma_j L^{(\alpha)}_k, \eeq which is the standard two-dimensional
massless Dirac operator in the presence of a monopole, but written
in an unusual basis. Then we have a ``chiral'' mass term
proportional to $(\sigma\cdot \hat{r})$, which plays the role of the
two-dimensional $\gamma_5$ (we have in fact $\{(\sigma\cdot
\hat{r}),\mathfrak{D}^{(\alpha)}\}$=0). Finally there is a constant
shift proportional to the charge $q_\alpha$.

Now, we can focus our investigation just on the operator
(\ref{spinor7}), since the spectrum of (\ref{spinor6}) follows from
that of $\mathfrak{D}^{(\alpha)}$. For each eigenfunction $\psi$ of
$\mathfrak{D}^{(\alpha)}$ with eigenvalue $\hat\rho\ne 0$ there
exists another eigenfunction $(\sigma\cdot \hat r)\psi$ with
eigenvalue $-\hat\rho$. The possible values of $\hat\rho$ can then
be computed by considering the eigenvalues of
$(\mathfrak{D}^{(\alpha)})^2$. This operator has the following
simple form \beq \label{spinor8} (\mathfrak{D}^{(\alpha)})^2=
\mu^2\left[(J^{(\alpha)})^2+\frac{1}{4}-{q^2_\alpha} \right], \eeq
and it is diagonal on the basis $\psi_{jm\pm}$ of the total momentum
eigenfunctions, which satisfy $(\sigma\cdot \hat
r)\psi_{jm\pm}=\psi_{jm\mp}$. The eigenvalues are
$\hat{\rho}^2_{j\alpha}= \mu^2((j+1/2)^2-q^2_\alpha)$. The
positivity of the operator $(\mathfrak{D}^{(\alpha)})^2$ imposes
$(j+1/2)^2-q^2_\alpha\ge 0$, and in turn $j\ge |q_\alpha|-
\frac{1}{2}.$ The degeneracy of each eigenvalue is $2(2j+1)$. On
this basis, the operator $\mathfrak{D}^{(\alpha)}$ is also diagonal
and it possesses the following spectrum \beq \label{spinor9}
\mathfrak{D}^{(\alpha)}\psi_{jm+}=\hat{\rho}_{j\alpha}\psi_{jm+} \ \
\ \ \ \mathrm{and}\ \ \ \ \
\mathfrak{D}^{(\alpha)}\psi_{jm-}=-\hat{\rho}_{j\alpha}\psi_{jm-}.
\eeq In (\ref{spinor9}) each eigenvalue  has degeneracy $(2j+1)$.
The above analysis does not directly extend to the kernel of the
operator $\mathfrak{D}^{(\alpha)}$, which is obtained for
$j=|q_\alpha|- \frac{1}{2}$. These zero-modes can be classified by
using the eigenvalues of the operator $(\sigma\cdot\hat{r})$: we
shall denote $\nu_\pm$ the number of zero modes  with eigenvalue
$\pm 1$. Then a simple application of the index theorem shows that
$\nu_+=|q_\alpha|-q_\alpha\ \ $ and $ \ \ \
\nu_-=|q_\alpha|+q_\alpha$, namely for positive $q_\alpha$ we have
only zero modes with negative chirality and viceversa.

We now turn back to the problem of diagonalizing the operator
$\mathcal{S}$ defined in (\ref{spinor6}). The operator $\mathcal{S}$
on the basis provided by the eigenvectors of
$\mathfrak{D}^{(\alpha)}$ is not diagonal. However, on the subspace
spanned by the eigenfunctions of non-vanishing eigenvalue, it
factorizes in an infinite series of two by two matrices. Each matrix
acts on the space generated by the eigenfunctions $\psi_{jm\pm}$ and
it has the form \beq \label{spinor10}
\begin{pmatrix} \rho_{j\alpha}+ i \mu q_\alpha & -\frac{2\pi}{\beta}\left(n+\frac{1}{2}\right)-a_\alpha+i\frac{\mu}{4}\\
-\frac{2\pi}{\beta}\left(n+\frac{1}{2}\right)-a_\alpha+i\frac{\mu}{4}&
-\rho_{j\alpha}+i \mu q_\alpha
\end{pmatrix}.
\eeq Since we  are only interested  in the determinant of the
operator $\mathcal{S}$, we shall not really need to convert this
matrix into a diagonal form, but it is sufficient the evaluate its
determinant \beq \label{spinor11}
\mu^2(j_\alpha+1/2)^2+\frac{4\pi^2}{\beta^2}\left(n+\frac{1}{2}+\frac{\beta
a_\alpha}{2\pi}-i\frac{\beta\mu}{8\pi}\right)^2 \ \ \mathrm{with}\
j_\alpha=|q_\alpha|+\frac{1}{2},|q_\alpha|+\frac{3}{2},\dots, \eeq
and to recall that there are $2j+1$ determinant with the same
value. Then by using the master formula of appendix
\ref{determinants} (with $\rho=2+2|q_\alpha|$, $\eta=1+|q_\alpha|$
and $w=\frac{1}{2}+\frac{\beta
a_\alpha}{2\pi}-i\frac{\beta\mu}{8\pi}$), the contribution of this
part of the spectrum gives \eqref{spinor12} and \eqref{spinor13}.
On the kernel of $\mathfrak{D}^{(\alpha)}$, the operator
$\mathcal{S}$ is instead diagonal and it has the following spectrum
\beq \label{spinor14}
\begin{split}
&\rho_{n\alpha+}=\frac{2\pi}{\beta}\left
[-n-\frac{1}{2}-\frac{\beta
a_\alpha}{2\pi}+i\frac{\beta\mu}{8\pi}+i \frac{\beta \mu
q_\alpha}{2\pi}\right],
\ \ \ \mathrm{with\ degeneracy}\  |q_\alpha|+ q_\alpha,\\
&\rho_{n\alpha-}=\frac{2\pi}{\beta}\left
[n+\frac{1}{2}+\frac{\beta
a_\alpha}{2\pi}-i\frac{\beta\mu}{8\pi}+i \frac{\beta \mu
q_\alpha}{2\pi}\right], \ \ \ \mathrm{with\ degeneracy}\
|q_\alpha|- q_\alpha.
\end{split}
\eeq

Now we have to deal with the regularization ambiguity discussed in
section \ref{fermions}. In our case, all the different choices for
the cuts in the $s$-plane are encoded in the two following
situations:

 \textit{(I)} we regularize  the determinants associated to
the  ``\textit{zero-modes}'' of negative and positive chirality by
choosing  opposite cuts in defining the complex power (one on the
real positive axis and the other on the real negative axis). With
the help of appendix (\ref{determinants}), we then obtain
(\ref{spinor151});

 \textit{(II)} we regularize  the determinants associated
to the  ``\textit{zero-modes}'' of negative and positive chirality
by choosing the same cut in  defining the complex power. A similar
analysis yields (\ref{spinor152}).

The appearance of a Chern-Simons term for case II and the total
fermionic contribution to the effective action for both cases are
discussed in section \ref{fermions} as well.

\sectiono{U(1) truncation of $\mathcal{N}=4$ super Yang Mills}

In this appendix we show that the previous results can be easily
recovered from ${\mathcal N}=4$ super Yang Mills theory on
$\mathds {R}\times S^3$ by a suitable $U(1)$ projection which
gives the maximally supersymmetric theory on $\mathds {R}\times
S^2$.

The single-particle partition function in the representation $R$,
$z^R(x)$, is given by
\begin{equation}\label{spz}
    z^R(x)=\sum_E x^E,
\end{equation}
where $E$ is the energy eigenvalue subtracted of the Casimir
energy, which can be derived for example with the procedure
described in the body of the paper. The eigenvalue $E$ can be
computed most directly by noting that the Laplacian on the sphere
may be written in terms of angular momentum generators which can
be diagonalized by means of generalized spherical harmonics on
$S^3$. The isometry group of $S^3$ is $SO(4)\simeq SU(2)_1\times
SU(2)_2$ and we will need the spherical harmonics for scalars,
vectors and fermions, which will be denoted by $S_{j,m,\bar
m}(\Omega)$, $V_{j,m,\bar m}(\Omega)$ and $F_{j,m,\bar
m}(\Omega)$, respectively. Here $m$ and $\bar{m}$ are the
eigenvalues of $J^3$ and $\bar J^3$ for $SU(2)_1$ and $SU(2)_2$
and $\Omega$ represents the coordinates of $S^3$. We follow here
the notation of \cite{Ishiki:2006rt}. Having determined the
single-particle partition functions on $S^3$ we may then perform a
$U(1)$ projection to derive the single-particle partition
functions on $S^2$. Such projection amounts in a consistent
truncation of $\mathcal{N}=4$ super Yang Mills as discussed in
\cite{Ishiki:2006rt}, and it can be realized by taking into
account that the only modes that actually contribute to the
partition function on $S^2$ are those for which the eigenvalue of
$\bar J_3$ is equal to half the monopole charge. The projection
onto $S^2$ can thus be performed introducing into the
$\mathcal{N}=4$ partition functions a $U(1)$ projection operator
of the form
\begin{equation}
\int^\pi_{-\pi} \, \frac{d\theta}{2\pi} e^{2 i \theta ( \bar
J^3-q)} \label{projector}
\end{equation}
where $2 q$ is the integer monopole charge of the BPS vacua on
$S^2$ and as a notation we shall assume $q\ge 0$.

The projection from $S^3$ to $S^2$ rescales the radius of the
sphere by $1/2$ thus giving an $S^2$ of radius $R=1/2$.

\subsection{Scalars} Scalars on $S^3$ can be expanded in scalar
spherical harmonics $S_{j,m,\bar m}(\Omega)$ where $m$ and
$\bar{m}$
 take the values $-j/2,\ -j/2+1,\dots,\ j/2-1,\ j/2$.
The energy of a scalar on $S^3$ with radius $R_{S^3}=1$,
conformally coupled to curvature, is $E=j+1$. The partition
function for a scalar on $S^3$ then is
\begin{equation}
z_4^{scal.}(x)=\sum_{j=0}^{\infty}\sum_{m=-j/2}^{j/2}\sum_{\bar
m=-j/2}^{j/2}x^{j+1}=\sum_{j=0}^{\infty}(j+1)^2
x^{j+1}=\frac{x(1+x)}{(1-x)^3}
\end{equation}
where the lower index on $z$ denotes the spacetime dimension.

Inserting the projector (\ref{projector}) we easily get the
partition function for a scalar on $S^2$. The scalar partition
function in the presence of a monopole of charge $q$ becomes
\begin{equation}\label{S2scalar}
{z^{scal.}} (x,q)=\sum_{j=0}^{\infty}\sum_{m=-j/2}^{j/2}\sum_{\bar
m=-j/2}^{j/2}\int^\pi_{-\pi} \, \frac{d\theta}{2\pi} e^{2 i \theta
(\bar m-q)}x^{j+1}.
\end{equation}
Performing the sums we end up with an integral
\begin{equation}
 {z^{scal.}} (x,q)=\int_0^\pi\frac{d\theta}{\pi}\frac{x(1-x^2)\cos(2 q\theta)}{(1+x^2-2 x
 \cos\theta)^2},
\end{equation}
that can be easily done and gives
\begin{equation}
\label{scal}
 {z^{scal.}} (x,q)= x^{2 q+1}\left[\frac{(1+x^2)}{(1-x^2)^2}+\frac{2
 q}{1-x^2}\right].
\end{equation}
We can now reintroduce the appropriate dependence on the radius
$R=1/\mu$. Keeping into account that the partition function
(\ref{scal}) is defined on an $S^2$ with radius $R=1/2$, to get
the one with a generic radius $R=1/\mu$ amounts in simply
replacing
\begin{equation}
\label{shift} x^2\to x\equiv e^{-\beta\mu}\
\end{equation}
without having to compute a single determinant.

\subsection{Vectors} Vectors on $S^3$ can be expanded in vector
spherical harmonics $V^{\pm}_{j,m,\bar m}(\Omega)$ which belong to
the representations $(j_1,j_2)=(\frac{j+1}{2},\frac{j-1}{2})$ and
$(j_1,j_2)=(\frac{j-1}{2},\frac{j+1}{2})$, respectively. The
energy for both the representations is given by $E=j+1$. The
partition function on $S^3$ for the $+$ vector component is then
\begin{equation}
z^{vect.}_{4
+}(x)=\sum_{j=1}^{\infty}\sum_{m=-(j+1)/2}^{(j+1)/2}\sum_{\bar
m=-(j-1)/2}^{(j-1)/2}x^{j+1}=\sum_{j=1}^{\infty}j(j+2)
x^{j+1}=\frac{x^2(3-x)}{(1-x)^3}\,; \label{zpv}
\end{equation}
for the $-$ vector component we obviously have the same result
$z^{vect.}_{+}(x)=z^{vect.}_{-}(x)$ and the sum of these two
quantities gives the partition function for a vector on $ S^3$
$$z_4^{vect.}(x)=z^{vect.}_{4+}(x)+z^{vect.}_{4-}(x)=\frac{x^2(6-2x)}{(1-x)^3}.$$

Inserting now the projector (\ref{projector}) into (\ref{zpv}) and
into the analogous one for $V^-$ we get for the $+$ and $-$ vector
components respectively
\begin{eqnarray}
&&z^{vect.}_{+}(x,q)=\sum_{j=1}^{\infty}\sum_{m=-(j+1)/2}^{(j+1)/2}\sum_{\bar
m=-(j-1)/2}^{(j-1)/2}\int^\pi_{-\pi} \, \frac{d\theta}{2\pi} e^{2
i \theta (\bar m-q)}x^{j+1} \cr&& =\int^\pi_{0} \,
\frac{d\theta}{\pi} \frac{\left(3 + x^2 - 4x\cos\theta\right)\cos
2 q\theta}{(1+x^2-2x\cos\theta)^2} =x^{2
q}\left[\frac{x^2(3-x^2)}{(1-x^2)^2}+2 q \frac{x^2}{1-x^2}\right]
\end{eqnarray}
and
\begin{eqnarray}
&&z^{vect.}_{-}(x,q)=\sum_{j=1}^{\infty}\sum_{
m=-(j-1)/2}^{(j-1)/2}\sum_{\bar
m=-(j+1)/2}^{(j+1)/2}\int^\pi_{-\pi} \, \frac{d\theta}{2\pi} e^{2
i \theta (\bar m-q)}x^{j+1} \cr&& =\int^\pi_{0} \,
\frac{d\theta}{\pi} \frac{\left(1+ x^2 - 4x\cos\theta+2\cos
2\theta\right) \cos 2 q\theta}{(1+x^2-2x\cos\theta)^2}.
\end{eqnarray}
For $q=0$ this integral gives
\begin{equation}
z^{(vec.)}_{-}(x,q=0)=\frac{x^2(1+x^2)}{(1-x^2)^2}, \label{zmv0}
\end{equation}
and for $q\ne 0$
\begin{equation}
z^{vec.}_{-}(x,q)=x^{2q}\left[\frac{x^2(1+x^2)}{(1-x^2)^2}-1+\frac{2
q}{1-x^2}\right]. \label{zmv1}
\end{equation}
The limit $q\rightarrow 0$ is discontinuous, in complete agreement
with the computations done in appendix \ref{VECTOR} . Therefore
the sums of the $+$ and $-$ vector partition functions for $q\ne
0$ give
\begin{equation}
z^{vec.}(x,q)=z^{vec.}_{+}(x,q)+z^{vec.}_{-}(x,q)= x^{2
q}\left[\frac{4 x^2}{(1-x^2)^2}-1+ 2
q\left(\frac{1+x^2}{1-x^2}\right)\right]
\end{equation}
whereas for $q=0$
\begin{equation}
z^{vec.}_{-}(x,q=0)=z^{vec.}_{+}(x,0)+z^{vec.}_{-}(x,0)= \frac{4
x^2}{(1-x^2)^2}.
\end{equation}
Again, with the substitution (\ref{shift}) we immediately get back
the results (\ref{vector191},\ref{vector211}).

\subsection{Fermions} Fermions on $S^3$ can be expanded in spinor
spherical harmonics $F^{\pm}_{j,m,\bar m}(\Omega)$ which belong to
the representations $(j_1,j_2)=(\frac{j}{2},\frac{j-1}{2})$ and
$(j_1,j_2)=(\frac{j-1}{2},\frac{j}{2})$, respectively. The energy
for both the representations is given by $E=j+1/2$. Therefore on
$S^3$ we get
\begin{equation}
z^{spin.}_{4
+}(x)=\sum_{j=1}^{\infty}\sum_{m=-j/2}^{j/2}\sum_{\bar
m=-(j-1)/2}^{(j-1)/2}x^{(j+1/2)}=\sum_{j=0}^{\infty}j(j+1)
x^{j+1/2}=\frac{2x^{3/2}}{(1-x)^3} \label{zpf}
\end{equation}
for the $+$ fermion component and the same result for the $-$
fermion component. The sum of these two quantities gives the
partition function for a fermion on $ S^3$
$$z_4^{spin}(x)=z^{spin.}_{4+}(x)+z^{spin.}_{4 -}(x)=\frac{4x^{3/2}}{(1-x)^3}.$$

Inserting the projector into (\ref{zpf}) and into the analogous
one for $F^-$ one gets the partition functions for a $+$ or $-$
spinor on $S^2$
\begin{eqnarray}
&&z^{spin}_{+}(x,q)=\sum_{j=1}^{\infty}\sum_{m=-j/2}^{j/2}\sum_{\bar
m=-(j-1)/2}^{(j-1)/2} \int^\pi_{-\pi} \, \frac{d\theta}{2\pi} e^{2
i \theta (\bar m-q)}x^{j+1/2}\cr&& =\int^\pi_{0} \,
\frac{d\theta}{\pi} \frac{2x^{3/2}\left(1-x\cos\theta\right) \cos
2 q\theta}{(1+x^2-2x\cos\theta)^2}=x^{2
q}\left[\frac{2x^{3/2}}{(1-x^2)^2}+\frac{2 q
x^{3/2}}{1-x^2}\right] \label{z3pf},\end{eqnarray}
\begin{eqnarray}
&&z^{spin.}_{-}(x,q)=\sum_{j=1}^{\infty}\sum_{
m=-(j-1)/2}^{(j-1)/2}\sum_{\bar m=-j/2}^{j/2}x^{(j+1)}
\int^\pi_{-\pi} \, \frac{d\theta}{2\pi} e^{2 i \theta (\bar
m-q)}x^{j+1/2}\cr&& =\int^\pi_{0} \, \frac{d\theta}{\pi}
\frac{2x^{3/2}\left(-x+\cos\theta\right) \cos 2
q\theta}{(1+x^2-2x\cos\theta)^2}=x^{2
q}\left[\frac{2x^{5/2}}{(1-x^2)^2}+\frac{q x^{1/2}}{1-x^2}\right]
\label{z3mf}.
\end{eqnarray}
Adding  (\ref{z3pf}) and (\ref{z3mf}) we get the partition
function for a fermion on $S^2$ in the non-trivial background
\begin{equation}
z^{spin.}(x,q)=z^{spin}_{+}(x,q)+z^{spin.}_{-}(x,q)= x^{2
q}\left[\frac{2 x^2(x^{1/2}+x^{-1/2})}{(1-x^2)^2}+2
q\left(\frac{x^{1/2}(1+x)}{1-x^2}\right)\right].
\end{equation}
With the substitution (\ref{shift}) we get back the result
(\ref{spinor13}).

The complete partition function for our theory can now be
constructed using (\ref{Zx2}). As we showed before the presence of
the monopole background (\ref{vacuabreak}) breaks the original
$U(N)$ invariance to the subgroup $\prod_{I=1}^k U(N_I)$ so that
the positive definite charge $2 q$, appearing in the
single-particle partition functions, is actually a function of the
integers labelling the sectors into which the monopole field
splits. It can be written here as
$$
q\to q_{IJ}=\frac{\left|n_I-n_J\right|}{2}.
$$
We easily get
\begin{equation}\label{Zf}
\mathcal{Z}_A(x)=\int[\prod_{I=1}^k
dU_I]\exp\left\{\sum^k_{I,J=1}\sum_{n=1}^\infty\frac{1}{n}
\left[z^{IJ}_B(x^n)+(-1)^{n+1}z^{IJ}_F(x^n)\right]{\rm
Tr}(U_I^n){\rm Tr}((U_J^\dagger)^n)\right\}.
\end{equation}
Here $k$ is the number of sectors into which the monopole field
splits and reintroducing the appropriate dependence on the radius
$R=1/\mu$ with the substitution (\ref{shift}), we recover for the
bosonic partition function
\begin{equation}
 z^{IJ}_{B}(x,q)= 6 x^{ q_{IJ}+1/2}\left[\frac{(1+x)}{(1-x)^2}+\frac{2
 q_{IJ}}{1-x}\right]+x^{
q_{IJ}}\left[\frac{4 x}{(1-x)^2}-1+ 2
q_{IJ}\left(\frac{1+x}{1-x}\right)\right],
\end{equation}
and for the fermionic one
\begin{equation}
z^{IJ}_{F}(x,q)= 4x^{q_{IJ}}\left[\frac{2
x(x^{1/4}+x^{-1/4})}{(1-x)^2}+2
q_{IJ}\left(\frac{x^{1/4}(1+\sqrt{x})}{1-x}\right)\right].
\end{equation}
We thus reobtain with a very simple and straightforward procedure
the result (\ref{matrix-model12}), up to the constant
(temperature-independent) Casimir contribution. Of course the path
integral approach has the advantages of giving to ${\rm Tr}(U_I)$
the meaning of matrix holonomy along the thermal circle and of
providing an explicit derivation of the Casimir energies.

\section{Solving the matrix model}
\label{D} 
\begin{figure}[htbp]
\begin{center}
\includegraphics[height=8cm,width=12cm]{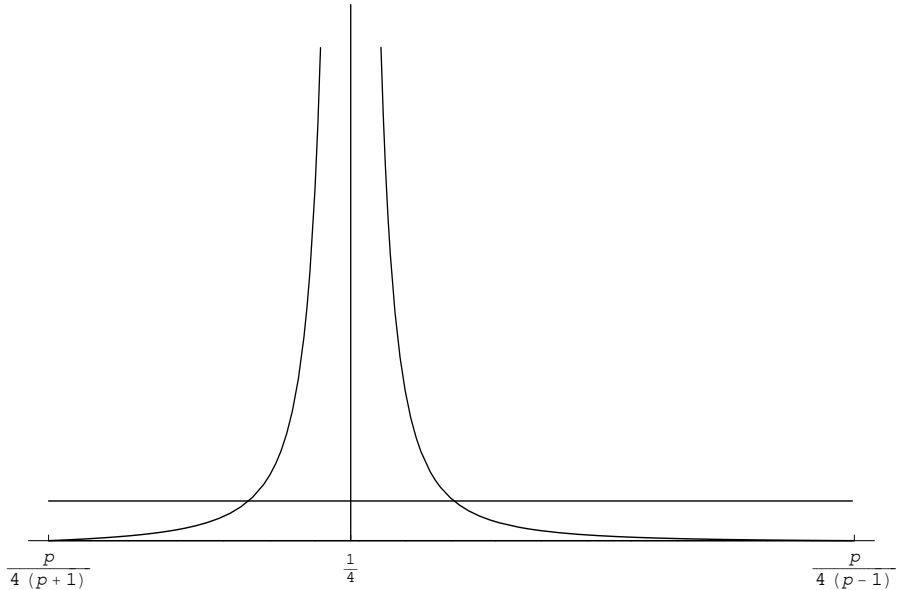}\\
\end{center}
\caption{\footnotesize Plot of the r.h.s of (\ref{caro}). It
diverges for $\rho'=1/4$. For any positive $t$ we have two
solutions. \label{11}}
\end{figure}

The solution of the matrix model in the presence of a
logarithmic interaction has been reduced, in section
\ref{Pennersec}, to solve the non-linear differential equation
(\ref{Pain2}) with a given set of boundary conditions. Surprisingly,
this equation can be explicitly integrated. To achieve this goal, we
first express $\rho(t)$ in terms of $t$ and $\rho'(t)$  by means of
(\ref{Pain2}) \beq \label{D1} \rho(t)=\frac{64 t \rho '(t)^3-16
\left(p^2+t-1\right) \rho '(t)^2+p^2}{16 \rho '(t) \left(4 \rho
'(t)-1\right)}. \eeq Subsequently  we  take the derivative of  with
respect to $t$ on both sides. The differential equation
(\ref{Pain2}) factorizes into two factors, which can be set
separately to zero. In fact, we obtain \beq \left(256 t \rho
'(t)^4-128 t \rho '(t)^3+16 \left(p^2+t-1\right) \rho '(t)^2-8 p^2
\rho '(t)+p^2\right) \rho ''(t)=0, \eeq which implies \beq
\label{sol1} \rho^{\prime\prime}(t)=0\ \ \Rightarrow \ \rho(t)=A t+B
\eeq and \beq \label{sol2} 256 t \rho'(t)^4-128 t \rho '(t)^3+16
\left(p^2+t-1\right) \rho '(t)^2-8 p^2 \rho '(t)+p^2=0. \eeq
Consider first (\ref{sol1}). This solution can only satisfy the
boundary condition $(s)$ associated to the strong-coupling region
(see section \ref{Pennersec}) and thus it seems the natural
candidate to generate ${\cal F}_0^s(t,p)$. This implies that the
integration constant $B$ is fixed to be $-\frac{1}{2} p (p+1)$. The
constant $A$ is instead determined by imposing that (\ref{sol1})
actually solves (\ref{Pain2})\footnote{Since we have taken a
derivative of \ref{Pain2}, we could have potentially added spurious
solutions}. We obtain \beq \rho_s(t)=\frac{p t}{4(p+1)}-\frac{1}{2}
p (p+1). \eeq The  free energy ${\cal F}_0^s(t,p)$ is evaluated by
integrating (\ref{rock}) with the boundary condition
\eqref{boundary0}
 \beq
\label{kkk} {\cal F}_0^s(t,p)= -\frac{1}{2} \left( (\log (4)-3)
p+(p+1)^2 \log (p+1)-p^2\log
(p)\right)+\frac{t}{4(1+p)}-\frac{p}{2}\log(t)~. \eeq As discussed
in section 7.1, (\ref{kkk}) is not the right solution at small $t$
because cannot reproduce the series obtained from the large $p$
expansions. We come now to (\ref{sol2}). It is an algebraic
quartic equation, which determines $\rho'(t)$ as a function of $t$
and $p$. We have four solutions, whose qualitative behavior can be
investigated by writing the inverse function \beq \label{caro}
t=-\frac{\left(4(p-1) \rho '(t)-p\right) \left(4(1+p) \rho '(t)
-p\right)}{16 \rho '(t)^2 \left(4 \rho '(t)-1\right)^2} \eeq and
by drawing its plot. Since we are interested in positive $t$ and
in real solutions, we can focus our attention just on the interval
$[\frac{p}{4(p+1)}, \frac{p}{4(p-1)}]$. The plot is given in fig.
\ref{11}.
We immediately recognize that there are two potential solutions in
this region. At small $t$, they are both finite and their values at
$t=0$ are respectively \beq \rho'_1=\frac{p}{4(p+1)}\ \ \
\mathrm{and}\ \ \ \rho'_2=\frac{p}{4(p-1)}. \eeq For large $t$, both
solutions approach $1/4$ but with opposite subleading term. In fact,
by setting $\rho\sim 1/4+ b t^\alpha$  in \eqref{caro}, we
immediately find \beq \rho'_1(t)=\frac{1}{4}-
\frac{1}{4\sqrt{t}}+O(t)\ \ \ \mathrm{and}\ \ \
\rho'_2(t)=\frac{1}{4}+ \frac{1}{4\sqrt{t}}+O(t). \eeq The actual
functions $\rho_{1,2}(t)$ can be easily recovered by exploiting
\eqref{D1}, which provides $\rho$ in terms of $\rho'$ and $t$ (and
$p$). It is easy to check that the solution $\rho_2(t)$  can be
dropped since its behavior at small and large $t$ is in contrast
with the boundary conditions. Instead, we can identify $\rho_1(t)$
with the weak-coupling solution $\rho_w(t)$ and by integrating
\eqref{FEdef} to evaluate ${\cal F}_0^w(t,p)$. The integration
constant is fixed by requiring that our free energy coincides with
that of the Gross-Witten model for large $t$. The logarithmic
interaction is in fact sub-leading for $t\gg 1$. Nicely the
integration over $t$ can be performed without an explicit knowledge
of $\rho_w(t)$. In fact \eqref{caro} defines an invertible mapping
in the range $\frac{p}{4(p+1)}\le\rho'\le \frac{1}{4}$ (see fig.
\ref{11}). Thus, by means of (\ref{caro}), we can write \beq
\begin{split}
\label{caro2}{\cal F}_0^w(t,p) &=f_w+\int dt \left(\frac{1}{4}-\frac{p^2}{2 t}-\frac{\rho_{w}(t)}{t}\right)=\\
   &=f_w+\int d\rho'_{w}\frac{\left(p^2 \left(4 \rho'_{w}-1\right)^3-64 \left(\rho'_{w}\right)^3\right) \left(p^2
   \left(4 \rho'_{w}-1\right)^3+16 \left(\rho'_{w}\right)^2 \left(4
   \rho'_w
   +1\right)\right)}{32 \left(1-4 \rho'_{w}\right)^2 \left(\rho'_{w}\right)^3 \left(p^2
   \left(1-4 \rho'_{w}\right)^2-16 \left(\rho'_{w}\right)^2\right)}=\\
&=f_w+\frac{1}{32} \left(\frac{8 p^2}{\rho'_{w}}-\frac{p^2}{2
\left(\rho'_{w}\right)^2}+16 \left(\log \left(\rho'_{w}\right)
p^2- 2 p\tanh ^{-1}\left(p+4\left(\frac{1}{p}-p\right)
   \rho'_{w}\right) +\right.\right.\\
   &+\left.\left.\log \left(1-4 \rho'_{w}\right)+\frac{2}{1-4 \rho'_{w}}\right)\right).
\end{split}
\eeq Here $f_w$ is the arbitrary constant of integration. Requiring
that we reobtain the usual Gross-Witten  model for $t\gg 1$ fixes
our constant to be \beq \label{caro3} f_w=-\frac{3}{4}+\frac{1}{4} p
((-3+\log (16)) p-2 \log (p-1)+2 \log (p+1)). \eeq With this choice
expansion of the free energy ${\cal F}_0^w(t,p)$ for large $t$ takes
the form \beq
\begin{split}
{\cal F}_0^w(t,p)=&{\sqrt{{t}}}+\frac{1}{4} \left(\log
\left(\frac{1}{t}\right)-3\right)-\frac{1}{2} p^2
\sqrt{\frac{1}{t}}-\frac{p^2}{4 t}+\frac{1}{24} p^2
   \left(p^2-4\right) \left(\frac{1}{t}\right)^{3/2}+\\
   &+\frac{1}{8} p^2 \left(p^2-1\right) \left(\frac{1}{t}\right)^2-\frac{1}{80} \left(p^2 \left(p^4-20
   p^2+8\right)\right)
   \left(\frac{1}{t}\right)^{5/2}+{\cal O}\left(\frac{1}{t^{3}}\right).
\end{split}
\eeq The leading  behavior is independent of $p$ and it coincides
with that of the Gross-Witten model. The above expression contains
also the result of the semiclassical approximation
(\ref{boundaryinfinity}), up to higher orders in
$p^{2n}/t^{n+m/2}$. We can also compute the small $t$ behavior of
this solution and it is given by \beq
\begin{split}
{\cal F}_0^w(t,p)=&\frac{1}{2} \left(\log (p) p^2+\left(3-\log 4\right) p-(p+1)^2 \log (p+1)\right)+\\
&+\frac{p}{2}\log(t)+\frac{t}{4 p+4}-\frac{p t^2}{32
(p+1)^4}+\frac{(p-1) p t^3}{96
   (p+1)^7}+O\left(t^4\right).
\end{split}
\eeq Surprisingly, we see that ${\cal F}_0^w(t,p)$  satisfies also
the boundary condition \eqref{boundary0} for small $t$ and
reproduces, in that regime, the result of the large $p$ expansion.
In other words, \eqref{caro2} and \eqref{caro3} provide a solution
which smoothly interpolates between the strong and the weak coupling
regime.

\end{document}